\documentclass[fleqn,usenatbib,letters]{mnras}

\usepackage[T1]{fontenc}

\DeclareRobustCommand{\VAN}[3]{#2}
\let\VANthebibliography\thebibliography
\def\thebibliography{\DeclareRobustCommand{\VAN}[3]{##3}\VANthebibliography}

\def\HI{H{\sc i}\, }

\def\Ms{$\textrm{M}_{\odot}$}
\def\kms{$\textrm{km~s$^{-1}$}$}

\def\green{\textcolor[rgb]{0.00,0.00,0.00}}
\def\red{\textcolor[rgb]{0.00,0.00,0.00}}

\usepackage{graphicx}	% Including figure files
\usepackage{amsmath}	% Advanced maths commands
\usepackage{amssymb}	% Extra maths symbols
\usepackage{float}
\usepackage{graphicx}
\usepackage{caption}
\usepackage{subcaption}
\usepackage{newtxtext,newtxmath}
\title[The volume density of gLSB]{The volume density of giant low surface brightness galaxies}

\author[A. S. Saburova et al.]{
Anna S. Saburova,$^{1,2}$\thanks{E-mail: saburovaann@gmail.com (AS)}
Igor V. Chilingarian,$^{3,1}$ Andrea Kulier,$^4$ Gaspar Galaz,$^5$ Kirill A. Grishin,$^{6,1}$  
\newauthor
Anastasia V. Kasparova,$^{1}$ Victoria Toptun,$^{1,7}$ Ivan Yu. Katkov$^{8,1}$
\\
$^1$ Sternberg Astronomical Institute, Moscow M.V. Lomonosov State University, Universitetskij pr., 13,  Moscow 119234, Russia\\
$^2$ Institute of Astronomy, Russian Academy of Sciences, Pyatnitskaya st., 48,  Moscow 119017, Russia\\
$^3$ Center for Astrophysics --- Harvard and Smithsonian, 60 Garden Street MS09, Cambridge, MA 02138, USA\\
$^4$ INAF-Padova Astronomical Observatory, Vicolo dell'Osservatorio 5, I-35122 Padova, Italy\\
$^5$ Instituto de Astrofísica, Pontificia Universidad Católica de Chile, Vicu\~na Mackenna 4860, Santiago 22, Macul, Chile \\
$^6$ APC,   AstroParticule   et   Cosmologie,   Universit\'e   Paris   Diderot,CNRS/IN2P3,  CEA/lrfu,  Observatoire  de  Paris,  Sorbonne  Paris  Cit\'e,\\ 10 rue Alice Domon et L\'eonie Duquet, 75205, Paris Cedex 13, France\\
$^7$ Department of Physics, M.V. Lomonosov Moscow State University, Leninskie gory 1, Moscow, 119991, Russia\\
$^8$ New York University Abu Dhabi, PO Box 129188 Abu Dhabi, UAE\\
}

\date{Accepted XXX. Received YYY; in original form ZZZ}

\pubyear{2022}

\begin{document}
\label{firstpage}
\pagerange{\pageref{firstpage}--\pageref{lastpage}}
\maketitle

% Abstract of the paper
\begin{abstract}
Rare giant low surface brightness galaxies (gLSBGs) act as a stress test for the \green{current} galaxy formation paradigm. To answer the question `How rare are they?' we estimate their volume density in the local Universe. A visual inspection of 120~sq.~deg. covered by deep Subaru Hyper Suprime-Cam data was performed independently by four team members. We detected 42 giant disky systems \red{(30 of them isolated)} at $z\leq0.1$ with either $g$-band 27.7~mag~arcsec$^{-2}$ isophotal radius or four disc scalelengths $4h \geq 50$~kpc, 37 of which \red{(including 25 isolated)} had low central surface brightness ($\mu_{0,g}\ge 22.7$ mag~arcsec$^{-2}$). This corresponds to volume densities of 4.70$\times 10^{-5}$ Mpc$^{-3}$ for all galaxies with giant extended discs and 4.04$\times 10^{-5}$ Mpc$^{-3}$ for gLSBGs, which converts to $\sim $12,700 such galaxies in the entire sky out to $z<0.1$. These estimates agree well with the result of the EAGLE cosmological hydrodynamical simulation. Giant disky galaxies represent the large-size end of the volume density distribution of normal-sized spirals, suggesting the non-exceptional nature of giant discs. We observe a high active galactic nucleus fraction among the newly found gLSBGs.
\end{abstract}

% Select between one and six entries from the list of approved keywords.
% Don't make up new ones.
\begin{keywords}
galaxies: kinematics and dynamics, galaxies: evolution, galaxies: formation
\end{keywords}

%%%%%%%%%%%%%%%%%%%%%%%%%%%%%%%%%%%%%%%%%%%%%%%%%%

%%%%%%%%%%%%%%%%% BODY OF PAPER %%%%%%%%%%%%%%%%%%

\section{Introduction}\label{intro}

Giant low surface brightness disc galaxies (gLSBGs) are rare objects with a low surface brightness disc (the central $B$-band disc surface brightness $\mu_0>22$ mag/arcsec$^2$) and radii up-to 130~kpc, dynamical and stellar masses on the order of $10^{12}$~{\Ms} and $10^{11}$~{\Ms} respectively ~\citep{Boissier2016, saburovaetal2021}. A remarkable case of this galaxy class,  Malin~1, was discovered by \citet{Bothun1987}, more than 30 years ago \citep[see a relatively recent description of the morphological features of Malin 1 in][]{Galaz2015}. Since that discovery, massive efforts were put into understanding of the possible formation scenarios of these extreme systems \citep[see][and references therein]{Penarrubia2006,Mapelli2008,Lelli2010,Reshetnikov2010,Kasparova2014,Galaz2015, Boissier2016, Hagen2016,Saburova2018,saburovaetal2019, saburovaetal2021}. According to \citet{saburovaetal2021}, in many cases, observational data suggest an external source of gas for the build-up of giant LSB discs. For example, some gLSBGs were likely formed in a two-stage process, in which the extended LSB disc grows by cold gas accretion from the cosmic web on a preexisting `normal' high-surface brightness (HSB) elliptical or early-type disc galaxy. Alternatively, an extended disc can be fostered by minor mergers with gas-rich satellites \citep{Penarrubia2006}. To preserve the angular momentum, they need to be orbiting the gLSBG progenitor in the form of a regular structure, i.e. a disc or a ring of satellites as those found around M~31 \citep{2013Natur.493...62I} and Cen~A \citep{2018Sci...359..534M}, \green{which was also found in cosmological simulations \citep{SantosSantos2022}}. Finally, in-plane major mergers of `normal' giant disc galaxies can form gLSB discs \citep{Saburovaetal2018}. Also, the formation of extended discs could be related to a peculiar dark matter (DM) halo having large radial scale and low central density \citep{Kasparova2014, Saburova2018}.

These enormous regularly rotating stellar systems act as a stress-test for the currently considered scenarios of galaxy formation: in the reference EAGLE cosmological hydrodynamical simulation \citep{eagleschaye}, galaxies with stellar masses of $\sim 10^{11}$~\Ms~(comparable to \green{those} of gLSBGs) have undergone significant mass growth from mergers \citep{Kulier2020}. Major mergers, however, tend to overheat and destroy discs \citep{Rodriguez-Gomez2017,Kulier2020} and they can survive only if the merger parameters (orbital configuration, energy, and spin) are fine-tuned (e.g. \citealp{Zhuetal2018} \green{studied} a gLSBG in the IllustrisTNG simulation). \green{Zhu et al. (subm.) found $\sim$200 gLSBGs in IllustrisTNG, demonstrating the capability of the new cosmological simulations to reproduce gLSBGs. However, the question whether the observed number of gLSBGs agrees with simulations still persists.}

\begin{figure}
\includegraphics[width=\hsize]{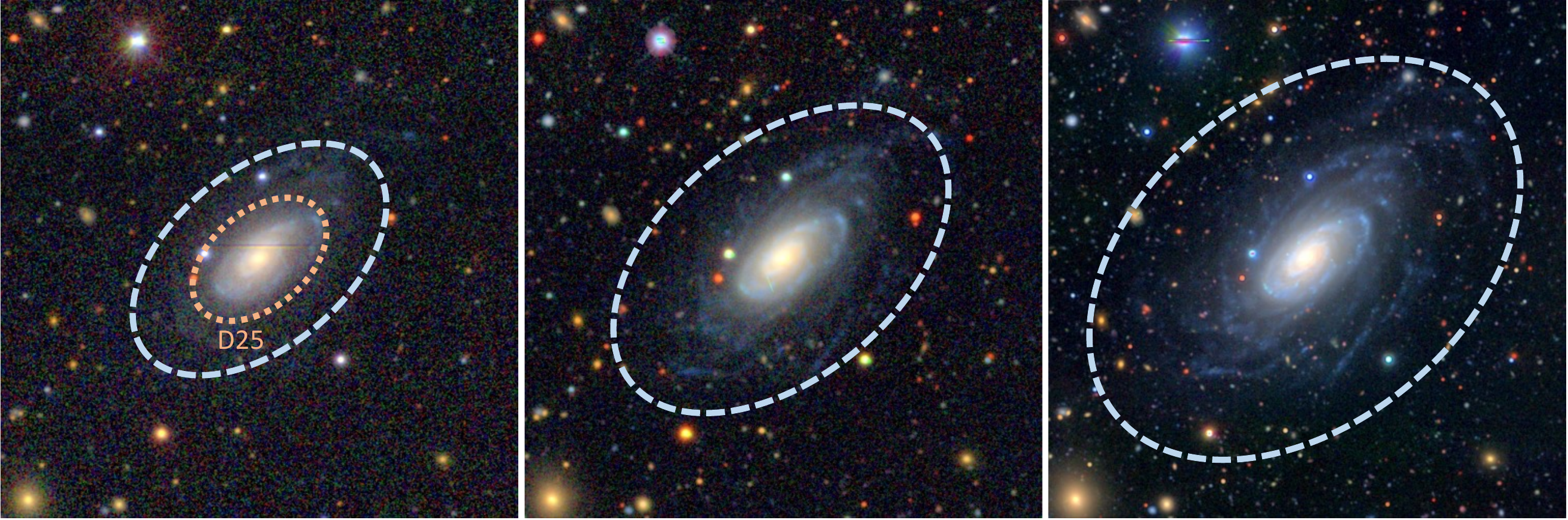}
\caption{$gri$ colour images of NGC~4202 from SDSS (left), DECaLS DR9 (middle), and HSC-SSP DR2 (right) demonstrating how a visually estimated diameter depends on the survey depth. The 25-th isophotal diameter ($B$ band) according to HyperLeda is displayed in light orange for comparison. Each panel has an angular size 250$\times$250~arcsec corresponding to 108$\times$108~kpc at the galaxy distance $d=86$~Mpc according to NED.\label{fig_ngc4202}}
\end{figure}

Presently, the number of known gLSBGs from observations remains of the order of several dozen \citep{saburovaetal2021} and no attempts have been made to compute the statistical frequency of these galaxies in the Universe. Moreover, only a few ($<10$) of them have been studied in depth, with conclusions made regarding their potential evolutionary channels. Even in this limited sample, no single formation scenario can explain the observed properties of all galaxies. To conclude which scenarios prevail over the others or, perhaps, which combinations of scenarios are relevant (e.g. peculiar DM halo + cold accretion), one needs to obtain a statistically significant sample of gLSBGs and compare it against cosmological simulations. 

Many gLSBGs remain undiscovered because of the low surface brightness of their giant discs: they look like passive elliptical or lenticular galaxies if the image is not deep enough. Two well-known examples of such systems are UGC~1382 and UGC~1922 \citep{Hagen2016, Saburovaetal2018}. With the advent of wide-field photometric surveys, the number of known gLSBGs is expected to increase \citep{Hagen2016}. So, now we can hope to answer the following questions: How many early-type galaxies do conceal extended discs of low surface brightness? How many giant LSB systems still remain undiscovered?

The adoption of an appropriate definition of galaxy size is crucial for LSB systems. The isophotal size represents a more robust approach than widely used half-light or Petrosian/Kron radii because it is directly linked to the surface brightness value rather than to the galaxy light profile, i.e. the inner, high-surface brightness areas of the light profile do not affect $r_{\mathrm{iso}}$ while they do affect $r_e$ and $r_{\mathrm{Petro}}$. The main reason why LSB galaxies are not detected in many surveys is because the limiting isophote is too bright. Another well-known factor precluding the detection of faint extended structures is the degree of uniformity of the background sky, which is reflected instrumentally on the accuracy of flat-fielding and background subtraction \citep[see, e.g.][]{Euclid2022}.

Deep images obtained with the 8-m Subaru telescope Hyper Suprime-Cam \citep[HSC,][]{2018PASJ...70S...1M} present the best opportunity to search for gLSBGs in a large area of the sky and study their structure. In Fig.~\ref{fig_ngc4202} we present colour composite cutouts of NGC~4202 from SDSS DR8 \citep{Aihara2011}, \green{DESI} Legacy Survey DR9 \citep{Dey2019} and HSC  \red{Subaru} Strategic \green{Program} DR2 \citep{hsc2019}, which clearly show that a low-surface brightness disc barely visible in SDSS images becomes more evident in DECaLS images and virtually impossible to miss in HSC data by visual inspection. Due to the low number of known gLSBGs in the HSC footprint, their automatic search with e.g. artificial neural networks is impossible because the training set is too small. Therefore, we decided to undertake a visual search for gLSBGs using HSC data by several team members to minimize the subjectivity of the selection. Our ultimate goal is to estimate the volume density of gLSBGs with disc radii $r_d\geq50$~kpc and compare it to the value derived from cosmological hydrodynamical simulations.

\section{Methodology} \label{method}
We chose a 120~sq.~deg.\ patch of the HSC \red{ Subaru} Strategic \green{Program} DR2 survey (28\degr < RA < 40\degr, -7\degr < Dec < 3\degr), which contains one historically known gLSBG (UGC~1382) and is relatively compact in the sky so that we can neglect the variations of the Galactic foreground extinction across the area. Four members of the team (AS, IC, KG, AK) visually inspected colour composite HSC-SS DR2 images available at the DESI Legacy Survey website. We used the zoom level 13 (0.5~arcsec~pix$^{-1}$) or higher. The characteristic features of gLSBG candidates compared to nearby `normal' spirals are thin and tight spiral arms and typically a redder bulge (the effects of an old metal-rich stellar population and a slightly higher redshift, which already at $z=0.06$ would cause a visual reddening of galaxies in colour images). We then looked for available redshifts in the Reference Catalog of galaxy SEDs\footnote{\url{https://rcsed2.voxastro.org}} \citep{Chilingarian2017}, NASA/IPAC Extragalactic Database (NED)\footnote{\url{https://ned.ipac.caltech.edu/}}, and SIMBAD\footnote{\url{https://simbad.u-strasbg.fr/Simbad}} \citep{Wenger2000} and retained the objects with visual radii $r>40$~kpc. For statistical analysis we restrain our sample to $z<0.1$ because at higher redshifts both cosmological dimming and $k$-corrections start affecting the completeness of the visual selection. Each of the 4 team members performed the search independently, then the lists of candidates were compared and merged. IC found 32 galaxies, 31 of which where found by at least one other participant, AS found 46 galaxies (31 matches with others participants), AK found 39 galaxies (30 matches) and KG used a lower size threshold ($\sim$30~kpc) and found 85 objects (51 matches). The visual estimates of galaxy radii varied by 15--21~per~cent among the participants and, hence, required a more quantitative approach using surface photometry.

We used the {\sc ellipse} task in the {\sc python photutils} library \citep{photutils}. We estimated the residual sky background using the {\sc Background2D} task and subtracted it from the data. We masked out the foreground objects with an iterative procedure. First, we generated the foreground masks based on the Legacy Survey tables ls\_dr9sv.tractor\_primary\_n, ls\_dr9sv.tractor\_primary\_s. Then we calculated the residuals, convolved \green{them} with a Gaussian and masked the regions where the residual was higher than the threshold (0.9 ADU~pix$^{-1}$). In the final, third iteration we lowered the threshold to 0.2 ADU~pix$^{-1}$. The initial guesses for the ellipticity and positional angles for the galaxies were extracted from the Legacy Survey tables ls\_dr9sv.tractor\_primary\_n and ls\_dr9sv.tractor\_primary\_s. We applied the correction for Galactic extinction according to \citet{Schlafly2011} and cosmological dimming by $(1+z)^4$. We assumed the following cosmological parameters: $H_{0}=71$~km~s$^{-1}$~Mpc$^{-1}$, $\Omega_{\rm{m}}=0.27$ and $\Omega_{\rm{\Lambda}}=0.73$.  At the end, we obtained the radial profiles of the surface brightness, ellipticity and positional angle. As a galaxy size, we adopt the isophotal radius of 27.7~mag~arcsec$^{-2}$ in the $g$-band isophote, which corresponds to the 28.0~mag~arscec$^{-2}$ in the $B$-band used in analysis of numerical simulations \citep{Kulier2020} assuming $g-r=0.3$~mag and the transformations from \citet{Jester2005}.

\begin{figure}
\centering
\includegraphics[width=0.32\hsize]{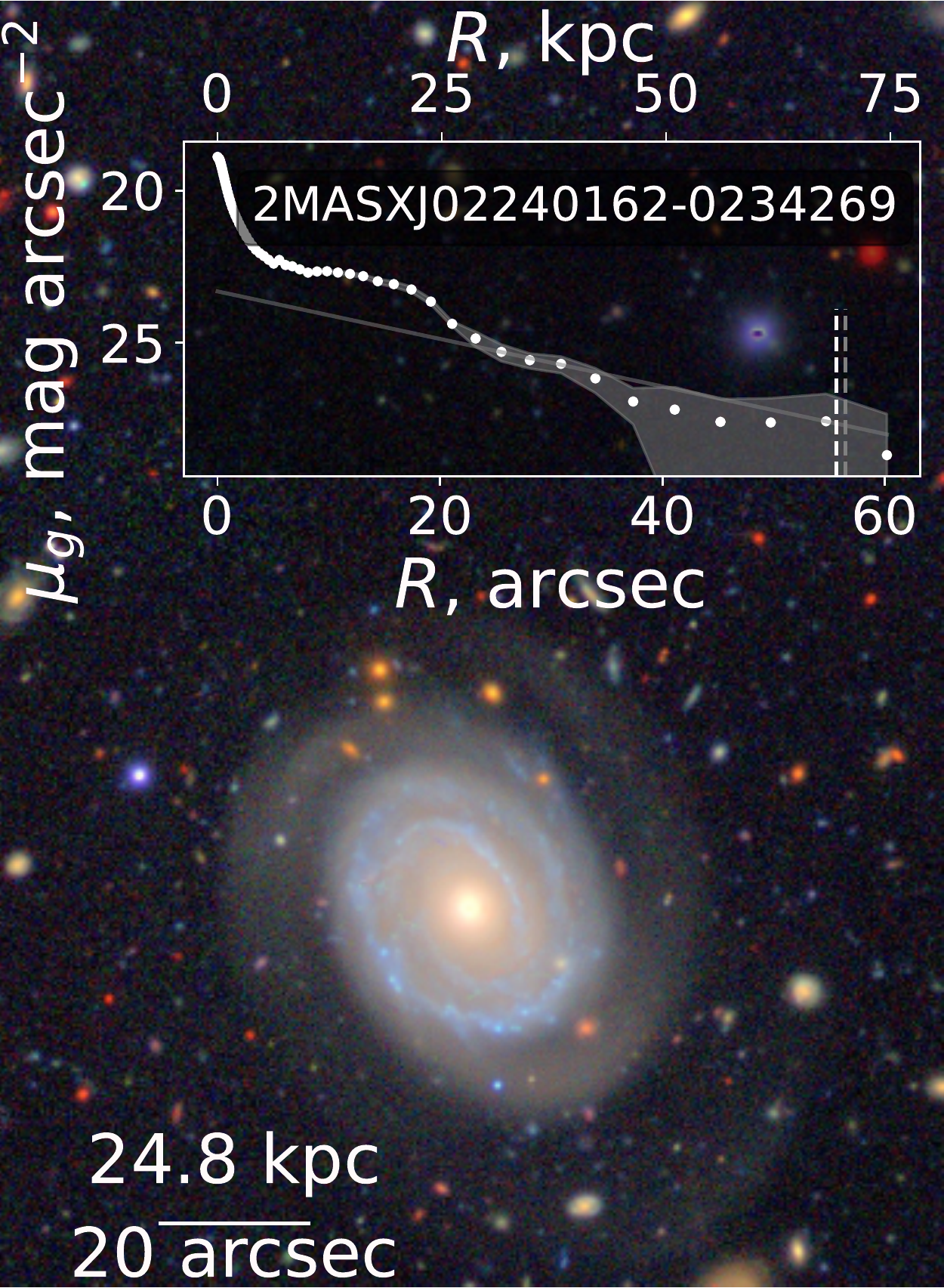}
\includegraphics[width=0.32\hsize]{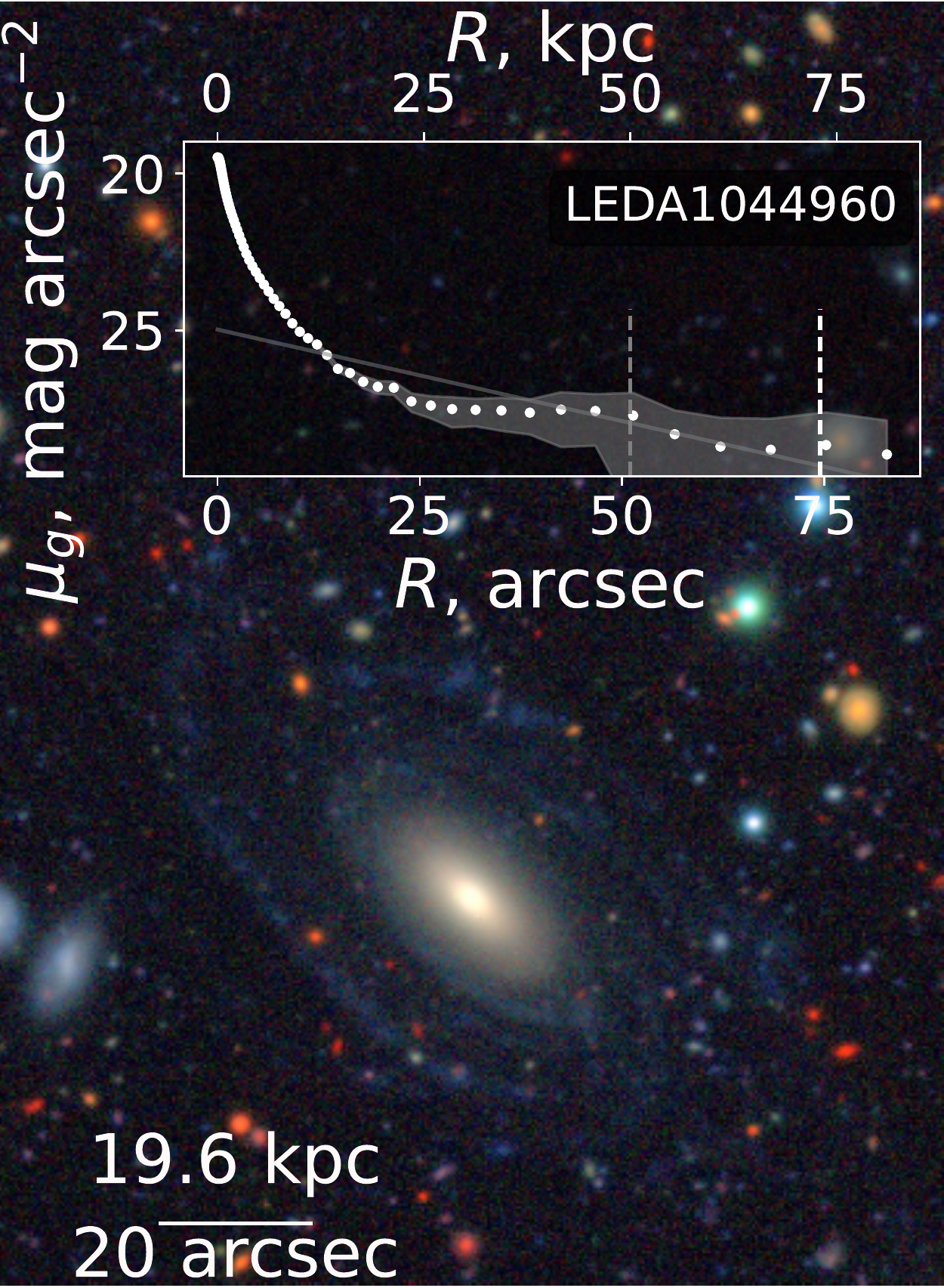}
 \includegraphics[width=0.32\hsize]{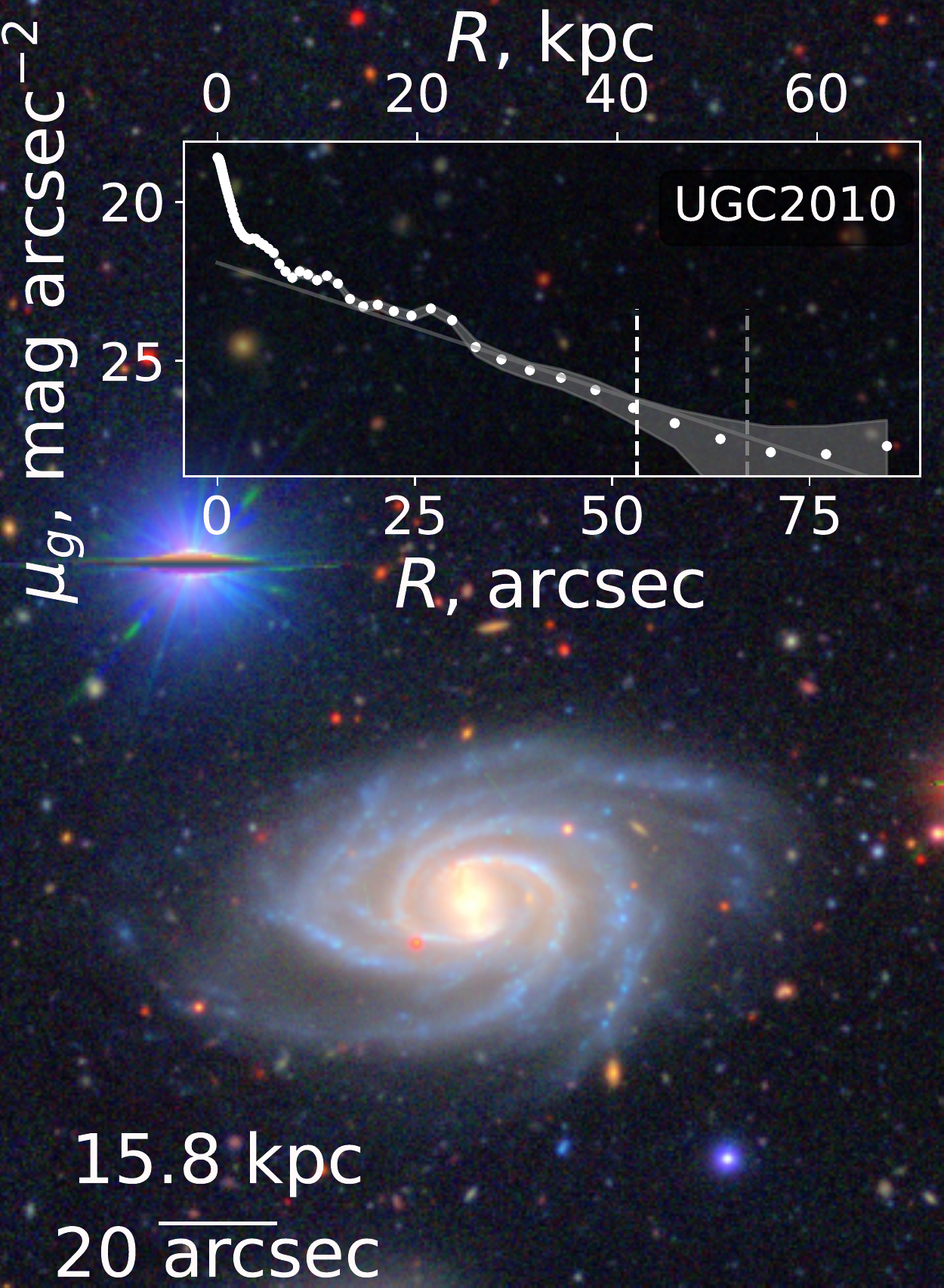}\\
\includegraphics[width=0.32\hsize]{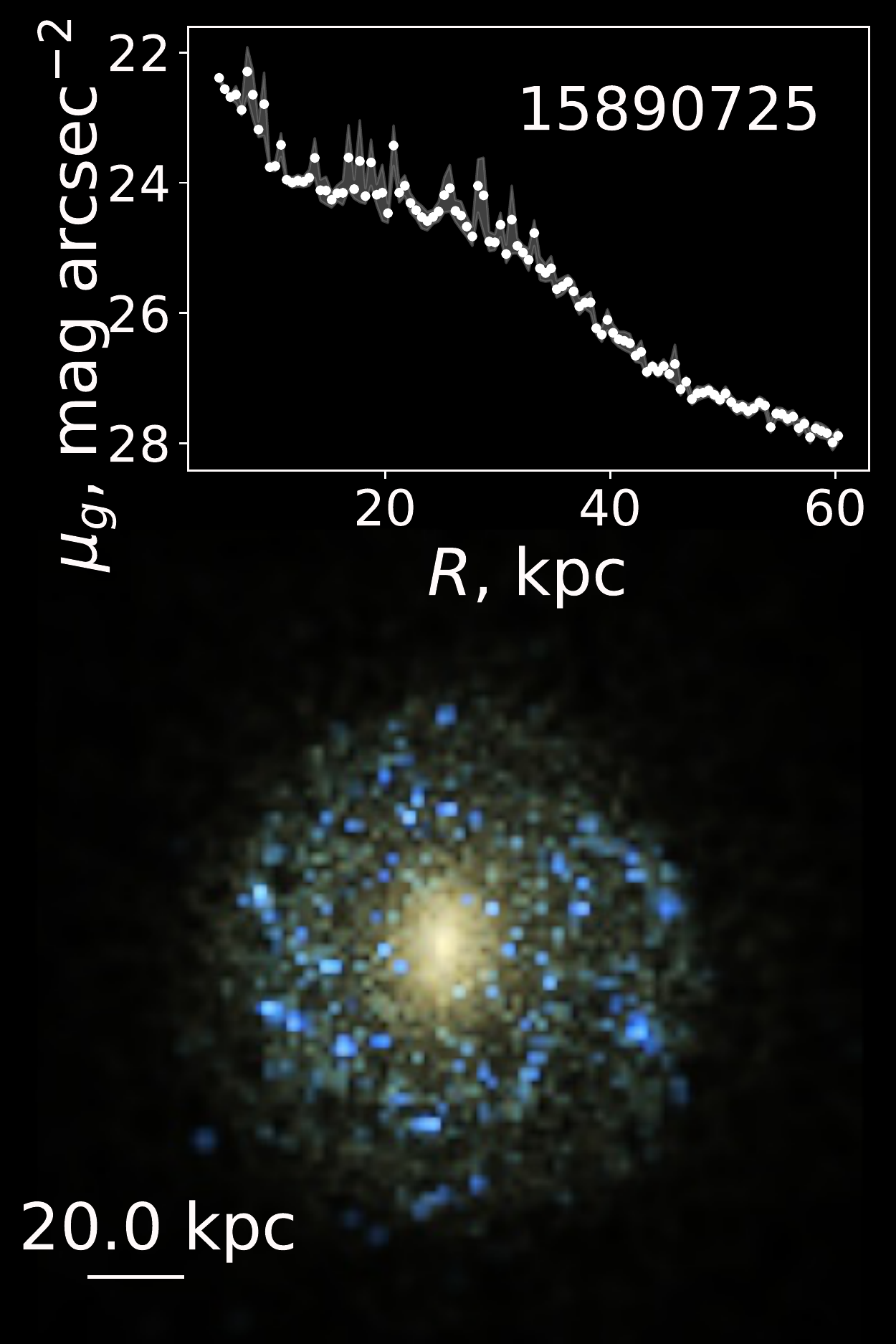}
\includegraphics[width=0.32\hsize]{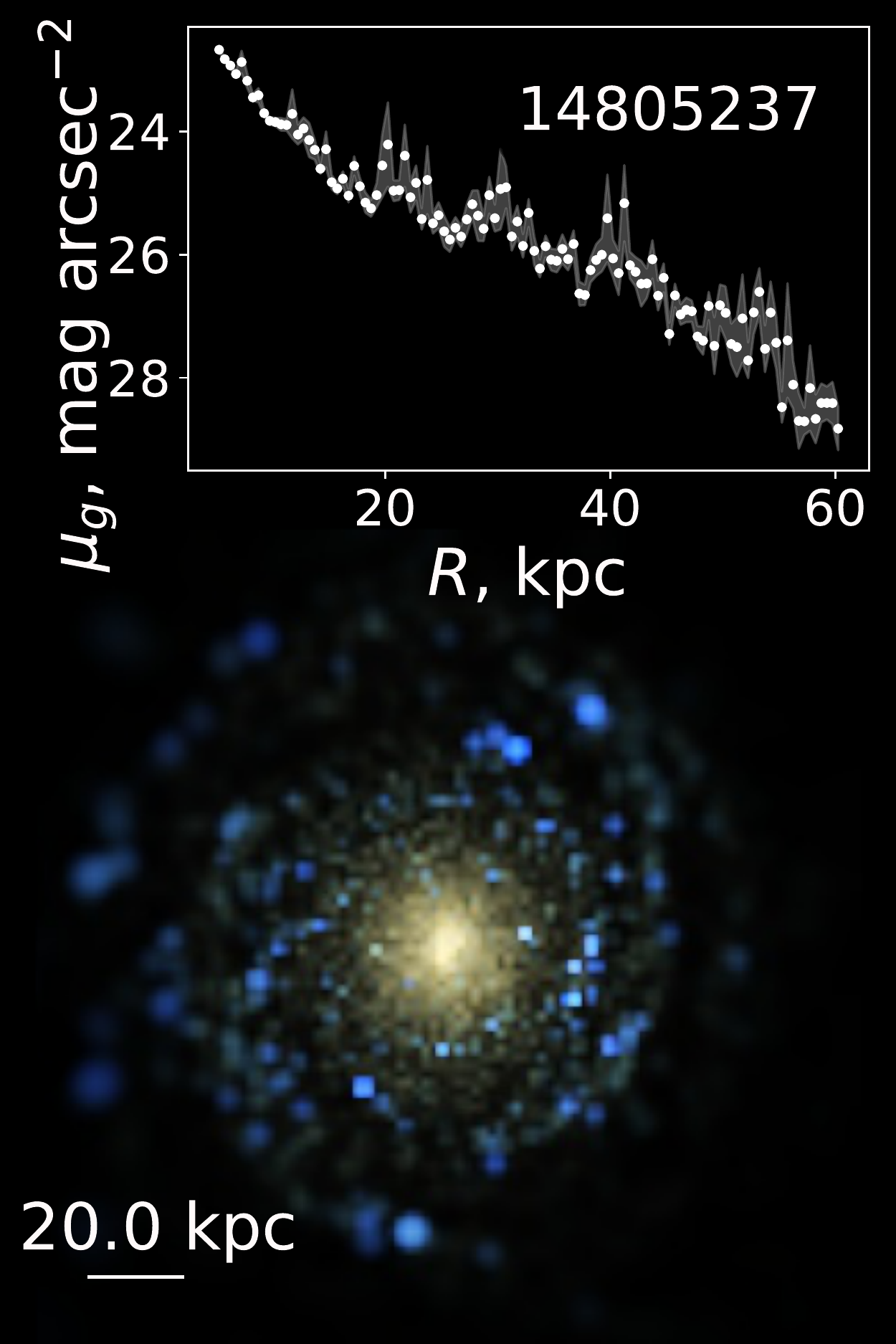}
\includegraphics[width=0.32\hsize]{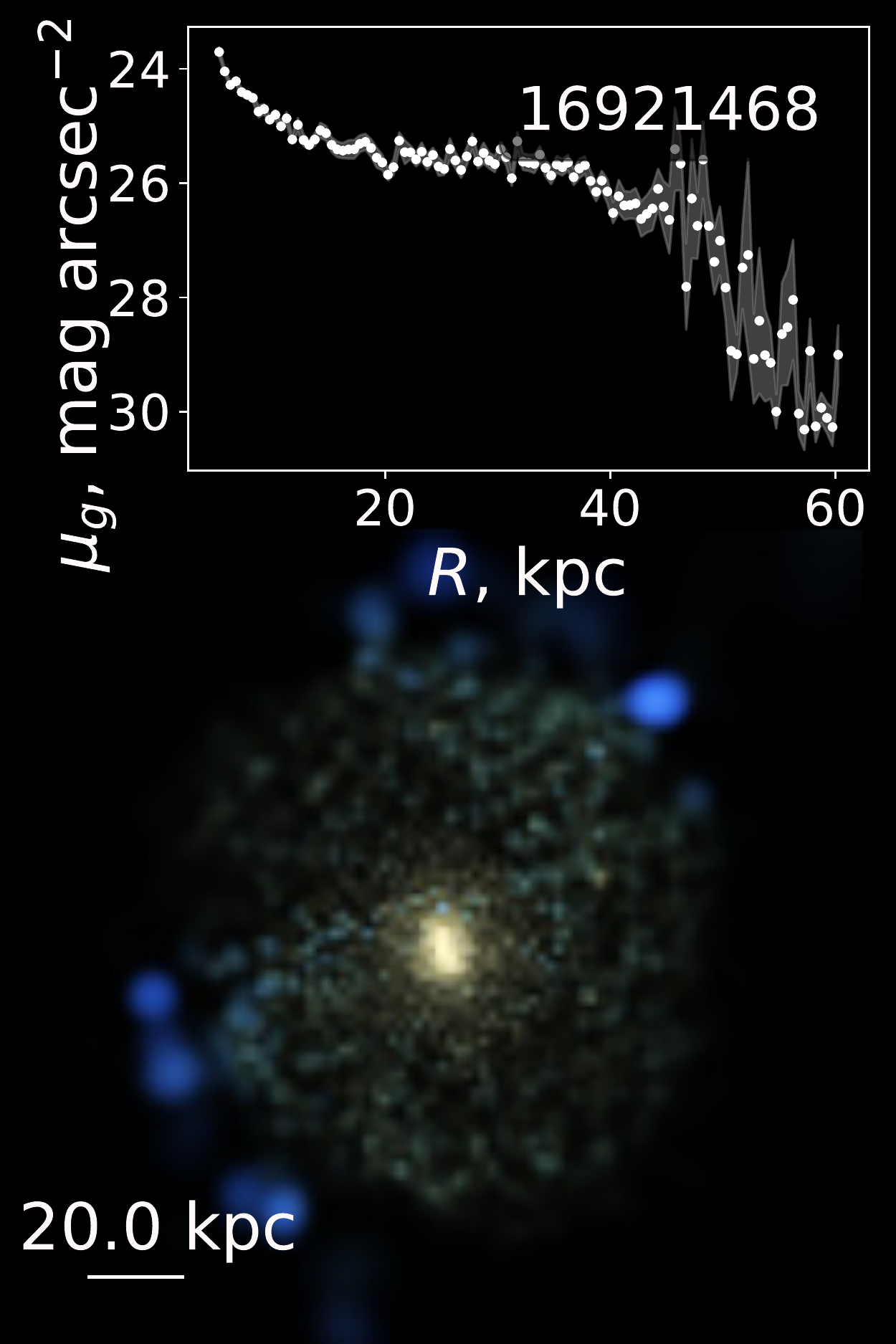}
 \caption{\textit{Top row:} HSC colour composite image of giant discs found by our method. Insets contain the reconstructed $g$-band light profiles with overploted exponential disc profile, interpolated from the outer parts of the galaxies. White and grey vertical lines give the positions of $R_D$ and $R_{27.7}$ correspondingly. The first two galaxies are gLSBGs and UGC~2010 is a giant HSB disc. \textit{Bottom row:} examples of models with extended discs from the EAGLE reference simulation, with inset panels showing {\it g}-band surface brightness profiles. The surface brightness maps were produced by smoothing over the $ugr$ luminosities of the star particles, which are $\sim 10^{6}~\mathrm{M_{\odot}}$ each. The inner 5 kpc of the surface brightness profiles is excluded to ensure that we are outside the convergence radius \citep{eagleludlow}.}
\label{fig_examples}
\end{figure}

\section{Results}\label{s_sample}

\subsection{The sample and the volume density of giant discs}
The isophotal analysis revealed 52 galaxies having $r_{\mathrm{iso,}28B}>50$~kpc.  The radii obtained from the analysis exceed those estimated by eye by our team members by 10--24~per~cent.  The next step was to derive the disc scalelength ($h$) and the central surface brightness ($\mu_0$) for each object, because some of the 52 galaxies resembled oversized `normal' HSB spirals. We assumed that the radial light profiles in the surface brightness range $\mu_g=27.6 \dots 25.1$~mag are adequately described by a single exponential profile, which then yields both $h$ and $\mu_0$ by extrapolation to the galaxy centre. This surface brightness range was chosen iteratively by visual inspection of the light profiles to find the best compromise between the effects of noise and background subtraction on the low end and the complexity of galaxy structure (e.g. spiral arms, H{\sc ii} regions) affecting the brighter end.

Since the galaxies in the sample have different inclination angles, we deprojected the surface brightness profiles using a simple geometric assumption for the thin disc without internal extinction, which is model dependent. We used the ellipticity from the isophotal 25th {\it g}-band isophote.  The correction \green{led} to the decrease of the isophotal radii in the more inclined galaxies. In some cases the galaxies came outside the limit $r_{\mathrm{iso,}28B}>50$~kpc. Some of the outliers still appeared to have giant discs if we used the alternative disc radius definition $r_D=4h$ \citep[][]{Kregel2004}, which is more suitable for LSB galaxies. Thus we decided to use the milder restriction $r_D\ge 50$~kpc or $r_{\mathrm{iso,}28B}\ge 50$~kpc to form the final sample of giant disky galaxies, and came up with 42 galaxies.   Among the sample of 42, we found 27 galaxies that satisfy both criteria and 37 galaxies with the outer discs having deprojected $\mu_{0,g}>22.7$~mag~arcsec$^{-2}$, which corresponds to $\mu_{0,B}>23.0$~mag~arcsec$^{-2}$. All 37 LSB galaxies have disc radii $r_D\ge50$~kpc. We considered the area of 120~sq.~deg. out to $z<0.1$, which corresponds to a volume of $9.2\times 10^{5}\, \mathrm{Mpc}^{3}$ and a volume density of $(4.7 \pm 0.64)\times 10^{-5}\, \mathrm{Mpc}^{-3}$ for all the galaxies having $r_{\mathrm{iso,}28B}\ge 50$~kpc or $r_D=4h\ge50$~kpc and $(4.04 \pm 0.7)\times 10^{-5}\, \mathrm{Mpc}^{-3}$ for gLSBGs. 

In Fig.~\ref{fig_examples} (top row) we present examples of 2 gLSBGs and 1 extended HSB disky galaxy discovered by our method and their light profiles. The 39 remaining galaxies  as well as the coordinates, redshifts, and structural properties of all 42 galaxies are presented in supplementary material in  Table~\ref{tab_sample_all} and Figs \ref{fig_examples_all},\ref{fig_examples_hsb}. We used the $g-r$ color and absolute magnitude $M_g$ from the Dark Energy Survey \citep{2018ApJS..239...18A}, $k$-correcting the magnitudes according to \citet{2010MNRAS.405.1409C}. HyperLeda reports the morphology for most galaxies as early-type spirals (Sbc or earlier), and some are mis-classified as ellipticals (e.g. UGC~1382), which indicates the presence of prominent bulges. Bars are noted in 3 out of 42 galaxies. However, further visual inspection of HSC images revealed 8 barred galaxies among 42 (a fraction of 1/5.25) including 6 gLSBGs (1/6.2), where bars are found half as frequently as in HSB galaxies \citep[1/3.2 as found by][]{Saburova2018}. We also searched for \HI data for the considered galaxies in ALFALFA \citep{Haynes2018} and HIPASS \citep{wong2006} and found measurements only for UGC~1382 and 2MASXJ02015377+0131087 corresponding to \HI masses $\log (M_{HI}/$\Ms$)$ of 10.2 and 10.3 respectively. The former value agrees with that from \cite{Hagen2016}.

\subsection{Numerical simulations}

We can now compare our results to cosmological hydrodynamical simulations to verify how well the models can reproduce the formation of extended galactic discs. \citet{Zhuetal2018} found  one analogue of Malin~1 in the (100~Mpc)$^3$ IllustrisTNG simulation, \green{ and very recently (Zhu et al. subm.) identified 203 galaxies resembling gLSBGs in IllustrisTNG}. %\citet{Kulier2020} found several gLSBG analogues in the (100 Mpc)$^3$ EAGLE reference simulation having sizes comparable to UGC~1382, proving that the formation of such systems can be reproduced by simulations, but a statistical frequency estimate was beyond the scope of their study. 
\green{ Because Zhu et al. (subm.) used a different selection criterion by first choosing the galaxies with extended gaseous discs,  which was not possible in our project because the \HI data for the considered candidates were not available, we apply the same selection to simulations that we used for observations.} Here we chose all $z=0$ galaxies in the EAGLE reference simulation with $r_{\mathrm{iso,}27.7g}>50$~kpc or  $4h>50$~kpc that have \green{visible spiral structure at $\sim50$ kpc} and significant rotation \green{(the median orbital circularity \citep{Abadi2003} was $>0.4$ in some 15~kpc range between 35 and 60~kpc)}. 44 galaxies satisfied these criteria, 15 of which satisfied both criteria. \green{If we neglected the rotation criterion it would add 10 \red{objects} with complex stellar kinematics and disturbed morphology to the sample.} Four out of 44 galaxies have bright haloes, which we do not observe, at least for highly inclined systems in which it is possible to detect such features, and four other galaxies do not have star formation which would make it impossible to detect such systems in observations.  The 44 model galaxies correspond to a volume density estimate of $4.4\times$ 10$^{-5}\, \mathrm{Mpc}^{-3}$ (and $1.5\times 10^{-5}\, \mathrm{Mpc}^{-3}$ for the systems which satisfy both criteria). In Fig. \ref{fig_examples} (bottom row) we demonstrate examples of the simulated galaxies with giant discs together with their {\it g}-band surface brightness profiles. %EAGLE thus reproduces well the number of galaxies with giant discs, but under-predicts by 2 times the number of systems that have both a giant isophotal radius and disc scalelength. 
%\green{It should be noted that the volume density according to the number of gLSBGs found by Zhu et al. is by an order higher $4.8\times$ 10$^{-4}\, \mathrm{Mpc}^{-3}$, however this number is more correctly to compare to the number of galaxies with available \HI data. The deep imaging of extended \HI discs might discover more gLSBGs. }

\subsection{Active galactic nuclei}\label{agn}
\citet{Schombert1998} and \citet{saburovaetal2021} noticed that gLSBGs tend to have higher AGN frequencies in comparison with LSB galaxies of moderate sizes and HSB disky galaxies. To check this possibility for the galaxies with spectral data we used the diagnostic diagrams \citep{BPT} available in the Reference Catalog of galaxy SEDs. We also searched for X-ray detections and estimated X-ray luminosity for 11 galaxies using X-ray fluxes from the 4XMM-DR10 \citep{2020A&A...641A.136W} and SWIFT 2SXPS \citep{2020ApJS..247...54E} catalogs. For the objects without catalog entries we extracted upper limits on the X-ray flux from XMM, Swift and ROSAT using the Upper Limit Server \citep{2022A&C....3800531S, 2022A&C....3800529K}.  In the 'BPT' column of Table~\ref{agnenv_all} we summarize the available spectral information (n/a designates galaxies without spectra). `AGN', 'Composite', and `SF' define the classification of a spectrum on a BPT diagram as pure non-photo-ionization, composite, and star-formation related. Galaxies where spectra were available but at least some emission lines required for the BPT classification were not detected are marked with `?'. According to Table~\ref{agnenv_all}, the AGN frequency varies from 19 to 40 per~cent for gLSBGs depending on which AGN criterion we use. The same conclusion is valid for the extended galaxies in general (both LSB and HSB), 17--36 per~cent. These frequencies are lower limits because the X-ray and optical data were not available for all our galaxies. The derived AGN frequency is slightly higher than that in HSB galaxies, 19~per~cent, and is significantly higher than that found for moderate-sized LSBGs, 5~per~cent \citep{galaz2011}. 

\subsection{Environment}\label{env}
In the last column in Table~\ref{agnenv_all} we provide a brief assessment of the environment of the galaxies in our sample. We checked the group membership according to 3 group catalogs \citep{tully2015, Saulder2016, Tempel2018}. We also visually inspected the adjacent sky regions searching for galaxies missed from the group catalogs having close velocities in NED. If we call a galaxy isolated when no companions are present out to 500~kpc in projection, 30 out of 42 galaxies with extended discs pass this definition (71~per~cent) including 25 of 37 gLSBGs (68~per~cent), which is close to that observed for normal HSB discs (70$\pm1$~per~cent) and somewhat lower than the value for moderate-sized LSBGs \citep[76$\pm2$~per~cent][]{galaz2011}. Another aspect of the environment is the membership in a close pair. We applied the close pair selection criterion from \citet{Ventou2019} and concluded that only UGC~1382 possesses a compact satellite projected on its disc with a close velocity \green{that} could be classified as a close pair. All remaining galaxies in the sample do not form close pairs. This appears to be substantially lower than reported by \citet{Ventou2019} for a sample of 2483 galaxies with spectroscopic redshifts, where they identified 366 close pairs spread over a large range of redshifts and stellar masses. In EAGLE, spirals with $4h < 50$~kpc have isolated fractions of 67 $\pm$2 per cent, where we consider a simulated galaxy isolated if it has no neighbors brighter than -19.8 mag in the $r$ band within 500~kpc in projection and 500~\kms ~in velocity difference. 70$\pm$5 per cent of gLSBGs are isolated using the same criterion. These numbers are in agreement with observations.

\section{Discussion}\label{discussion}

It turns out that gLSBGs are not as rare as was thought when only a handful of such systems were known. According to our volume density estimate, about 13 thousands of such galaxies are expected in the entire sky out to $z\leq0.1$. The obtained volume density for galaxies with large isophotal radii is in agreement with that expected from the EAGLE reference simulation. %\green{At the same time there are even by an order more gLSBGs candidates in TNG100, Zhu et al. in prep.}

\begin{figure}
\centering
%\hspace{-0.5cm}
\includegraphics[width=0.87\hsize]{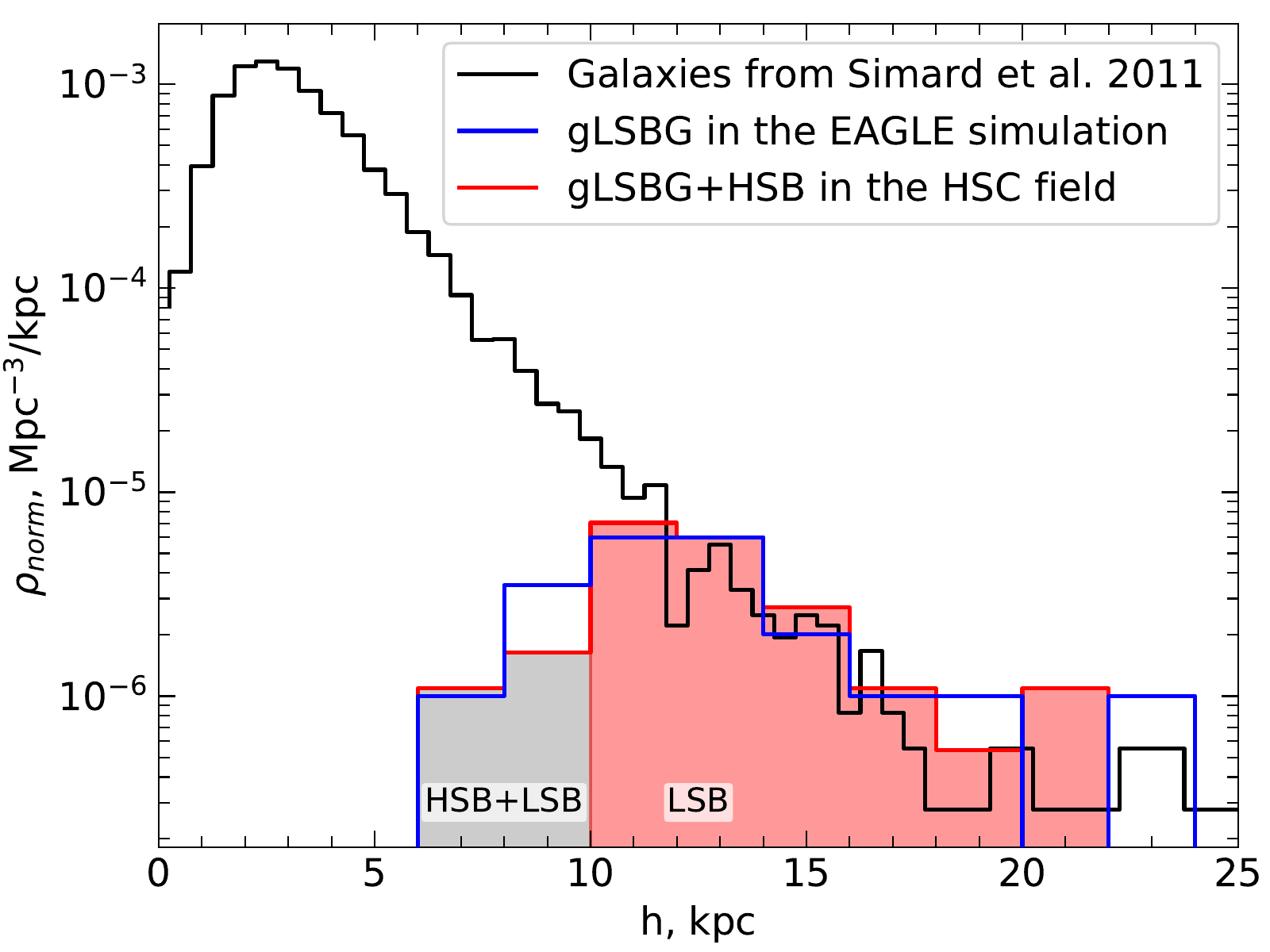}
\caption{The distribution of spiral galaxies by disc scalelength. The black histogram shows the statistics for spiral galaxies according to \citet{Simard2011}. The red histogram denotes giant disky galaxies (both LSB and HSB) found in the current paper. The blue line gives the normalized volume density of giant disky galaxies according to the results of the EAGLE simulation.}
\label{vol_density}
\end{figure}

Do giant disky galaxies represent a separate class of systems or just an extension of normal-sized spirals toward higher radii? In Fig.~\ref{vol_density} we compare the volume densities of galaxies with giant discs we found versus a large sample of disc galaxies from \citet{Simard2011} binned by the disc scale-length $h$. For this comparison we chose galaxies classified as spirals (probability $>0.7$) by the Galaxy Zoo project \citep{Lintott2011} by at least 10 members (NVOTE $>10$). We kept the objects at redshifts $0.02<z<0.06$ in the area of the sky $0^\circ<$~Dec~$<60^\circ$; $9$h~<~R.A.~<~$15$h, which corresponds to the volume $7.2\cdot 10^6$ Mpc$^3$. Fig.~\ref{vol_density} shows that our extended discs follow the trend established by `normal' spirals at $h<10$~kpc, \red{which can indicate non-exeptional formation histories of gLSBGs}. \red{A similar conclusion can be drawn from the gLSBG mass, specific angular momentum, and gas fraction \citep{ManceraPina2021}.} \red{At the same time, we found} a higher number of very large discs ($h=17$~kpc); so do the simulations. One should keep in mind that at least some `large' galaxies from \citet{Simard2011} have their sizes overestimated by the automatic analysis because of foreground or background objects. Also, SDSS images do not probe the surface brightnesses of outer LSB discs and therefore $h$ values are estimated in the inner regions of the discs which might not represent the entire light profiles (Fig.~\ref{fig_examples}).

EAGLE reproduces quite well the observed volume density of extended discs. Cosmological hydrodynamical simulations also allow one to explore the formation history of galaxies with giant discs and pin down their formation mechanisms. \green{We did it for the sample of gLSBGs models that we selected in a similar way that was done in \cite{Kulier2020}}. Compared to moderate-size counterparts, galaxies with extended discs in simulations have: (i) a~higher fraction of star particles at $z=0$ originating from mergers ($25.8\pm2.5$~per~cent vs $19.8\pm0.1$~per~cent); (ii) a higher number of both major (mass ratio 1:4 or higher) and minor mergers in their formation history, with a more pronounced difference in major mergers; (iii) minor mergers at $z<1$ have angular momenta better aligned with that of the pre-existing host system. Major mergers tend to be well aligned in both giant and moderate-sized discs. Otherwise a disc would likely be overheated and destroyed. 27$\pm$8~per~cent of simulated galaxies with giant extended discs experienced zero major mergers since $z=3$ indicating that gLSBGs can be formed via minor mergers and cold gas accretion. Our results agree with \red{\cite{2022MNRAS.514.5840P}, suggesting that massive LSBs have larger ex-situ stellar fractions than HSBs}.

The apparently high AGN frequency in gLSBGs is in line with the origin of extended discs from minor mergers and accretion: the nuclear activity is efficiently triggered by gas infall on retrograde orbits \citep{Khoperskovetal2021}, suggesting that some fraction of accreted satellites did not have their spins aligned with that of the main disc. On the other hand, undermassive central BHs found in gLSBGs \citep{saburovaetal2021} are unlikely to produce enough feedback to clear out the central region of the galaxy from gas and stop AGN fueling, which might prolong the period of the activity and, hence, increase the AGN fraction in the observed gLSBG sample.

So the observational data do not contradict the merger scenario at least for some of gLSBGs. However, as it was noted before, even simulations leave room for other possible scenarios, like e.g. gas accretion from the filament, which could be valid for the gLSBGs without enhanced stellar velocity dispersion \citep[like UGC~1378 and Malin~2;][]{saburovaetal2021}. However we still lack data on the new gLSBGs (e.g.  the stellar velocity dispersion). The new upcoming observational estimates of stellar and gas kinematics and metallicity will shed light on this question.

\section*{Acknowledgements}
\green{We thank the anonymous referee for valuable comments and Q.~Zhu for sharing the submitted manuscript on gLSBGs in TNG100.}
The search of the gLSBGs was done with the support of RScF grant No. 19-12-00281. The
surface photometry was done with the support
of the RScF grant No. 19-72-20089. IC's research is supported by the SAO Telescope Data Center. GG gratefully acknowledges support by the ANID BASAL projects ACE210002 and FB210003.

%%%%%%%%%%%%%%%%%%%%%%%%%%%%%%%%%%%%%%%%%%%%%%%%%%
\section*{Data Availability}
No new data were generated or analysed in support of this research.

%%%%%%%%%%%%%%%%%%%% REFERENCES %%%%%%%%%%%%%%%%%%

% The best way to enter references is to use BibTeX:

\bibliographystyle{mnras}
\bibliography{saburova_lsb} % if your bibtex file is called example.bib
\appendix

\section{Supplementary material}
\begin{figure*}
\centering
\includegraphics[width=0.21\hsize]{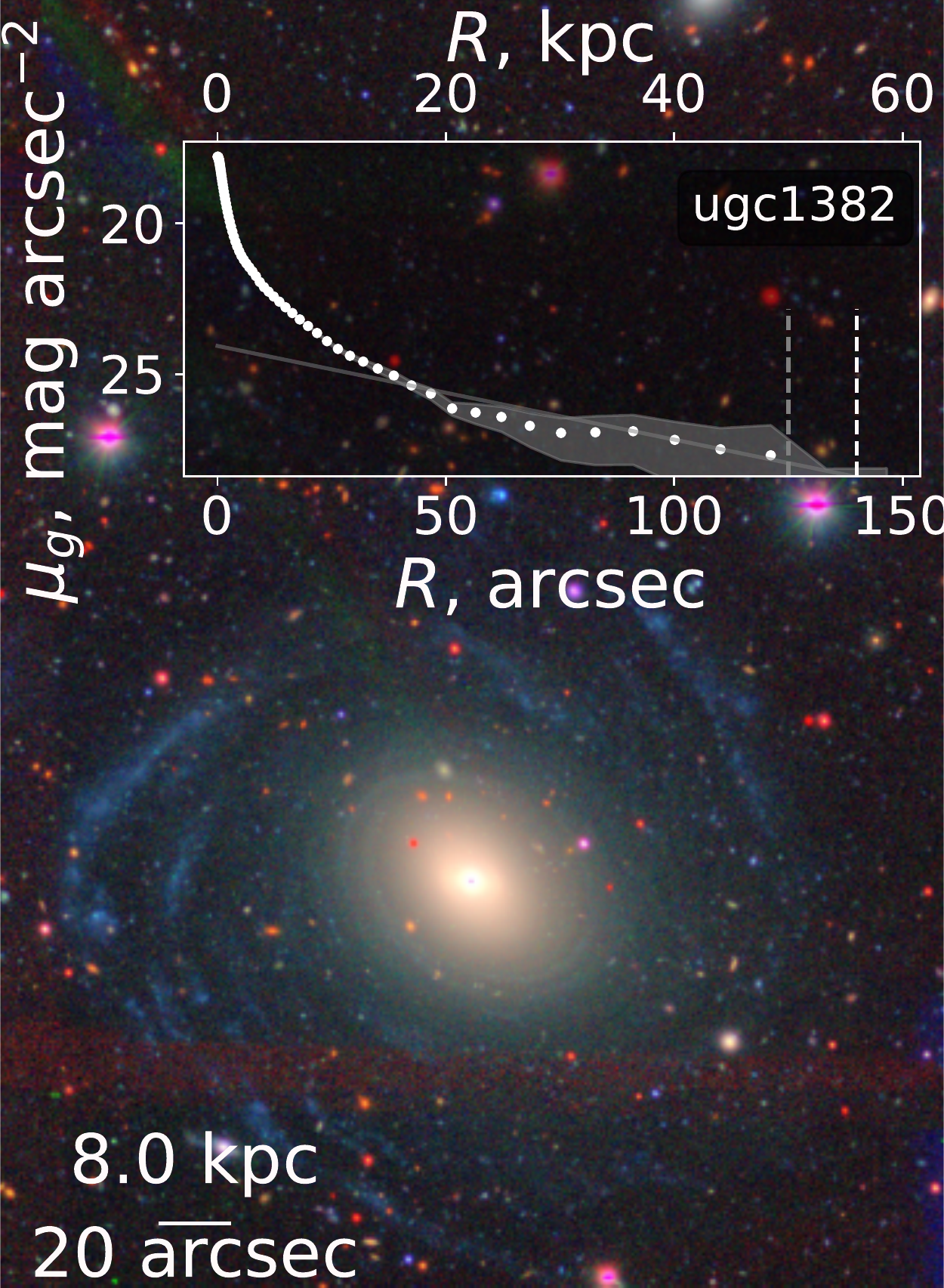}
\includegraphics[width=0.21\hsize]{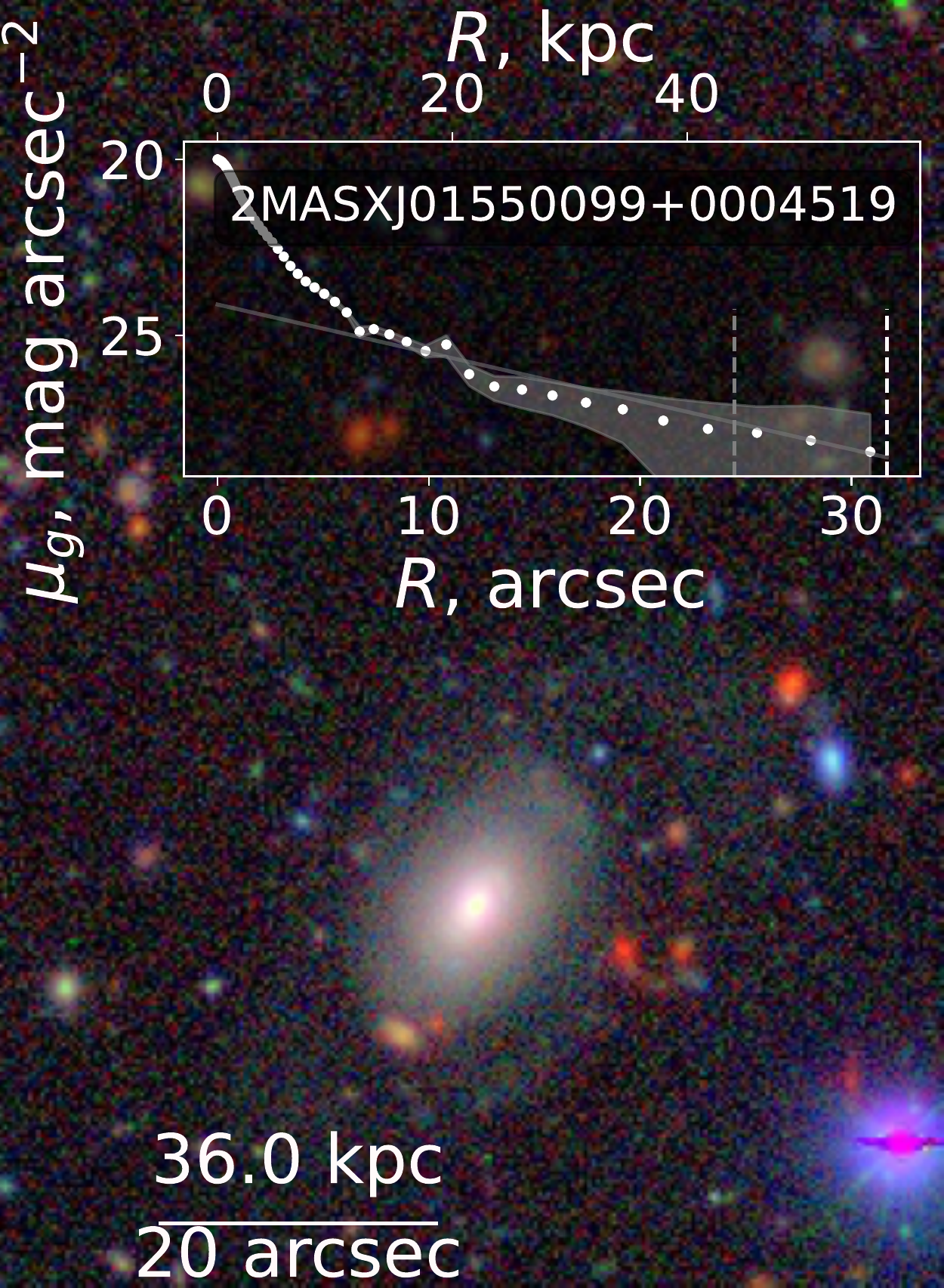}
\includegraphics[width=0.21\hsize]{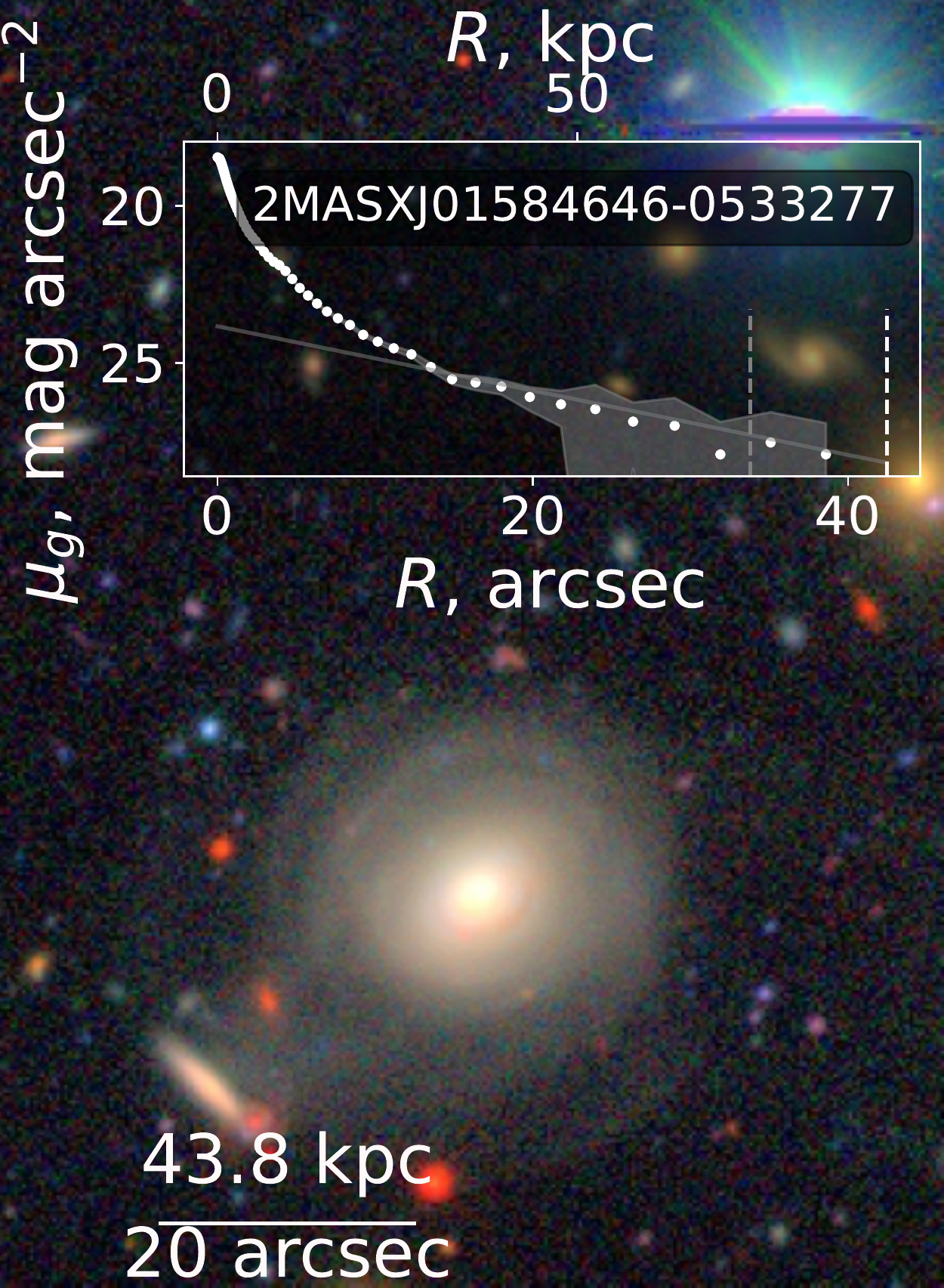}
\includegraphics[width=0.21\hsize]{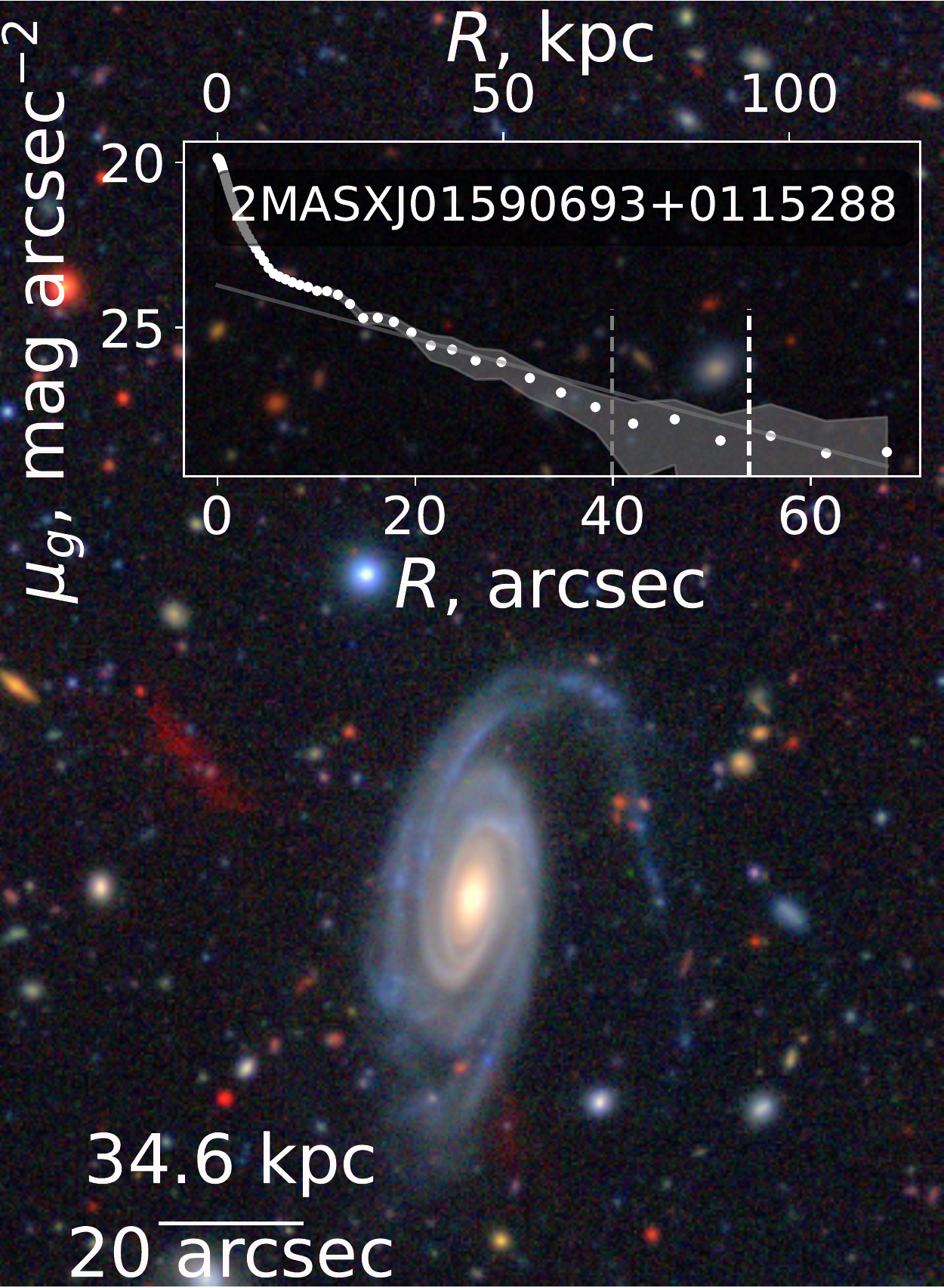}
\includegraphics[width=0.21\hsize]{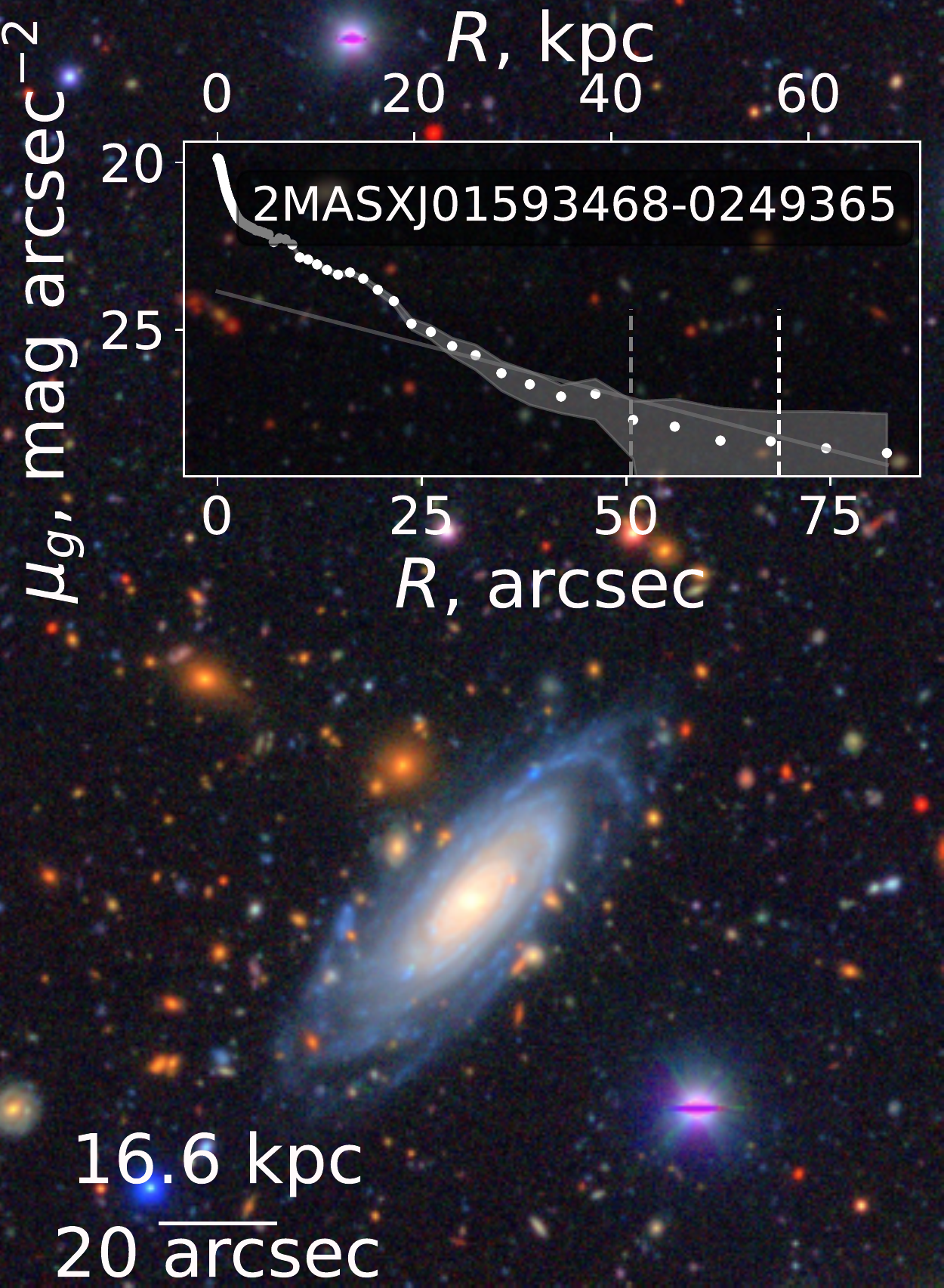}
\includegraphics[width=0.21\hsize]{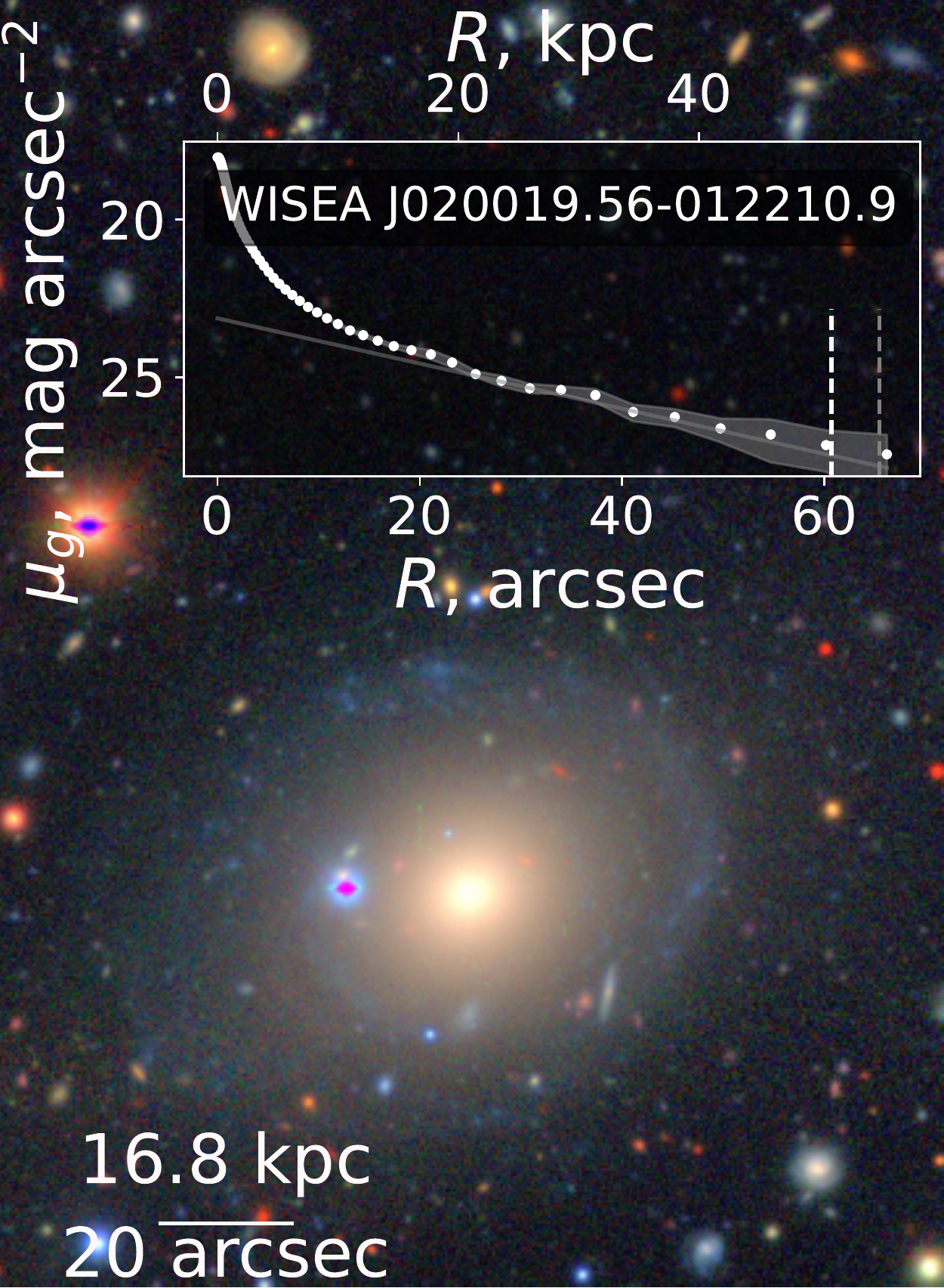}
\includegraphics[width=0.21\hsize]{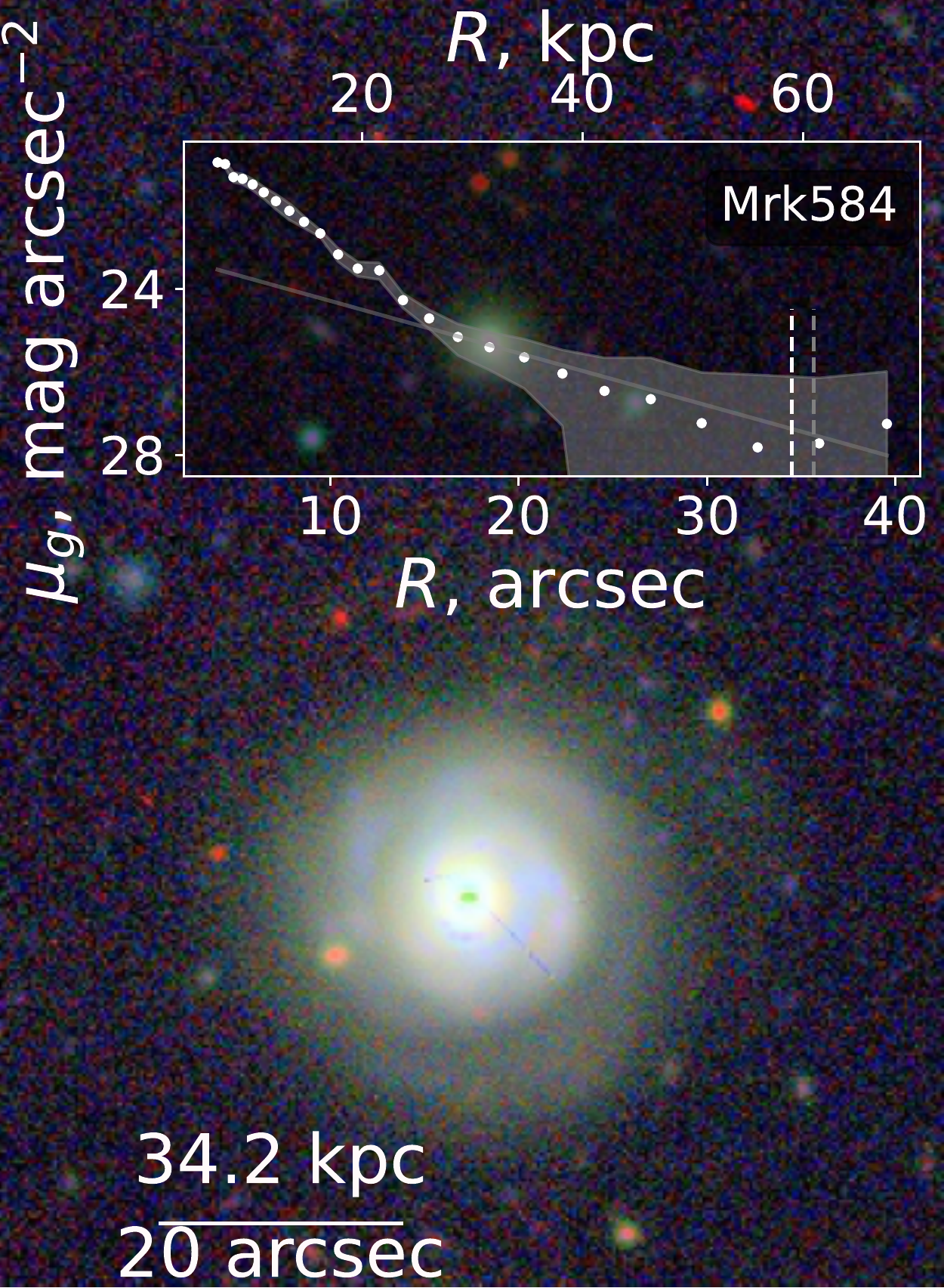}
\includegraphics[width=0.21\hsize]{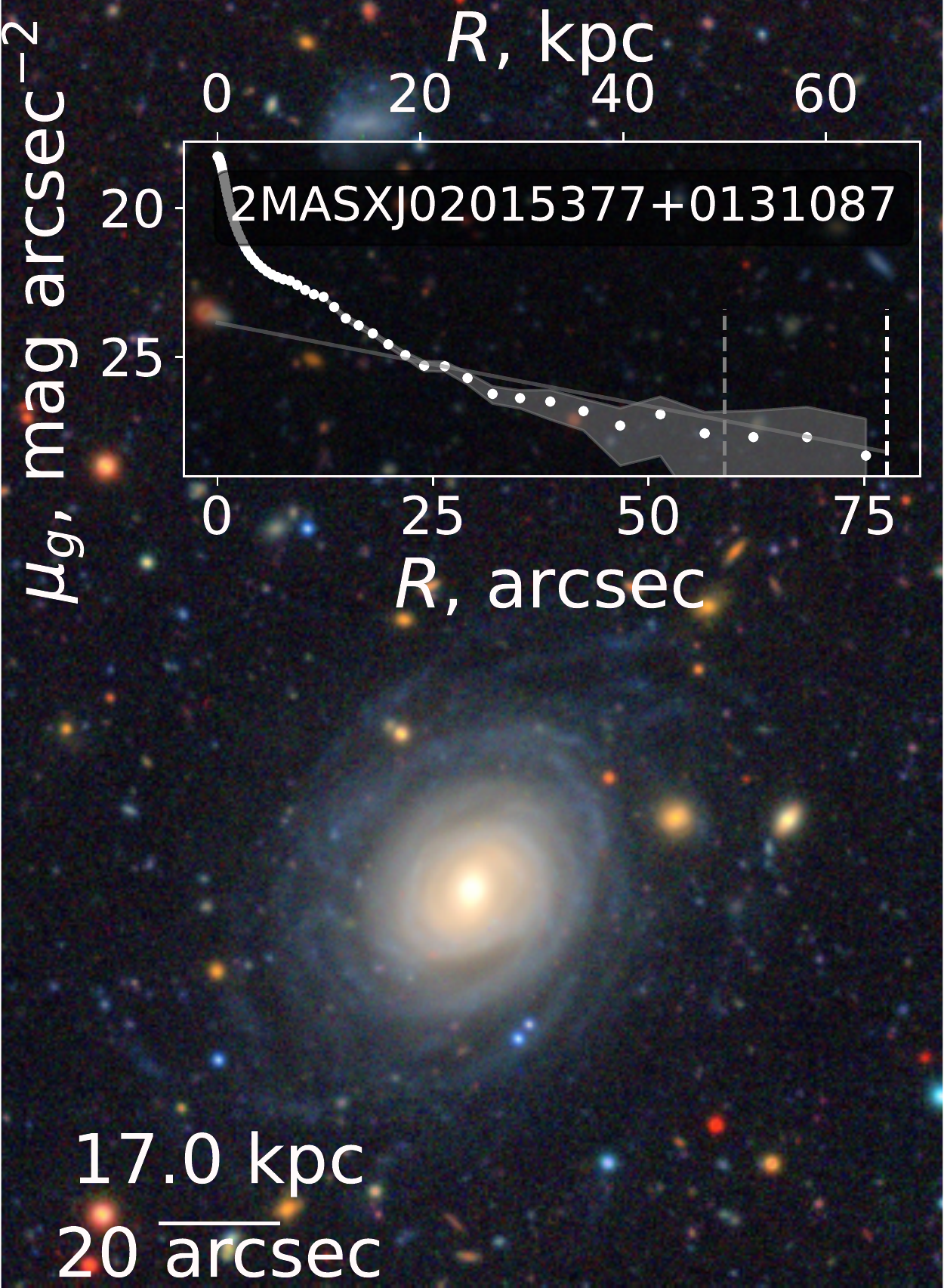}
\includegraphics[width=0.21\hsize]{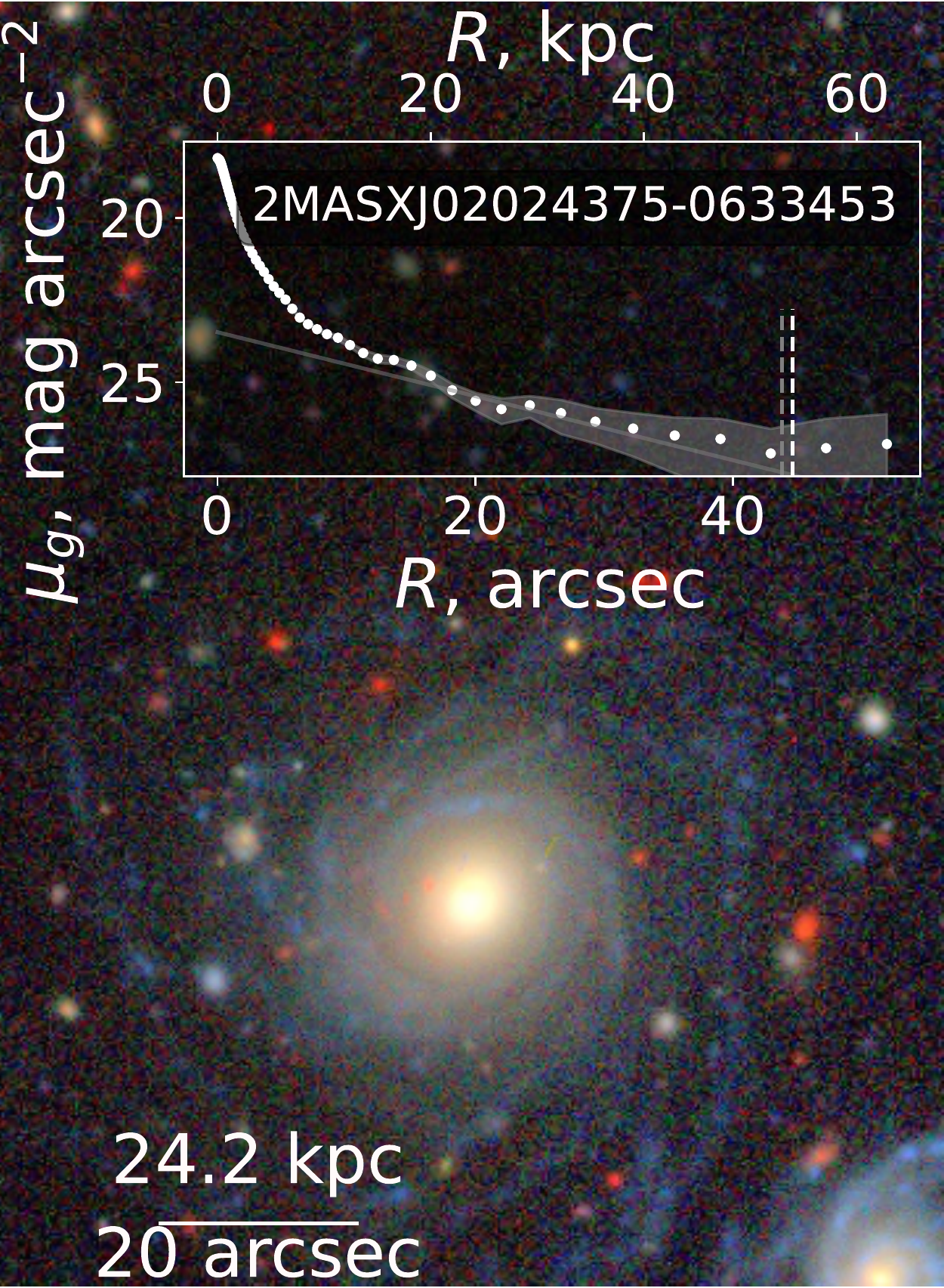}
\includegraphics[width=0.21\hsize]{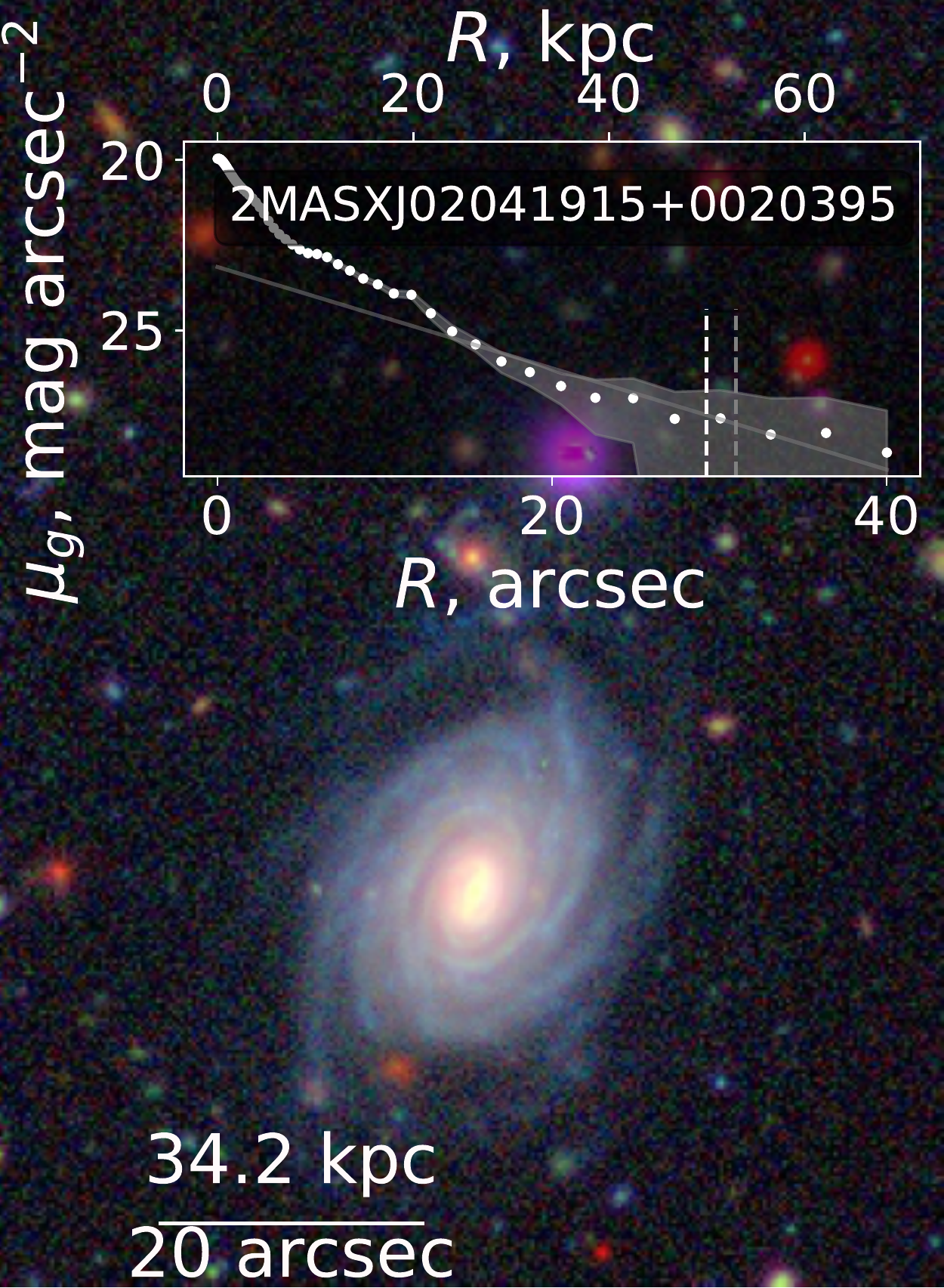}
\includegraphics[width=0.21\hsize]{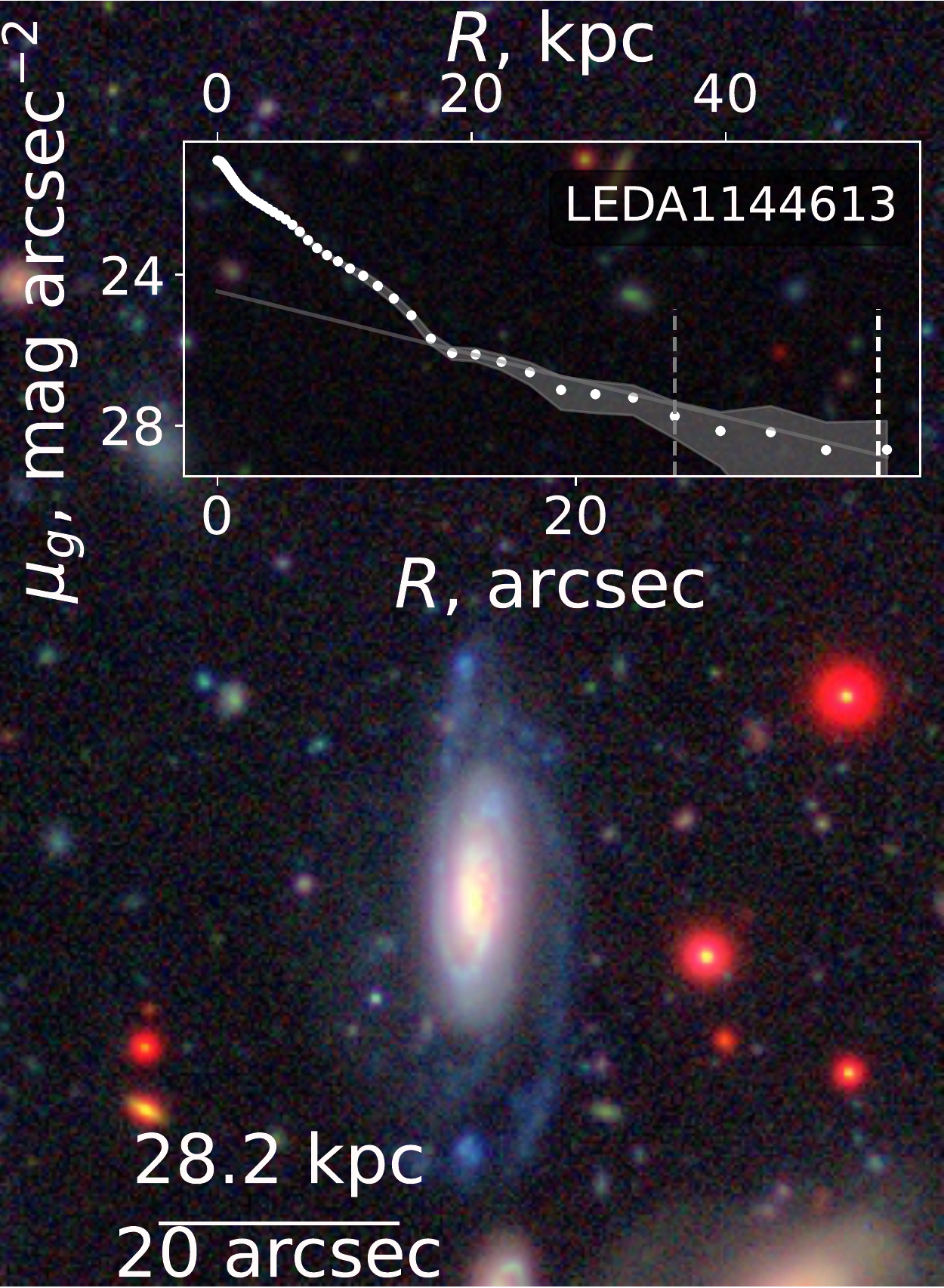}
\includegraphics[width=0.21\hsize]{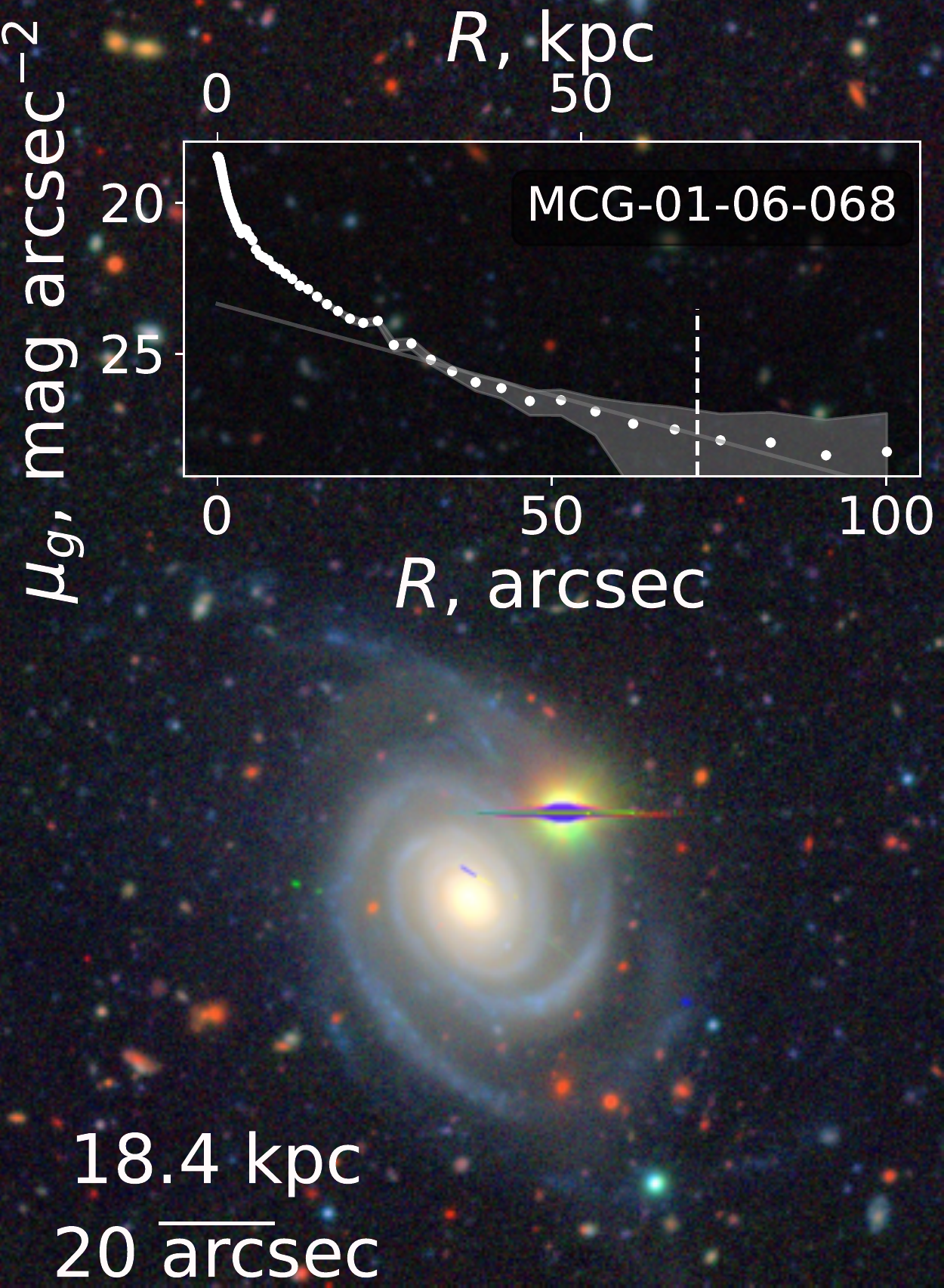}
\includegraphics[width=0.21\hsize]{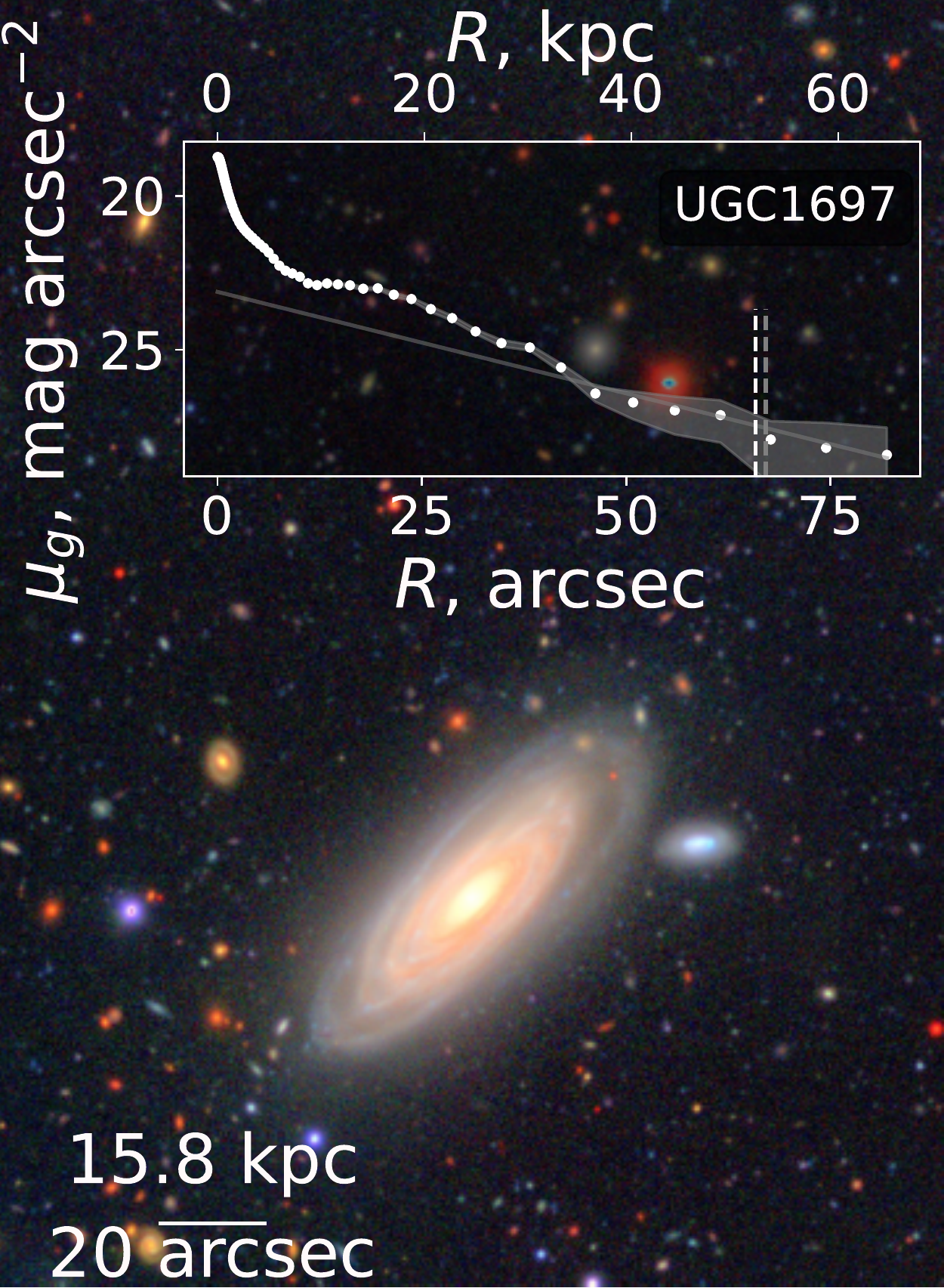}
\includegraphics[width=0.21\hsize]{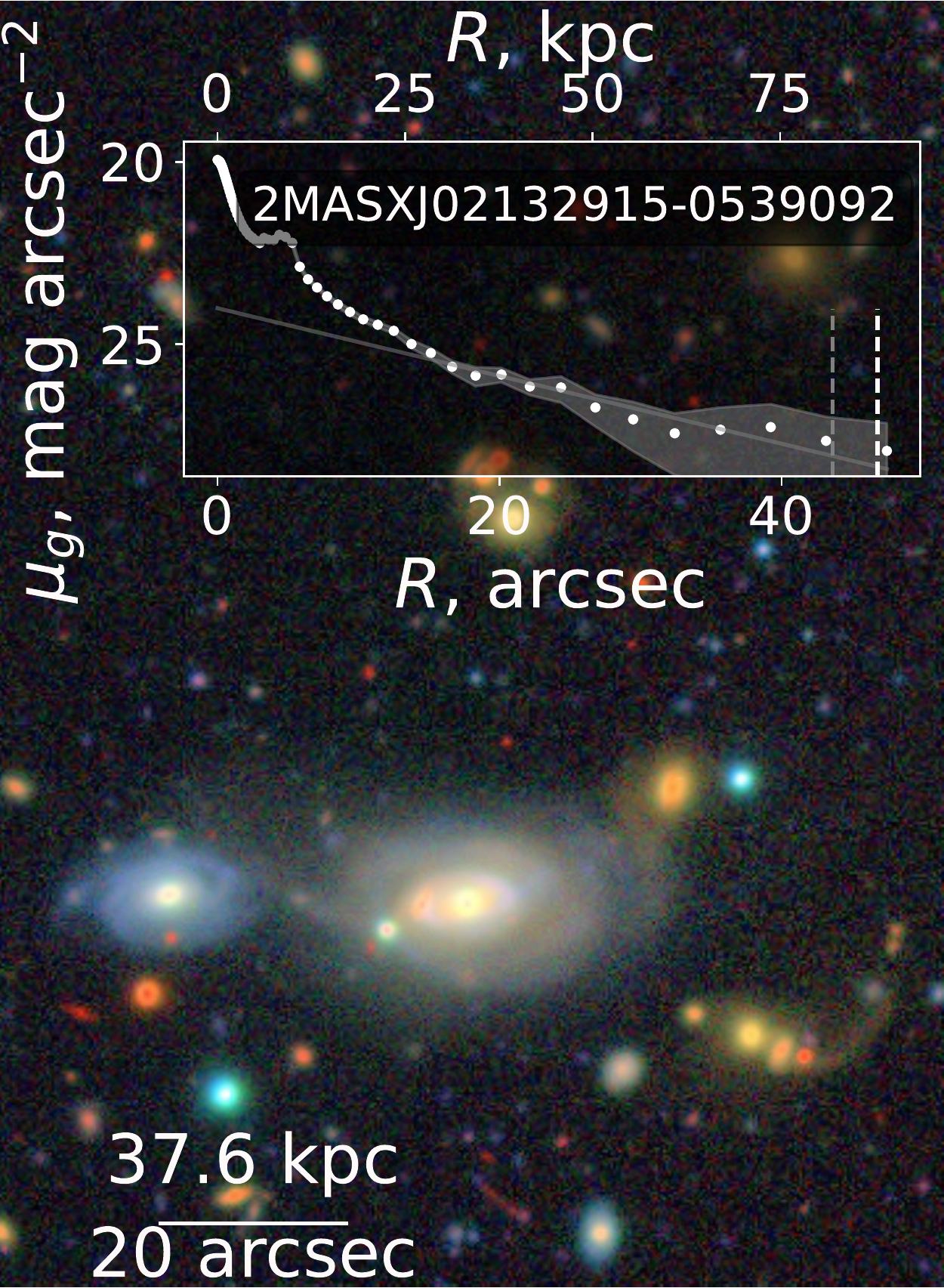}
\includegraphics[width=0.21\hsize]{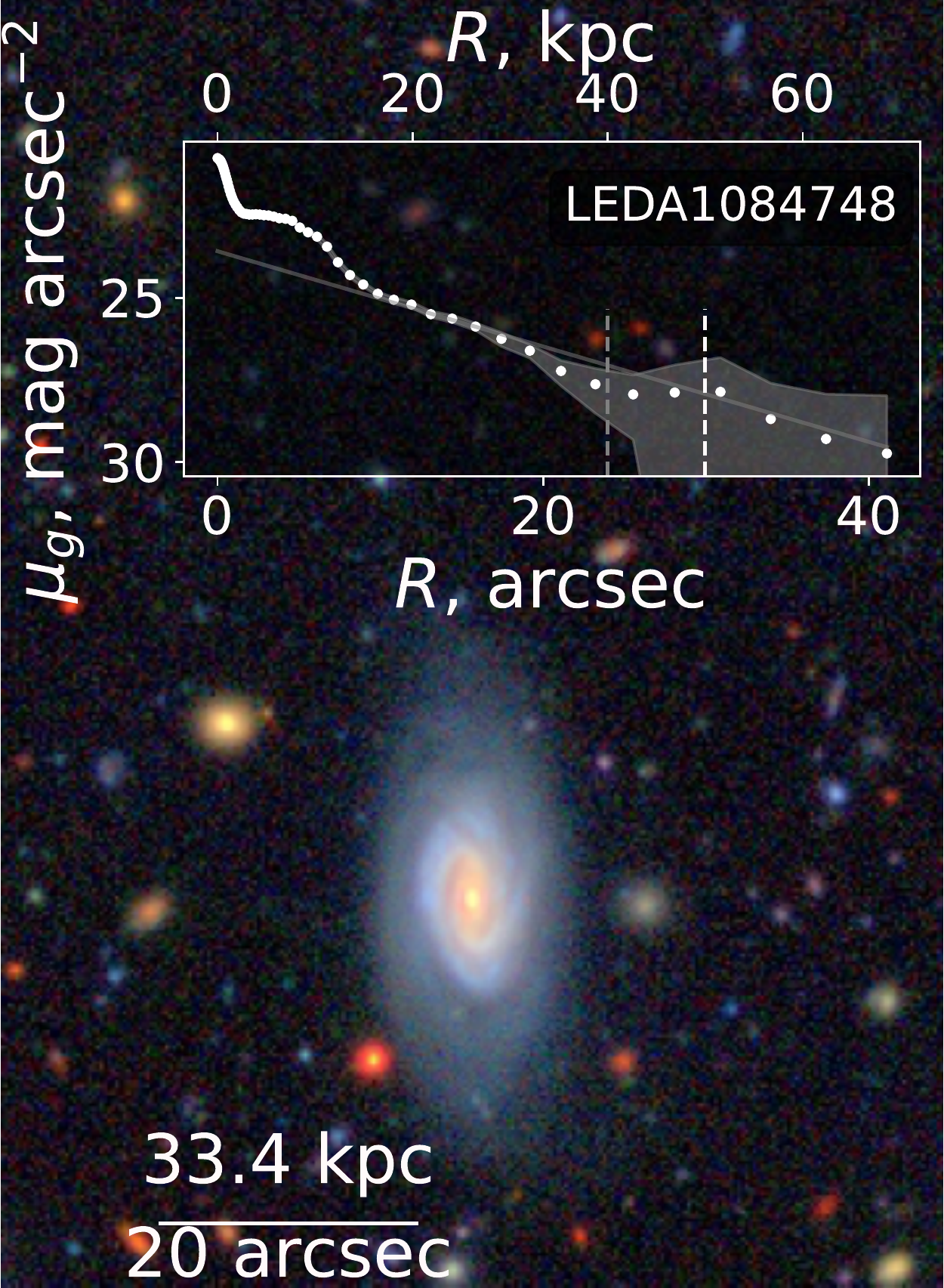}
\includegraphics[width=0.21\hsize]{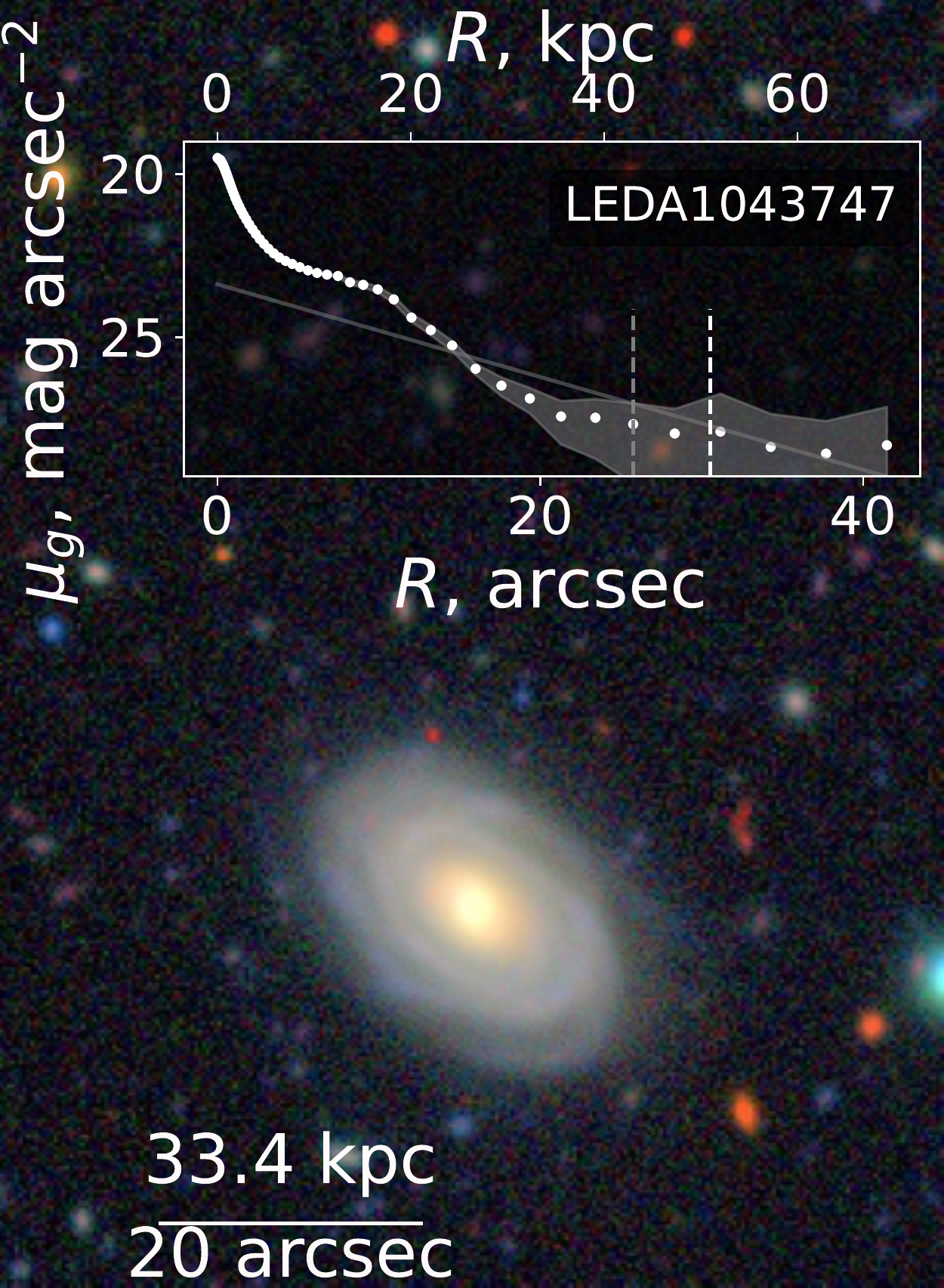}
\caption{The colour composite HSC images and light profiles of the gLSBGs found in this study.}
\label{fig_examples_all}
\end{figure*}

\begin{figure*}
\centering
\includegraphics[width=0.21\hsize]{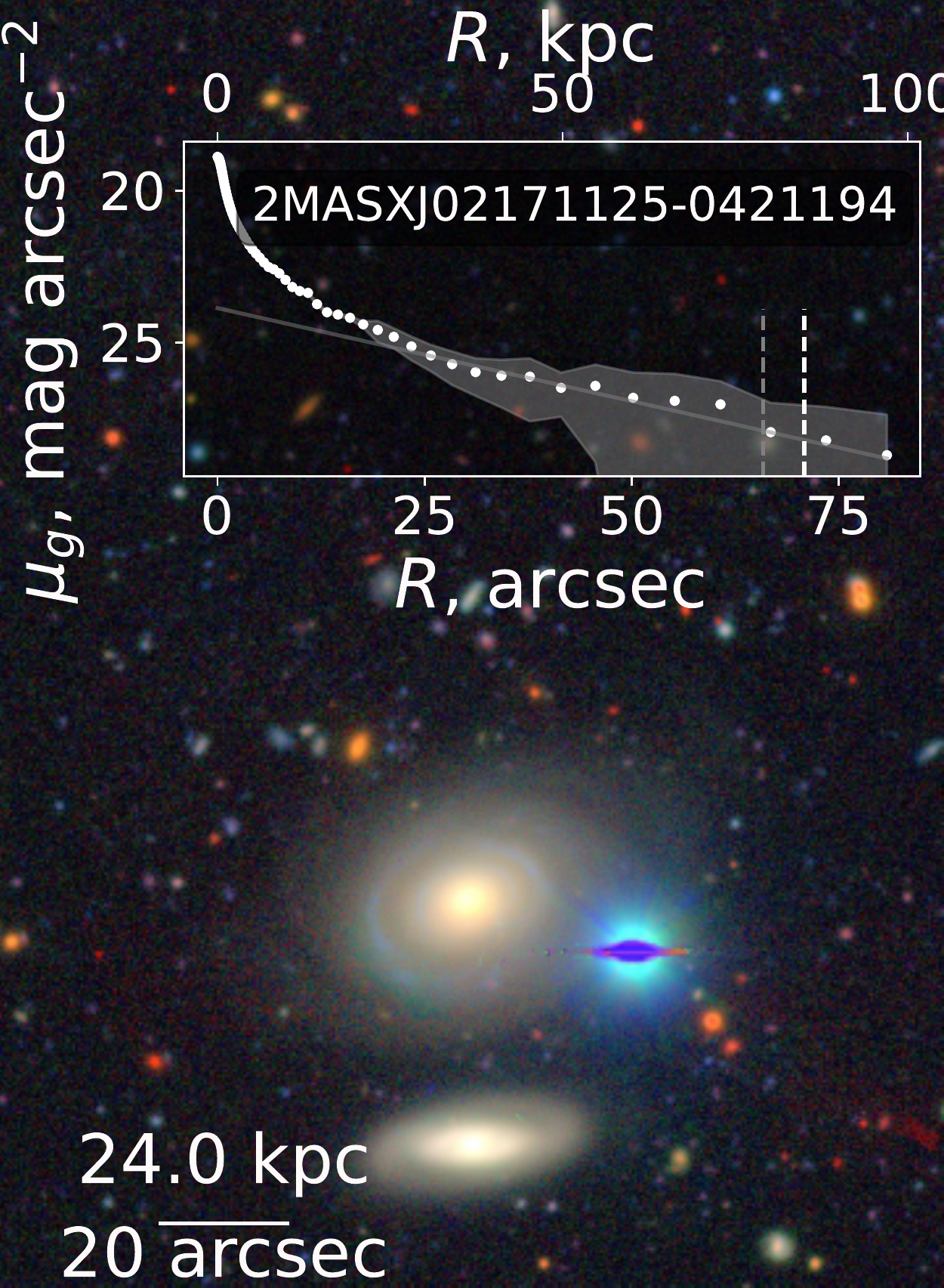}
\includegraphics[width=0.21\hsize]{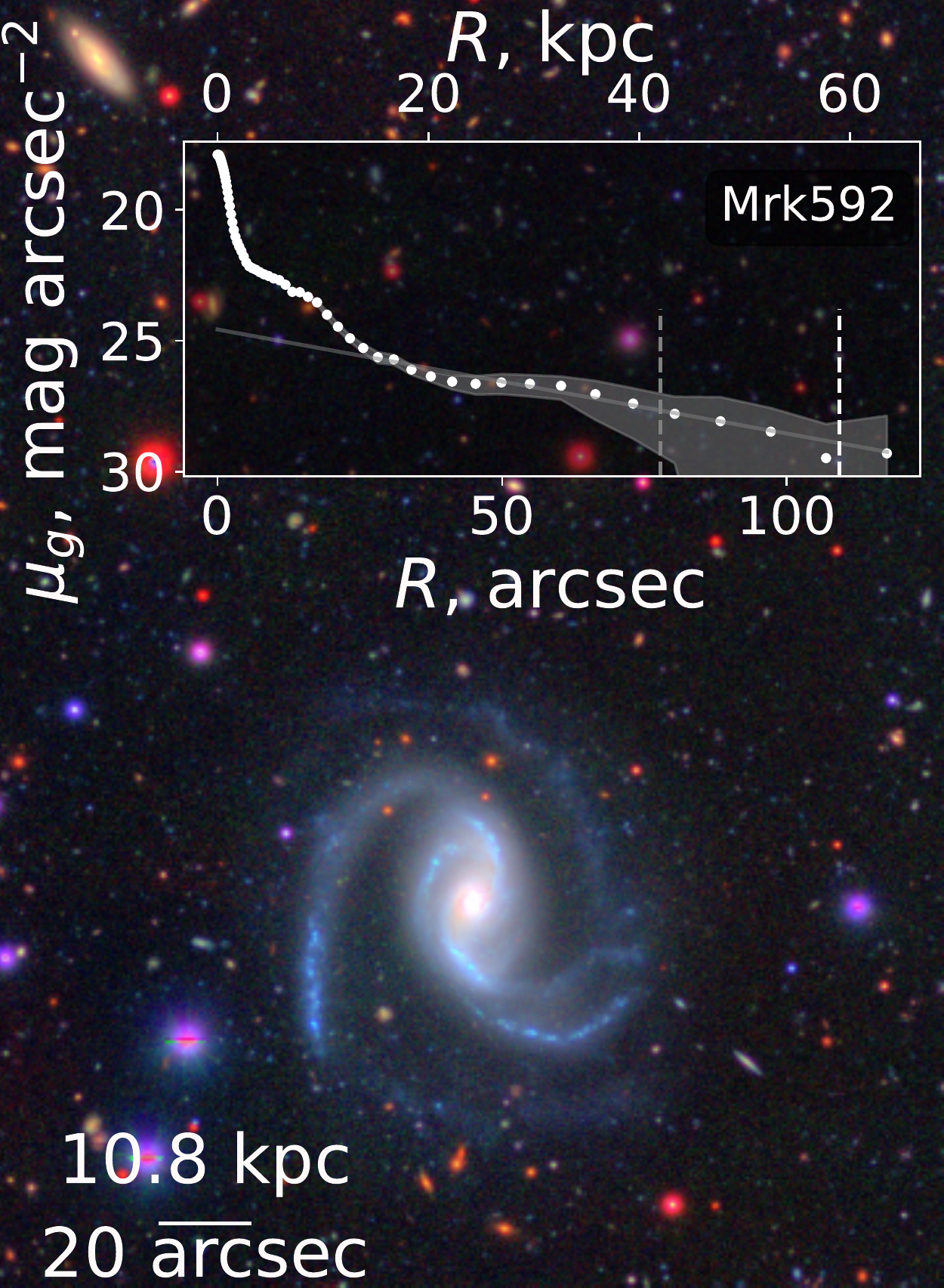}
\includegraphics[width=0.21\hsize]{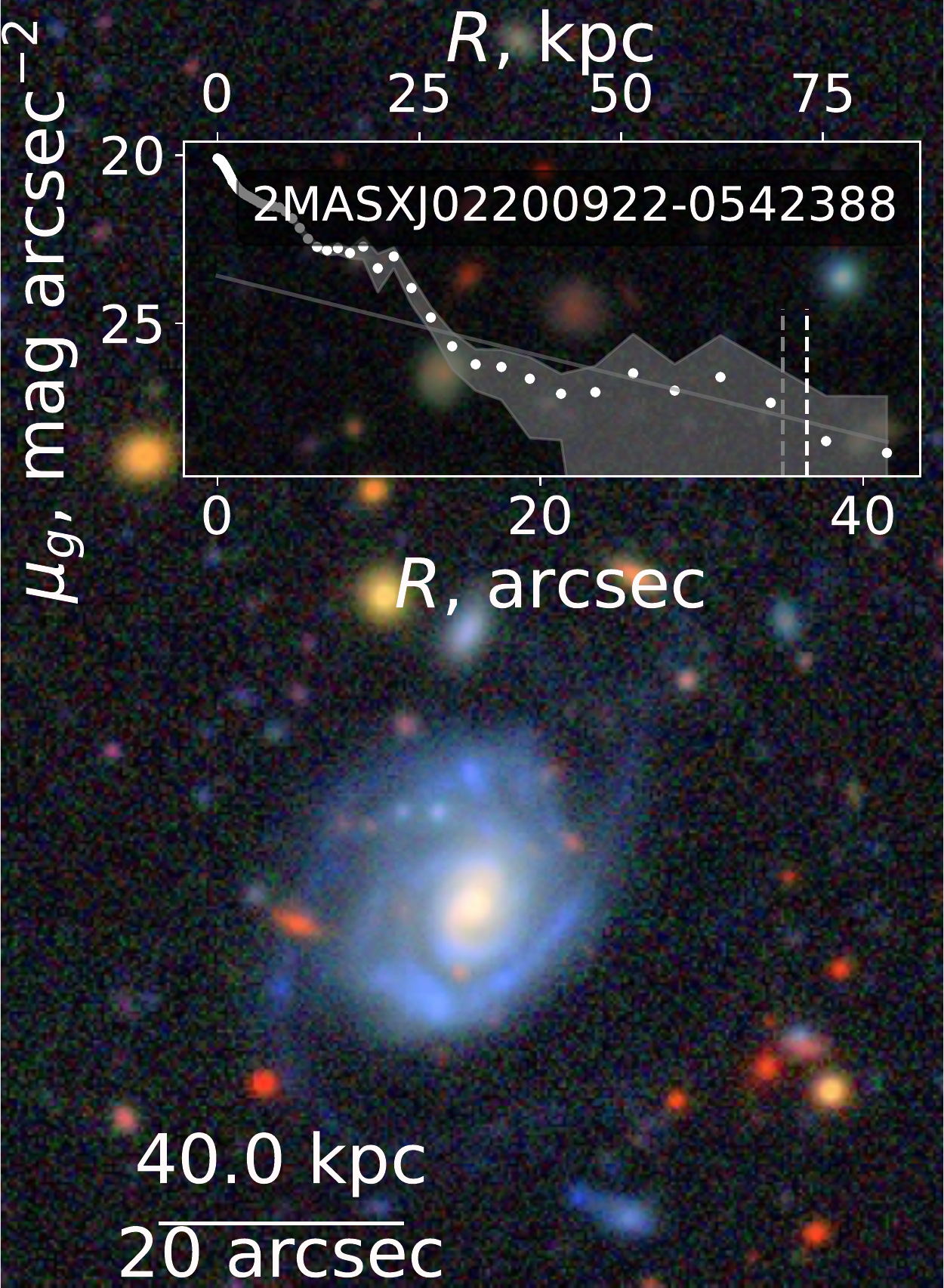}
\includegraphics[width=0.21\hsize]{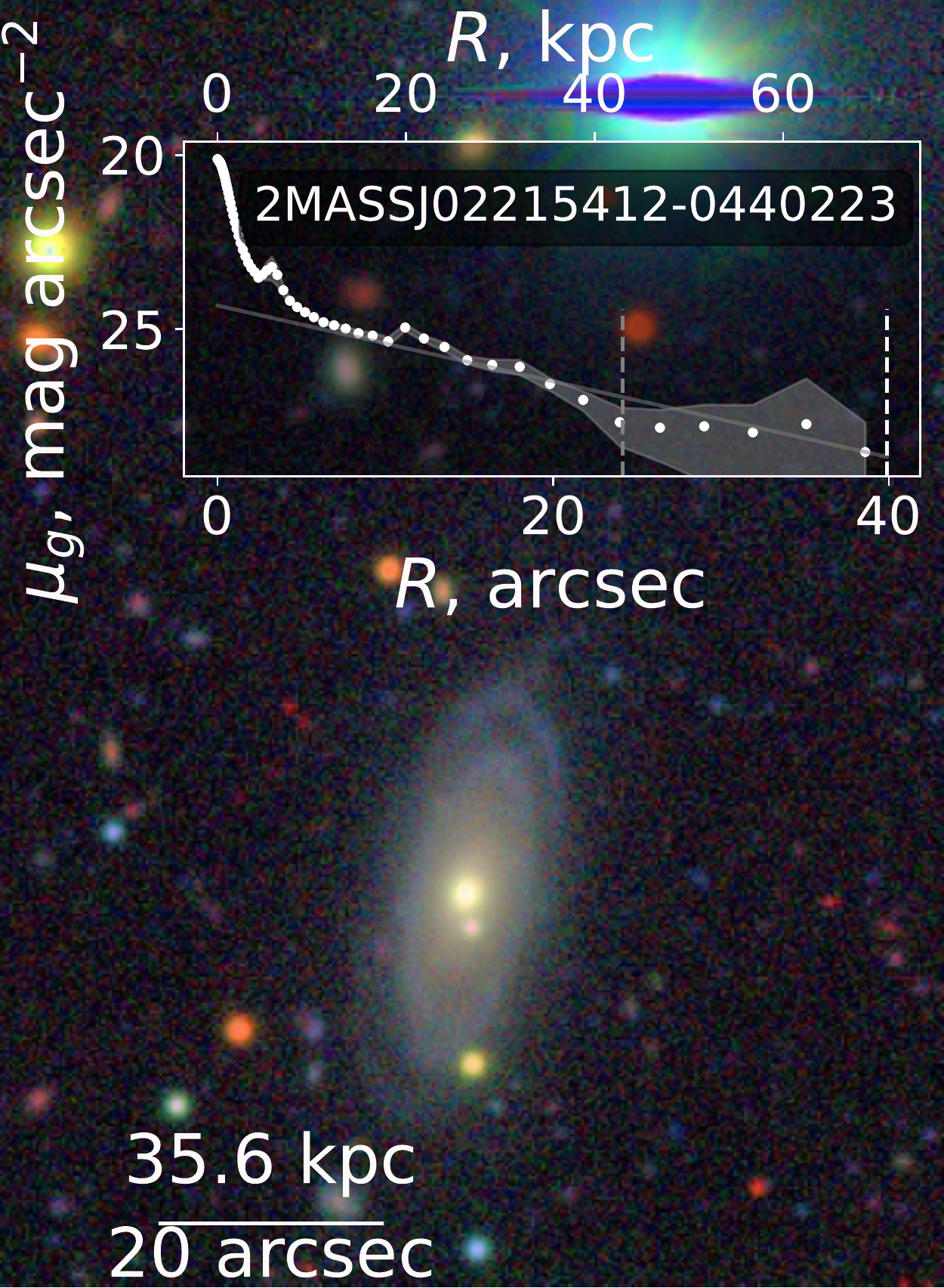}
\includegraphics[width=0.21\hsize]{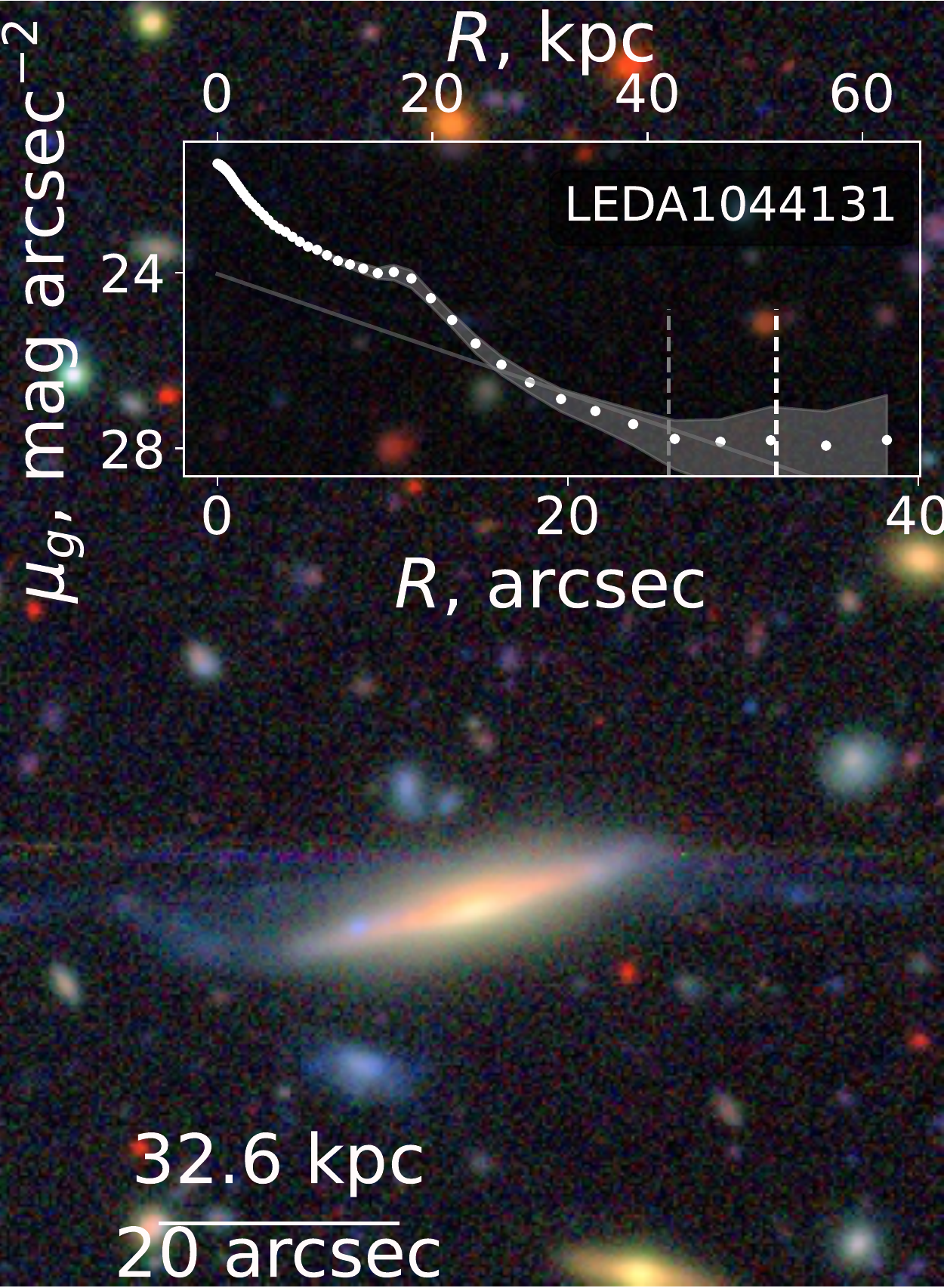}
\includegraphics[width=0.21\hsize]{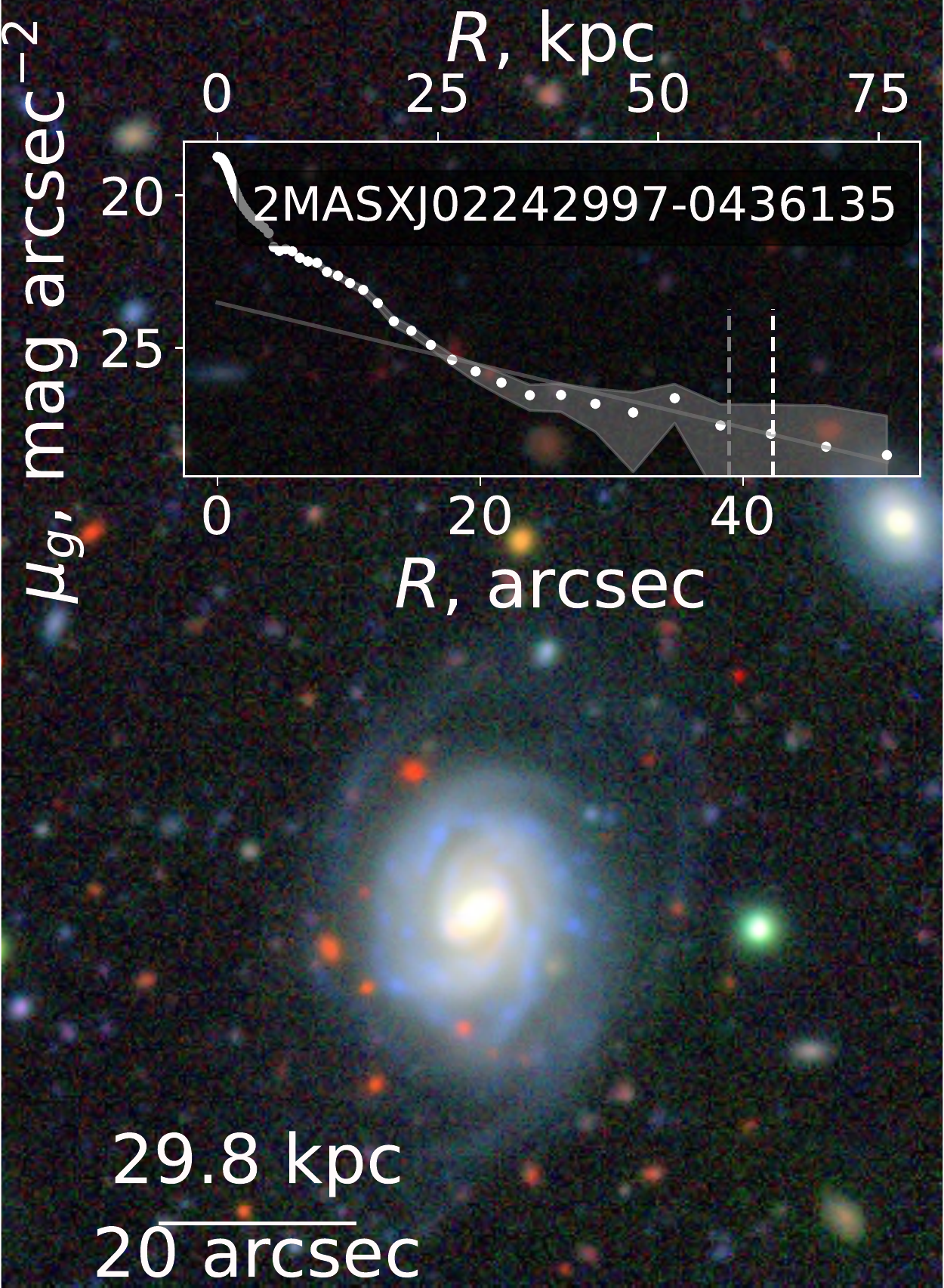}
\includegraphics[width=0.21\hsize]{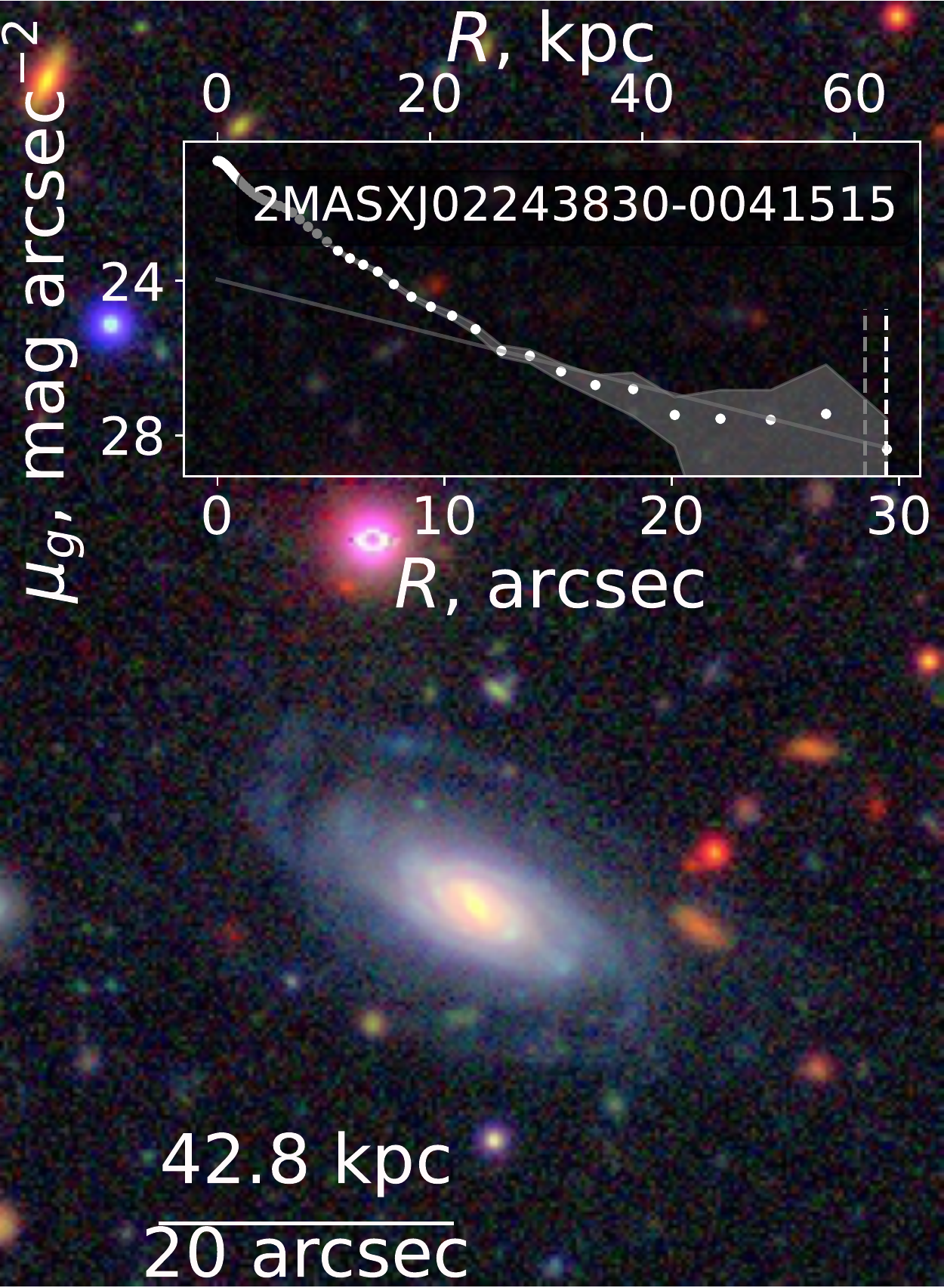}
\includegraphics[width=0.21\hsize]{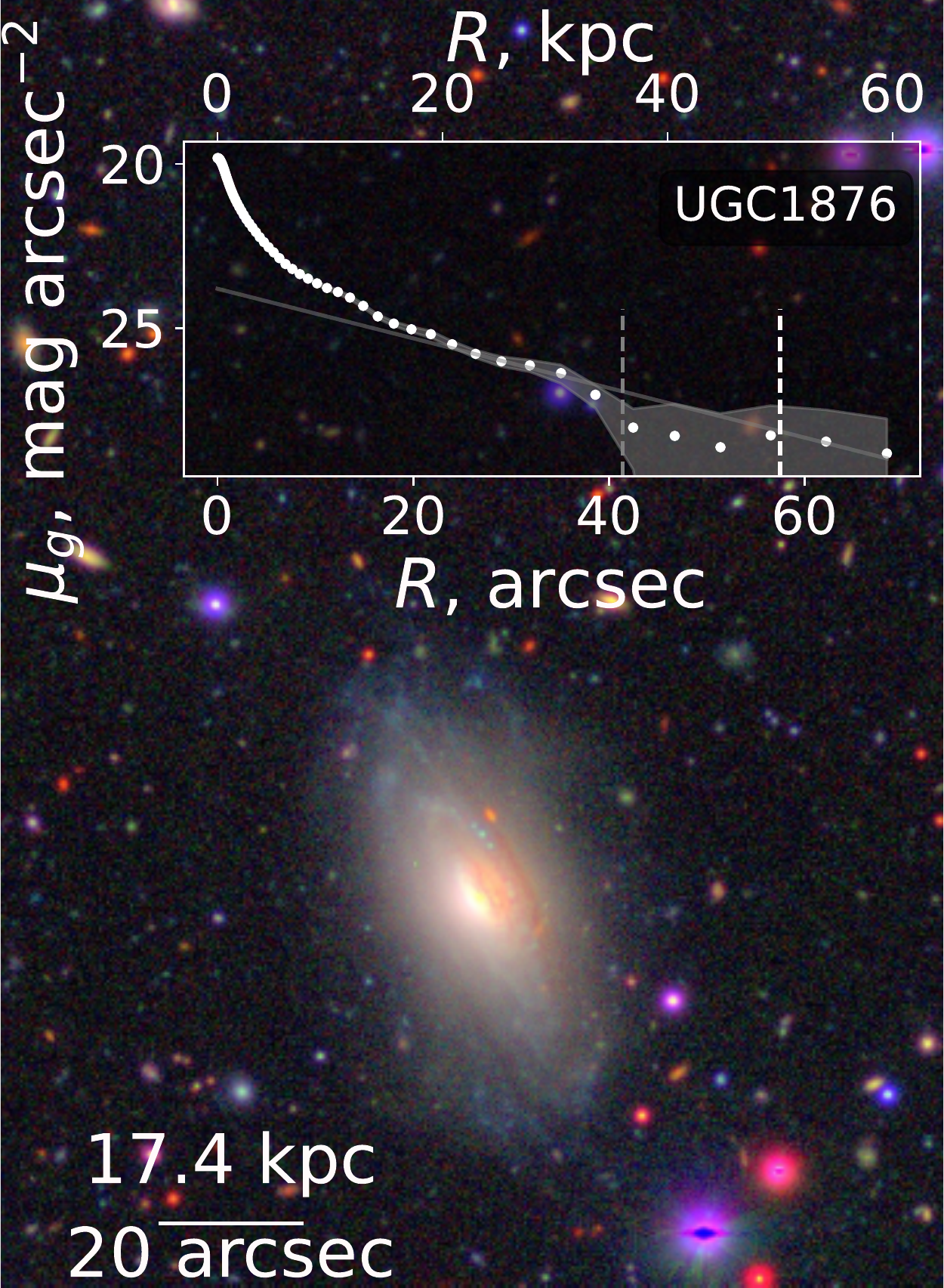}
\includegraphics[width=0.21\hsize]{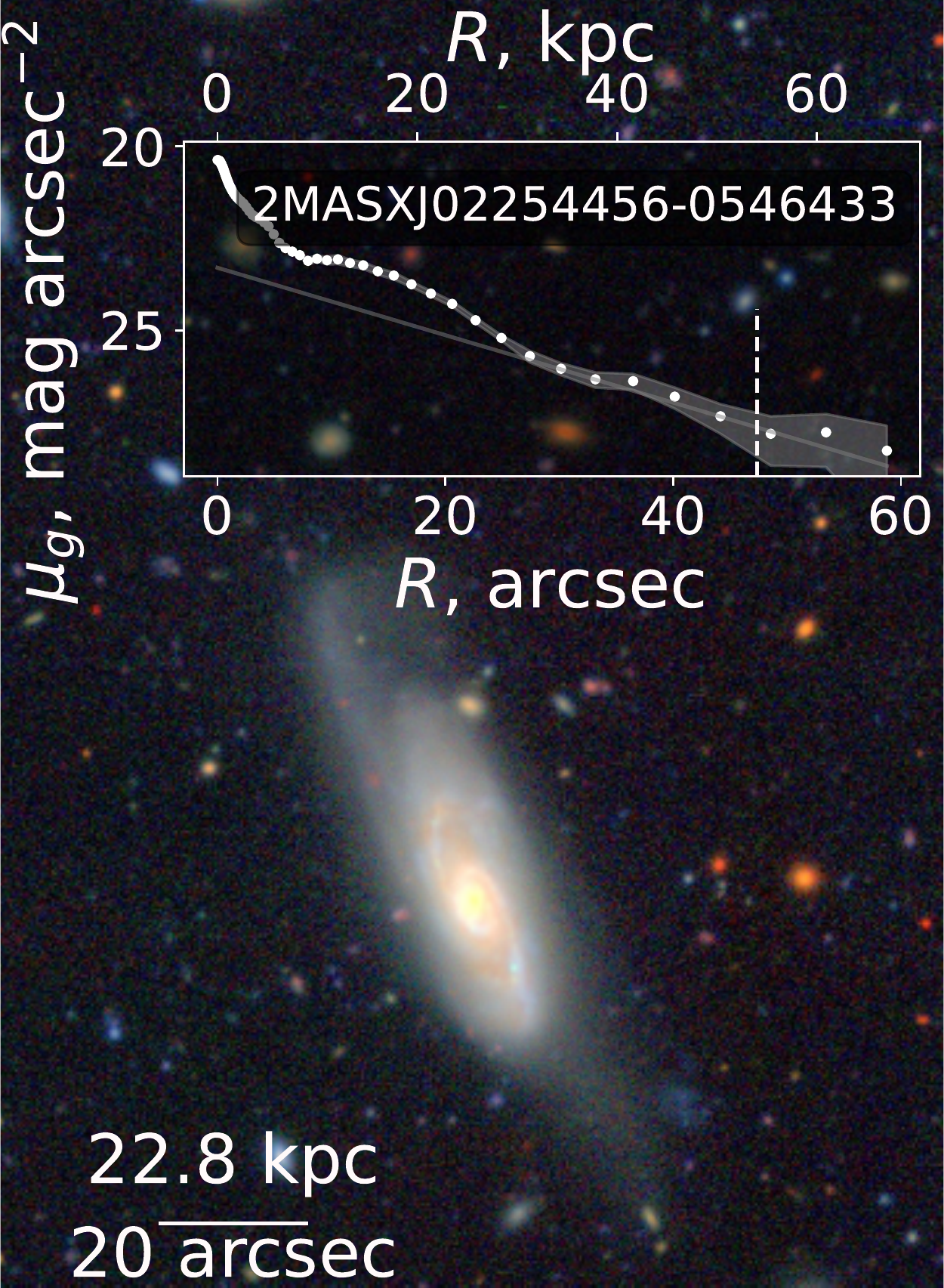}
\includegraphics[width=0.21\hsize]{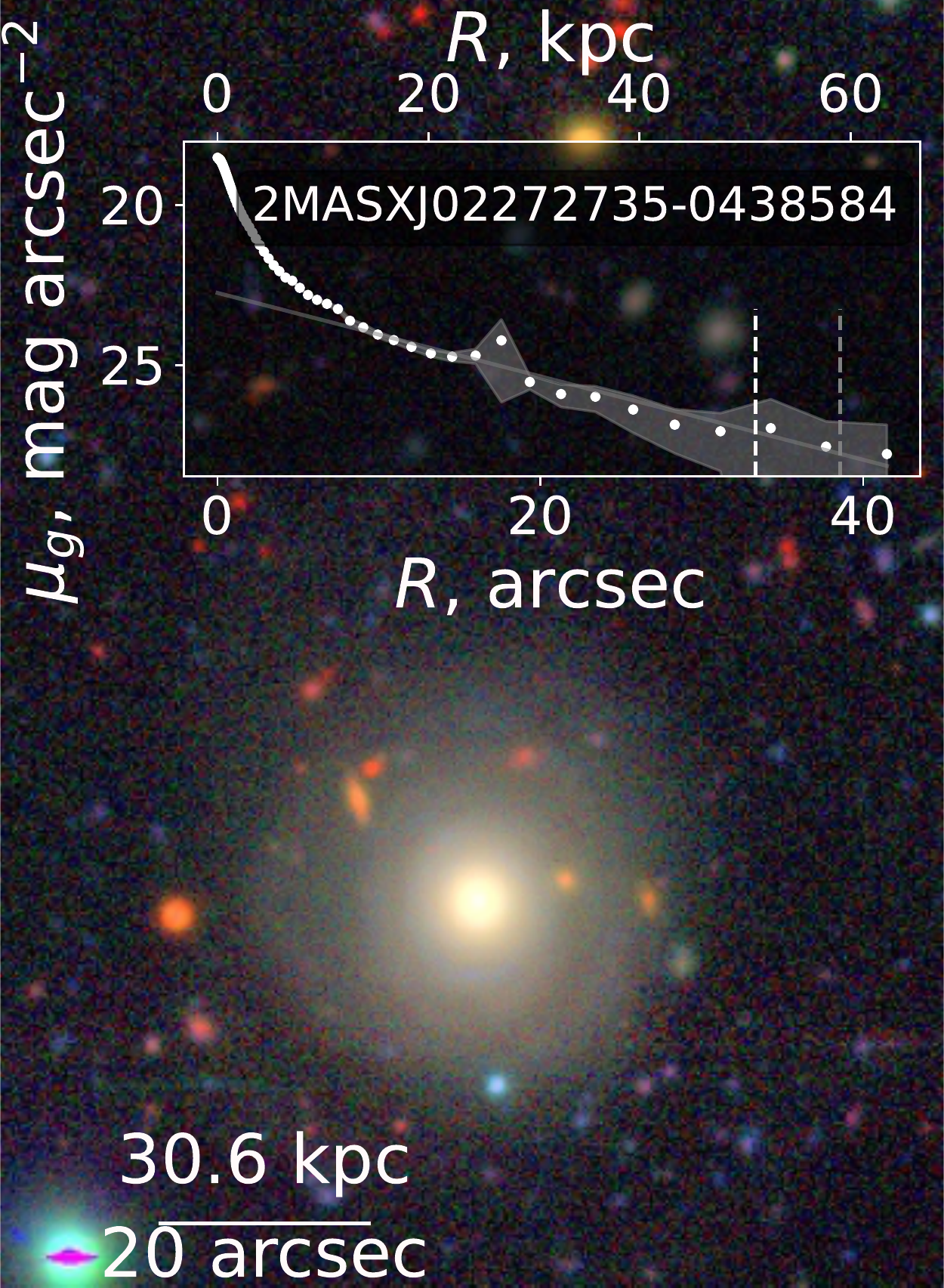}
\includegraphics[width=0.21\hsize]{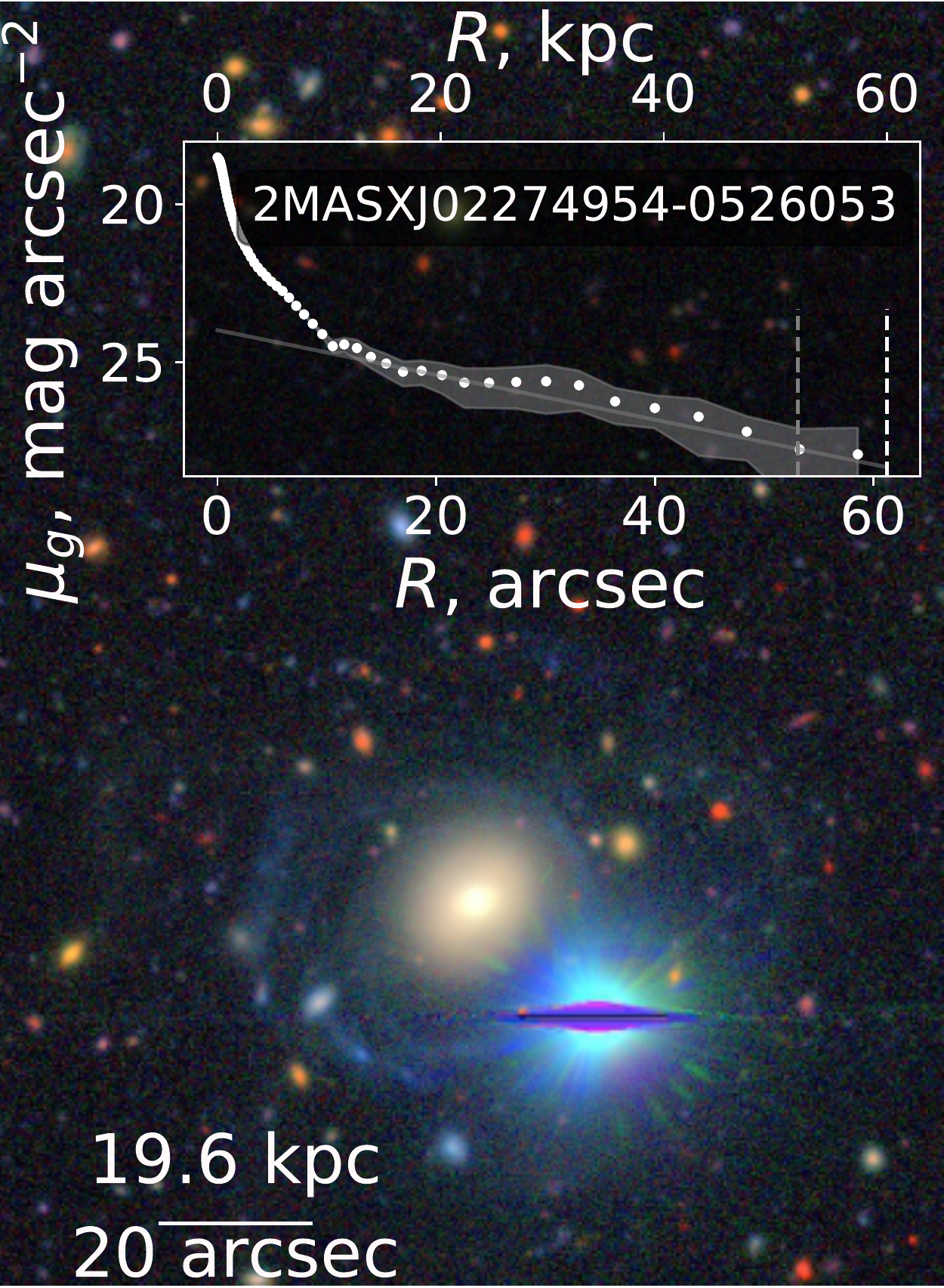}
\includegraphics[width=0.21\hsize]{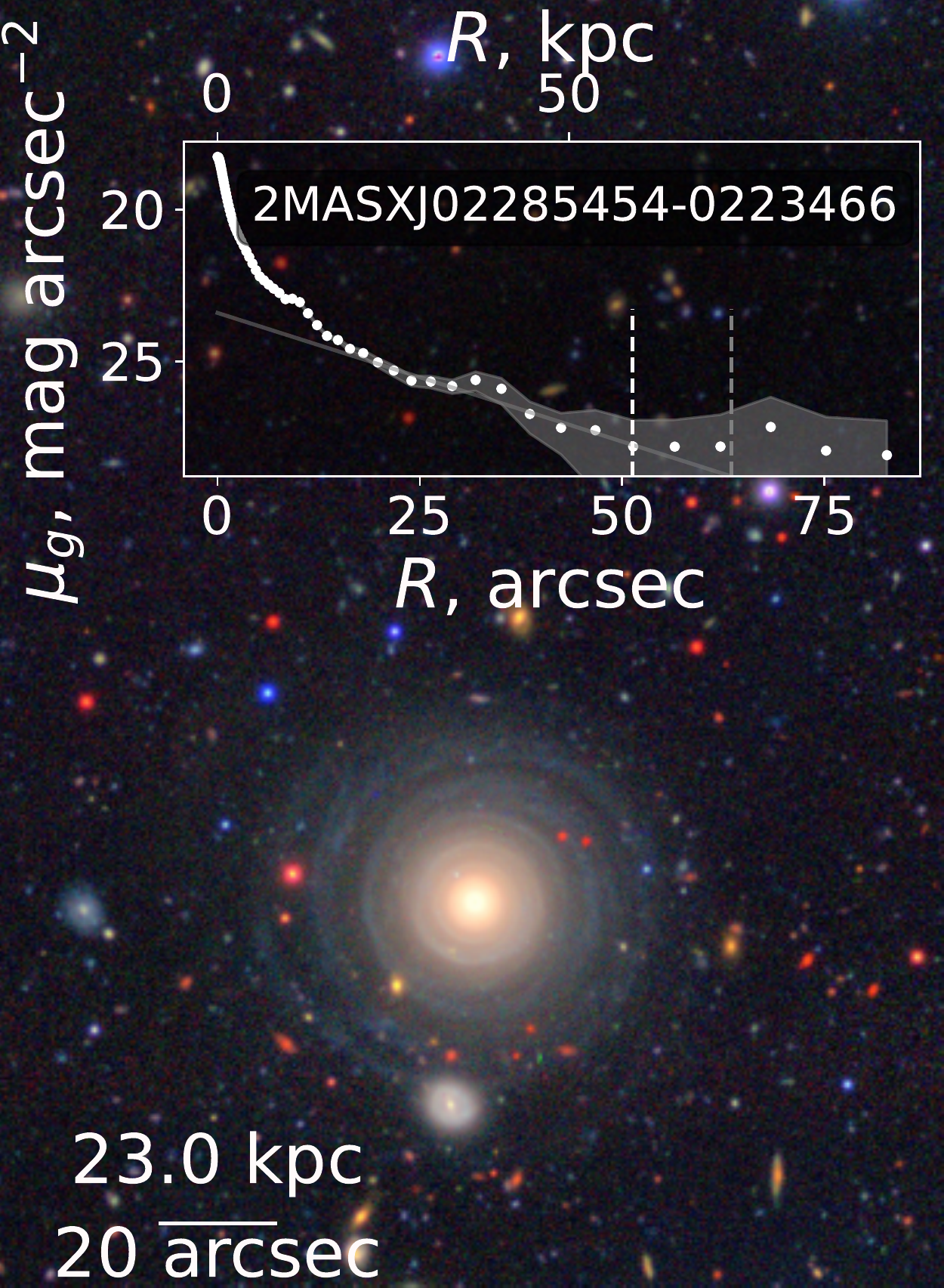}
\includegraphics[width=0.21\hsize]{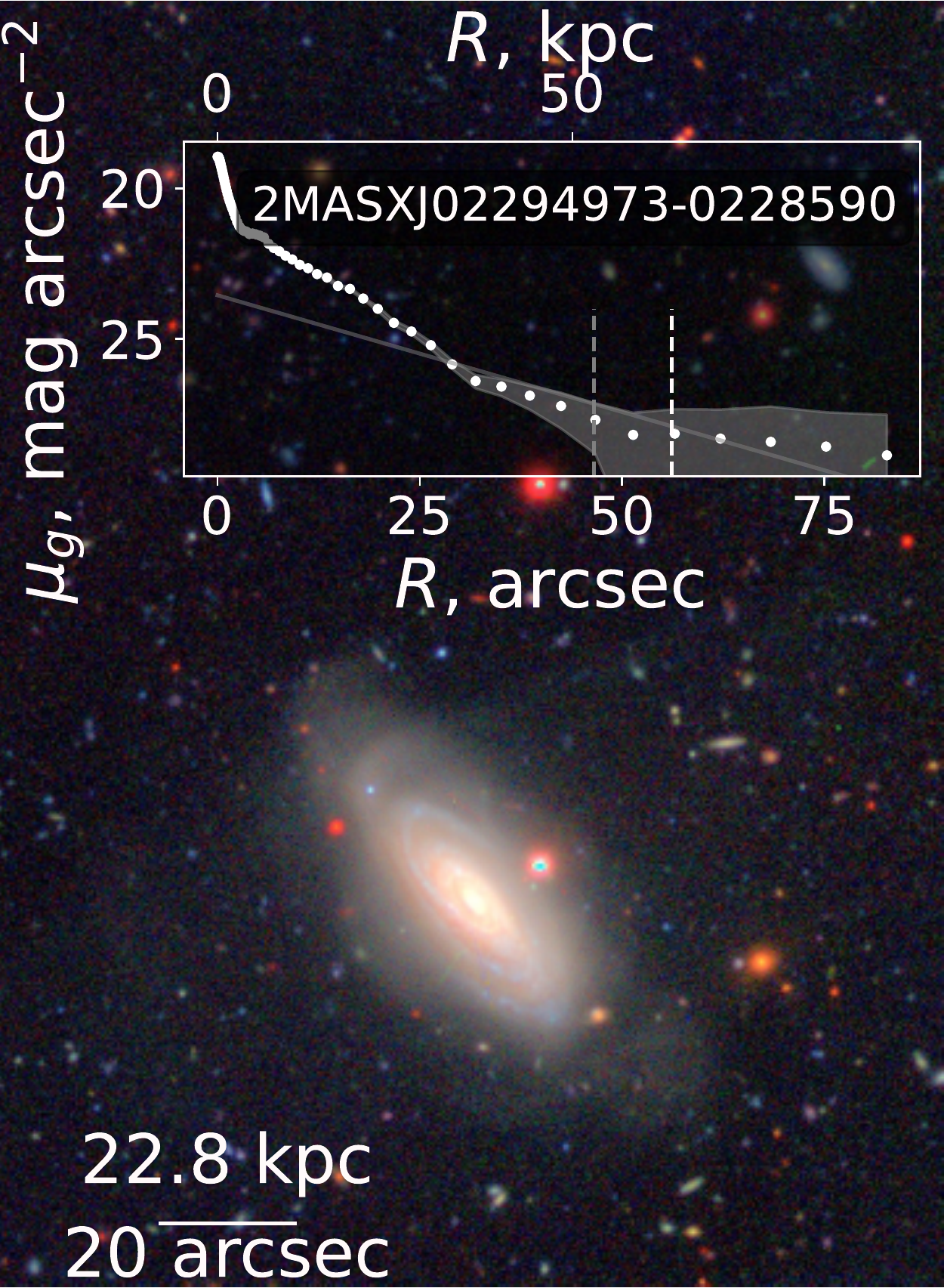}
\includegraphics[width=0.21\hsize]{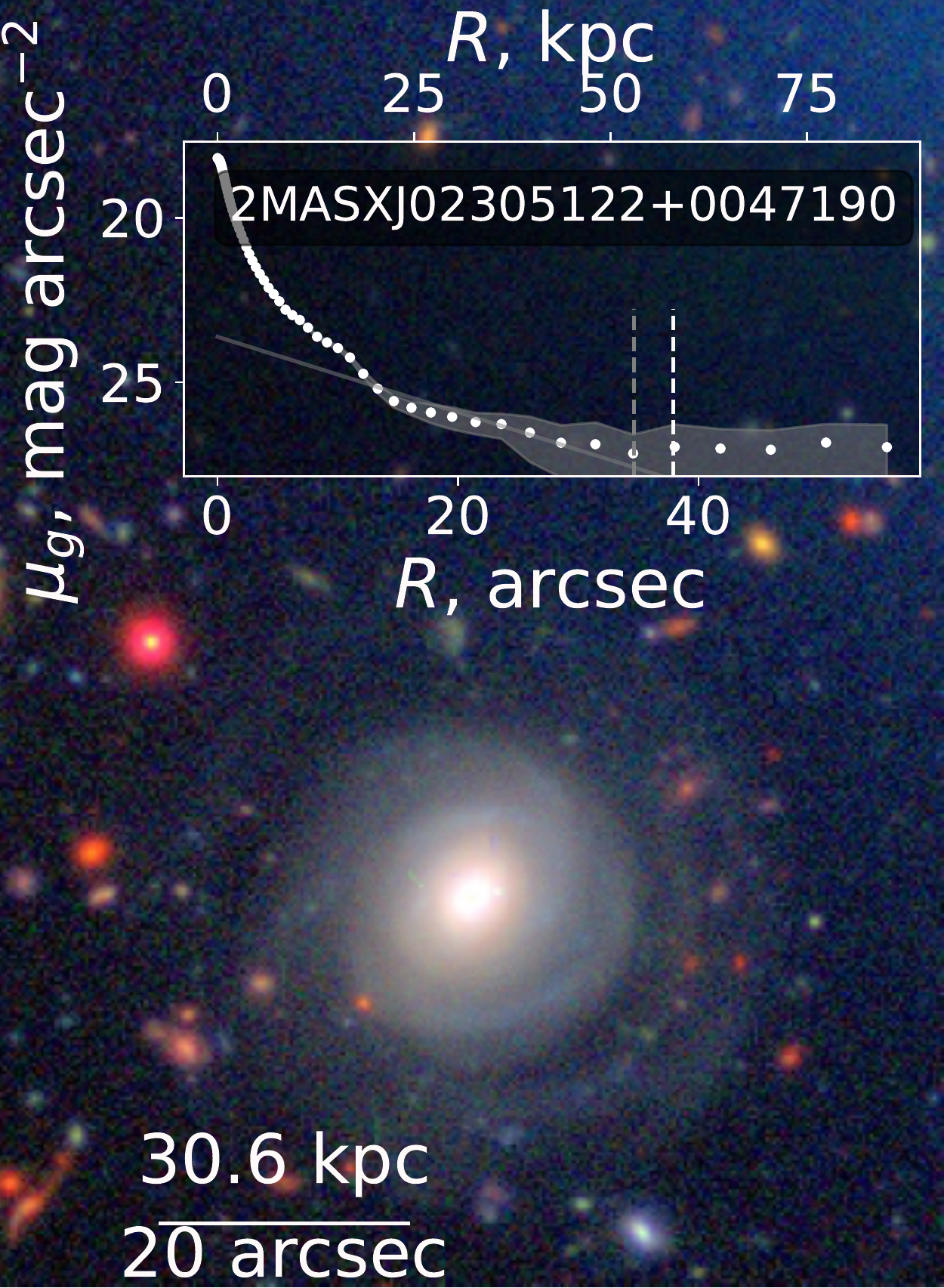}
\includegraphics[width=0.21\hsize]{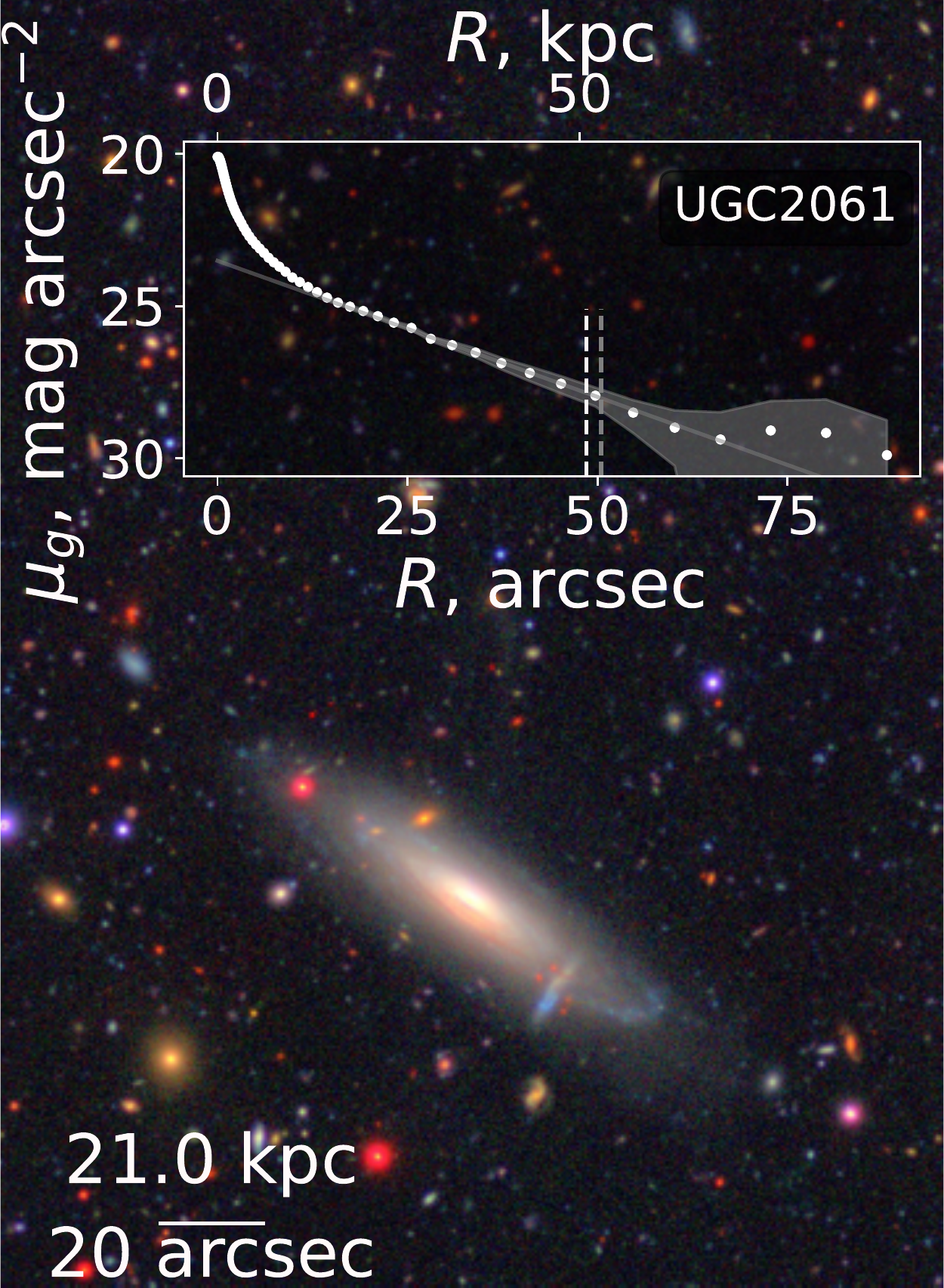}
\includegraphics[width=0.21\hsize]{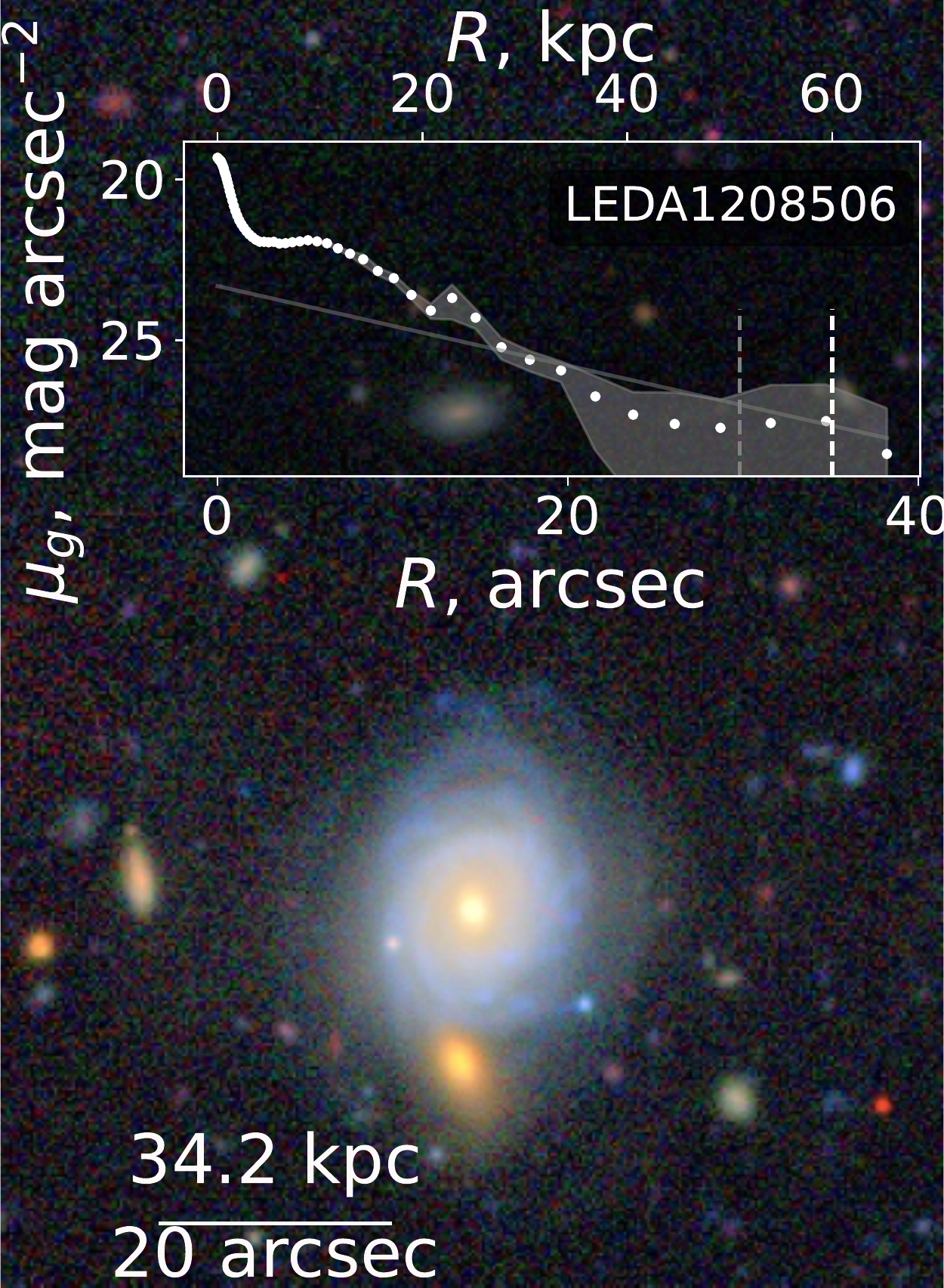}

\caption*{ continued}
\end{figure*}
\begin{figure*}
\centering

\includegraphics[width=0.21\hsize]{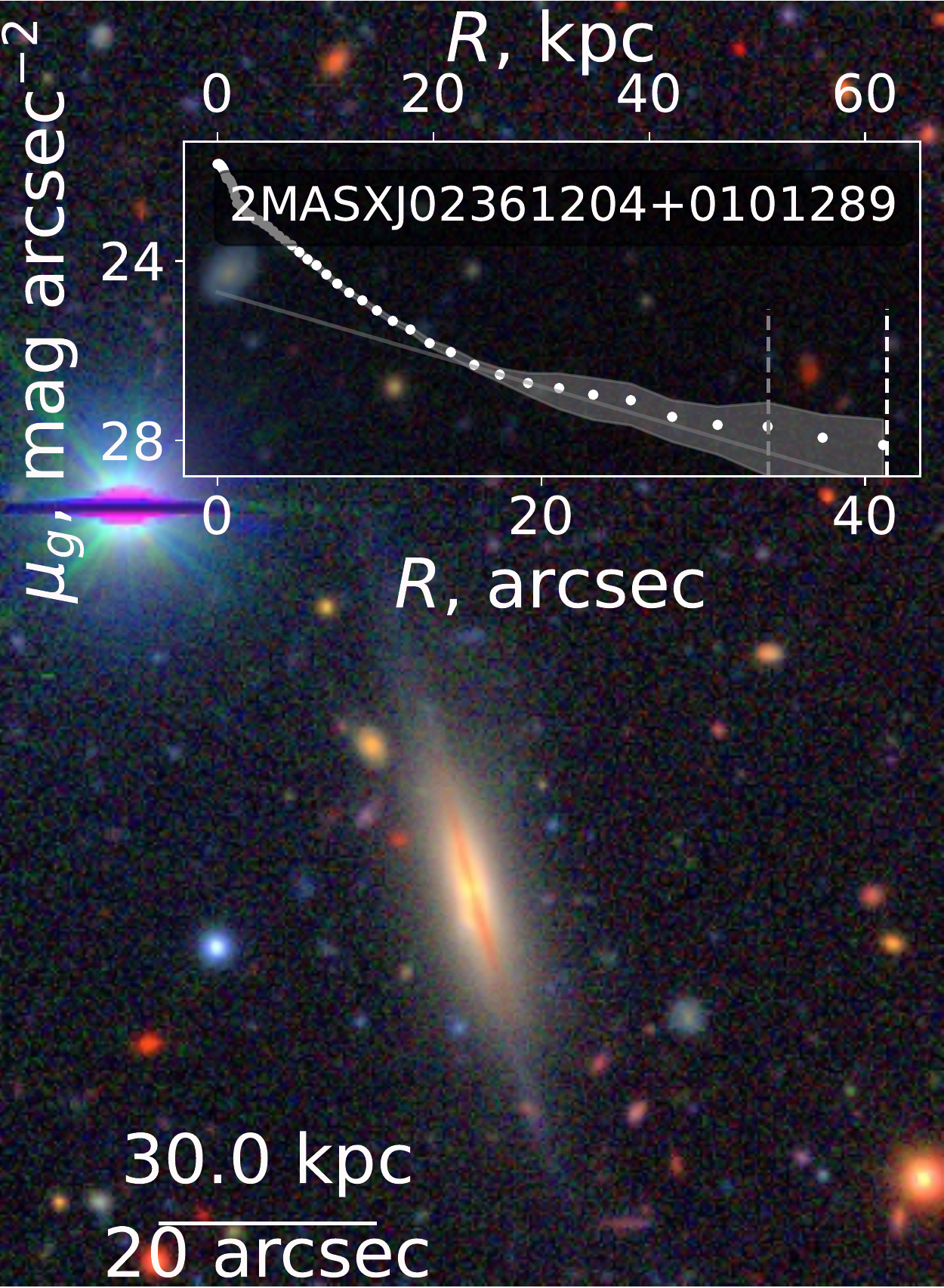}
\includegraphics[width=0.21\hsize]{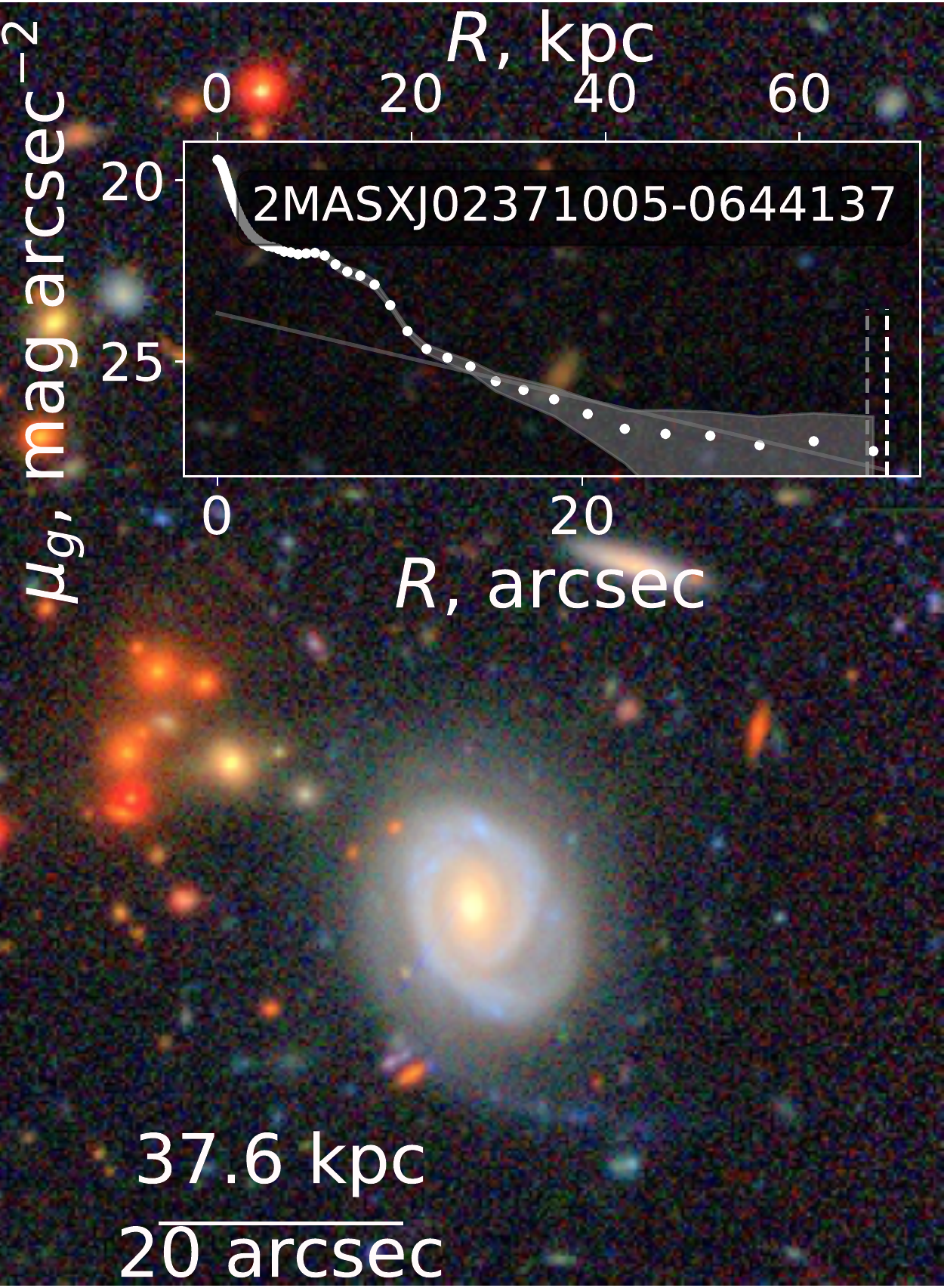}
\includegraphics[width=0.21\hsize]{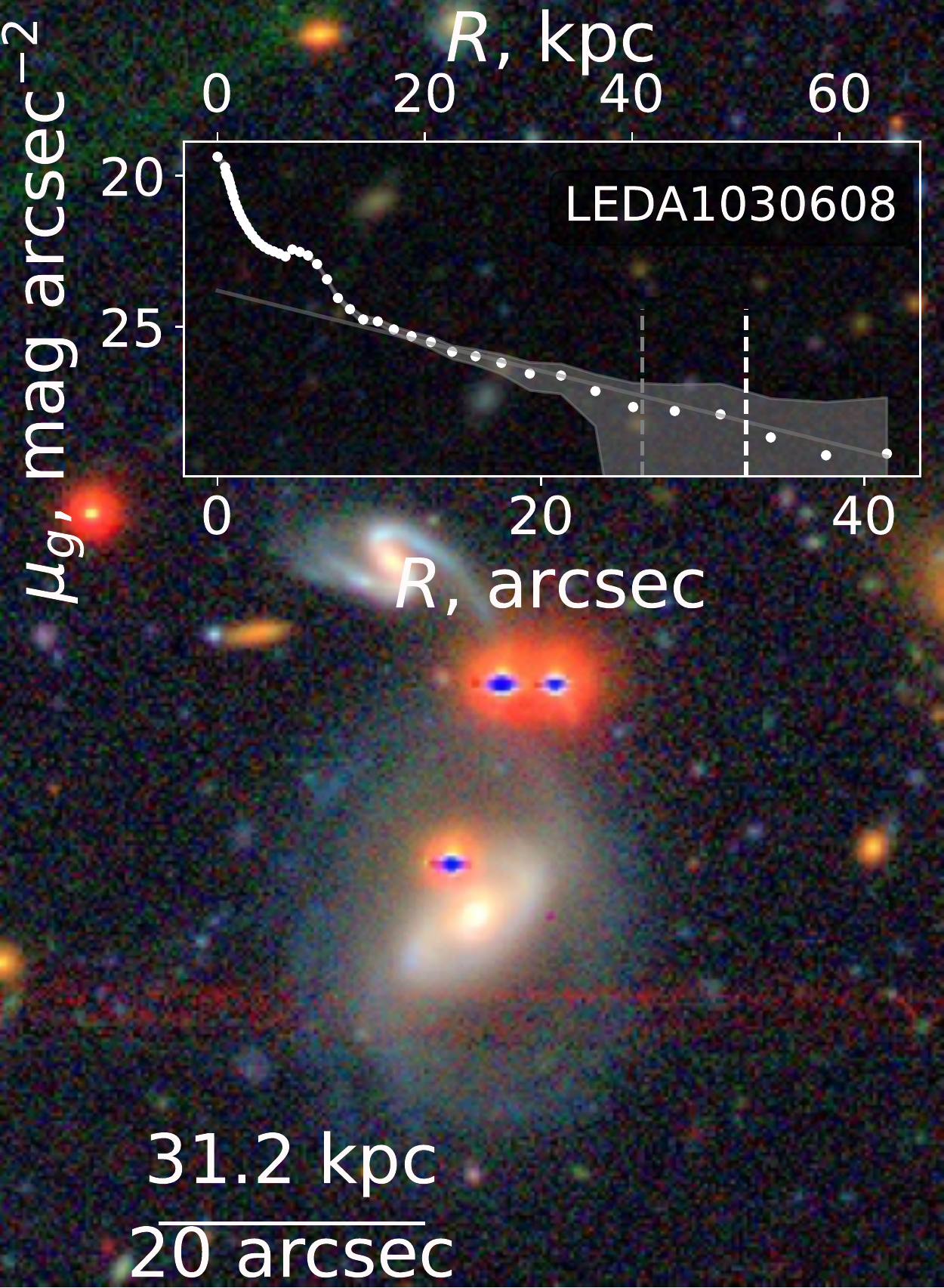}
\caption*{ continued}
\end{figure*}

\begin{figure*}
\includegraphics[width=0.195\hsize]{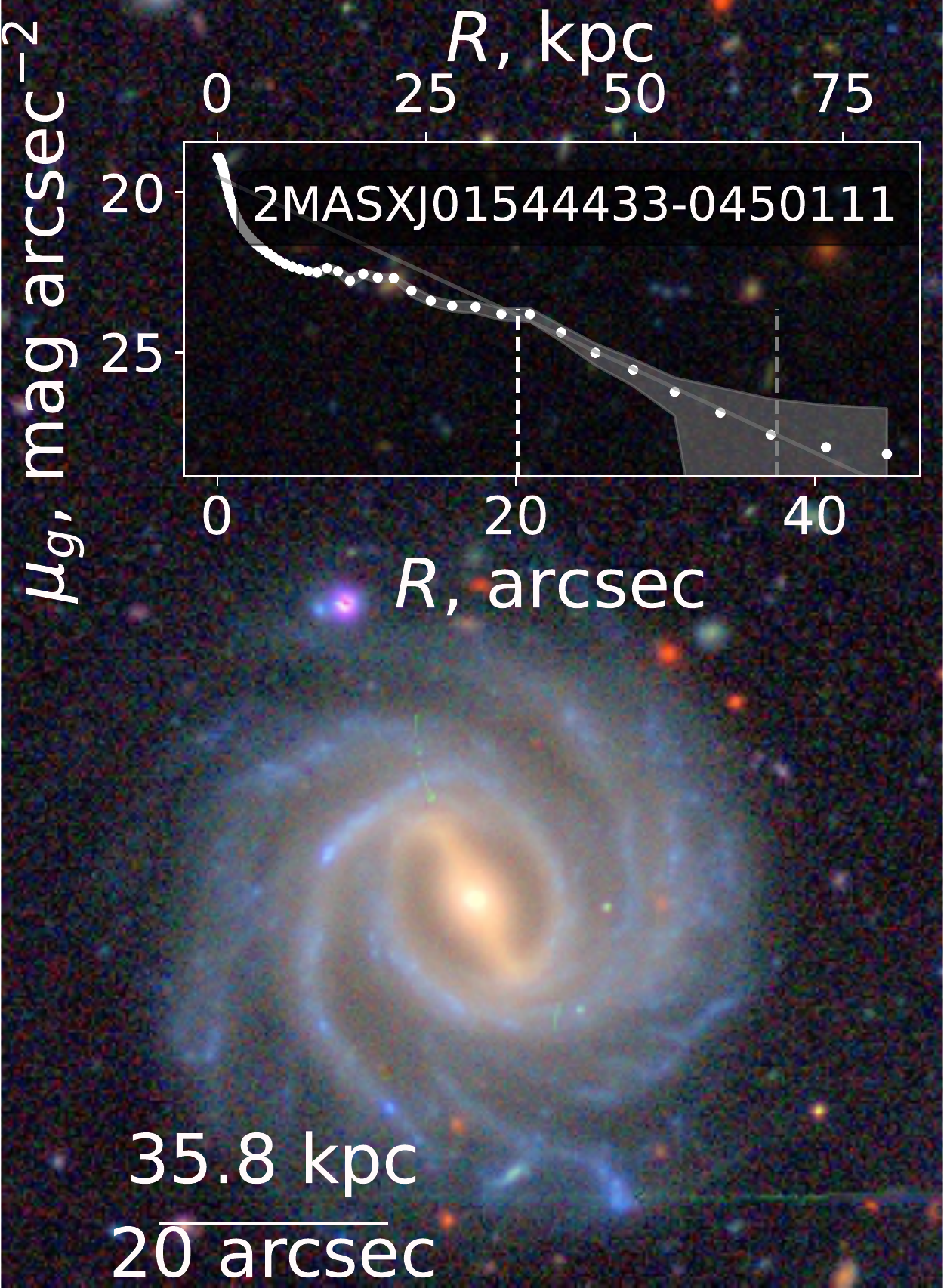}
\includegraphics[width=0.195\hsize]{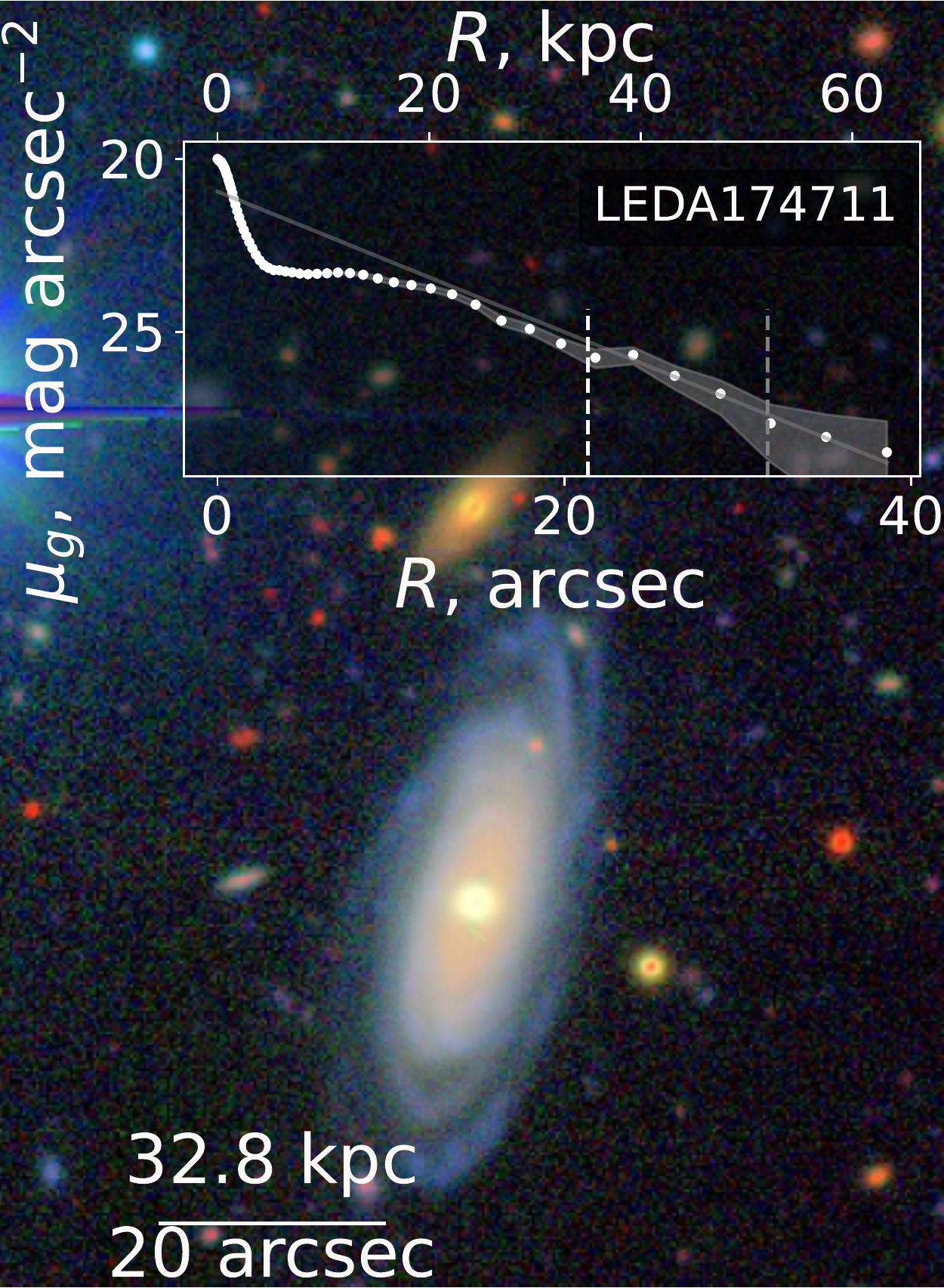}
\includegraphics[width=0.195\hsize]{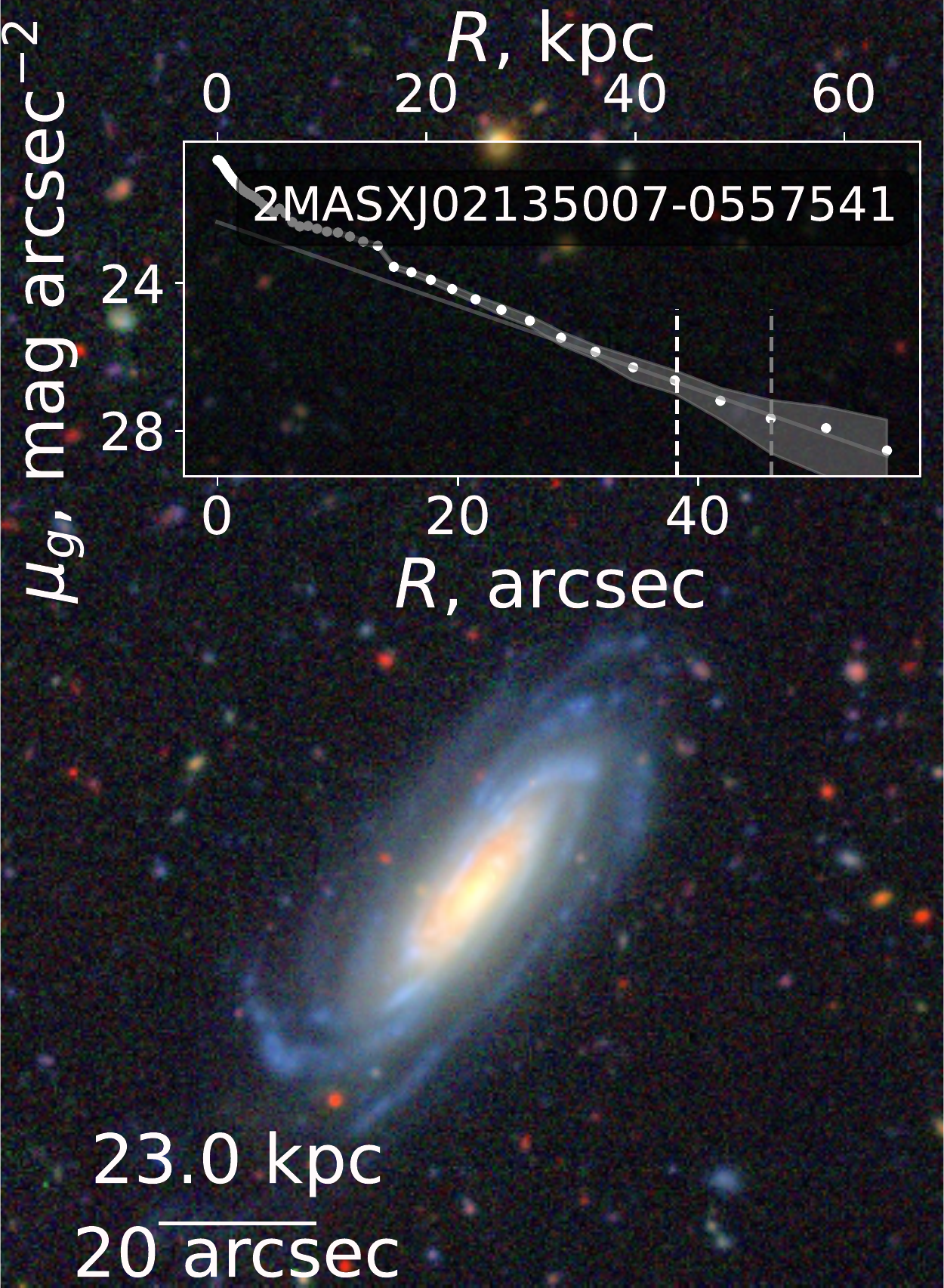}
\includegraphics[width=0.195\hsize]{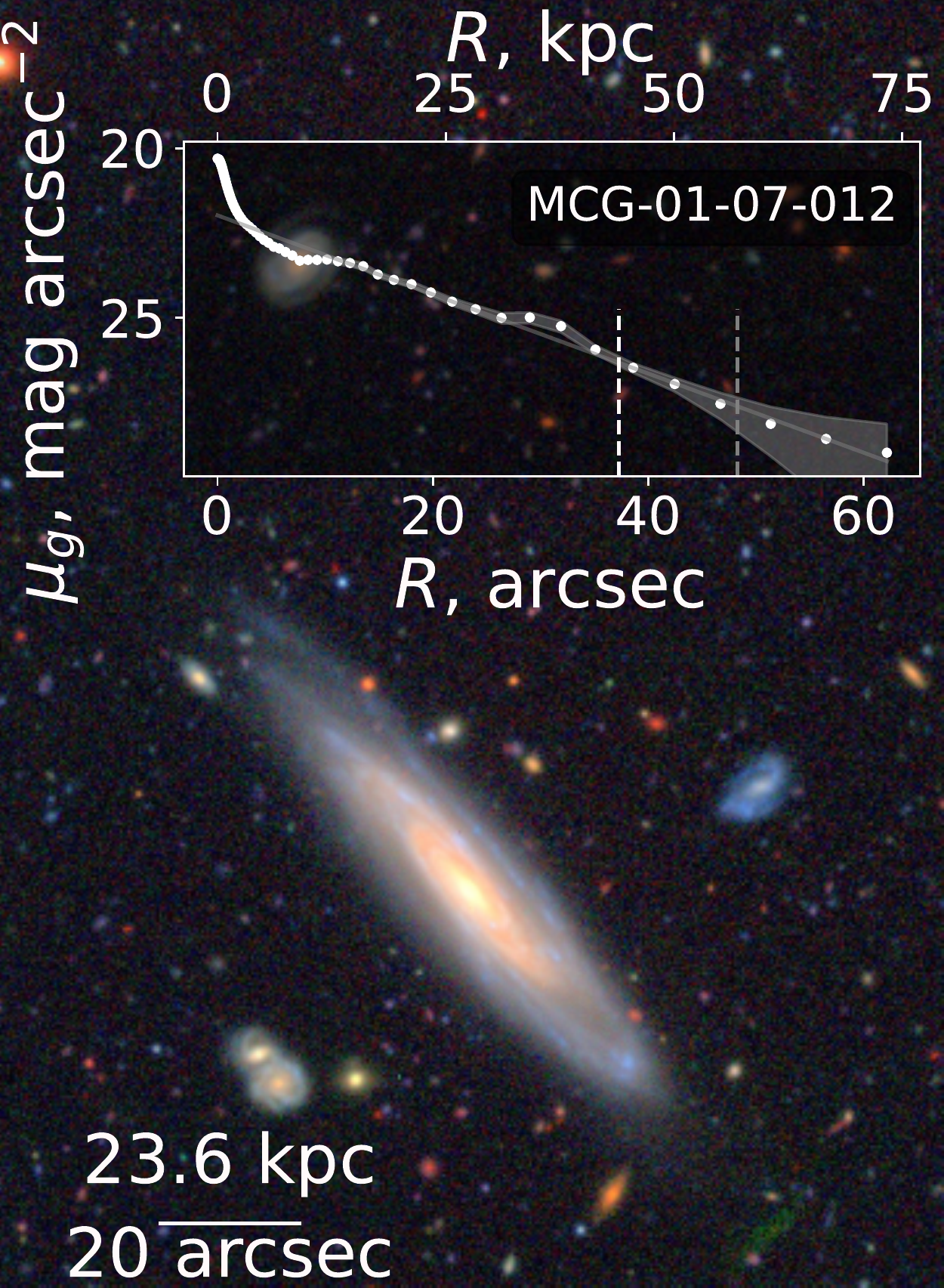}

\caption{The colour composite HSC images and light profiles of the giant HSB disky galaxies found in this study.}
\label{fig_examples_hsb}
\end{figure*}
\begin{table*}
\caption{The giant disc galaxies discovered in the HSC imaging data by visual inspection. The top part of the table contains 37 gLSBGs and the bottom part gives the information on the 5 remaining giant HSB discs.\label{tab_sample_all}}
\begin{center}
\begin{tabular}{llllllllllll} 
 \hline
 Galaxy& R.A.&  Dec.& $z$& Type& Scale& $R_{g=27.7}$& $\mu_{0,g}$& 4$h$ & $g-r$ & $M_g$&b/a\\
& ICRS, deg& ICRS, deg& & & kpc~arcsec$^{-1}$& kpc& mag~arcsec$^{-2}$& kpc & mag & mag& \\ 
 \hline 
  UGC1382&28.6709&-0.145&0.019& E               &0.4&50&24.06&56&0.80&-20.41&0.77\\
  2MASXJ01550099+0004519&28.7543&0.08107&0.083& E?          &1.8&44&24.12&57&0.61&-20.38&0.58\\
  2MASXJ01584646-0533277&29.6938&-5.5579&0.099&             &2.19&74&23.83&93&0.68&-21.78&0.88\\
  2MASXJ01590693+0115288&29.7788&1.258&0.079& Sab            &1.73&69&23.72&93&0.61&-21.65&0.40\\
  2MASXJ01593468-0249365&29.89458&-2.82681&0.039& Sab             &0.83&42&23.86&57&0.57&-20.93&0.46\\
  WISEA J020019.56-012210.9&30.0814&-1.3701&0.04&&0.84&55&23.13&51&0.81&-21.19&0.81\\
  Mrk584&30.10958&2.66938&0.079&                 &1.71&61&23.03&59&0.46&-22.67&0.97\\
  2MASXJ02015377+0131087&30.4739&1.5186&0.041&                 &0.85&50&23.85&66&0.69&-21.09&0.77\\
  2MASXJ02024375-0633453&30.6824&-6.5626&0.056&                 &1.21&53&23.48&54&0.75&-21.19&0.87\\
  2MASXJ02041915+0020395&31.08&0.3439&0.079& Sab             &1.71&53&23.14&50&0.56&-21.33&0.74\\
  LEDA1144613&31.87217&-0.38683&0.065& S?              &1.41&36&24.44&52&0.59&-20.31&0.34\\
  MCG-01-06-068&32.945&-5.9156&0.043& Sb              &0.92&66&23.35&66&0.67&-21.67&0.83\\
    UGC1697&33.08504&-2.14156&0.038& Sb              &0.79&53&23.12&52&0.80&-21.75&0.42\\
  2MASXJ02132915-0539092&33.37125&-5.65244&0.086&           &1.88&82&24.01&88&0.67&-21.32&0.52\\
  LEDA1084748&33.92729&-2.70933&0.077&                 &1.67&40&23.58&50&&&0.47\\
  LEDA1043747&34.21897&-5.51539&0.077&                 &1.67&43&23.38&51&0.67&-21.16&0.66\\
  2MASXJ02171125-0421194&34.29669&-4.35543&0.056& S0-a            &1.2&79&23.87&85&0.74&-21.35&0.71\\
  Mrk592&34.9214&-0.2557&0.026& Sb              &0.54&42&24.57&59&0.52&-20.75&0.50\\
  2MASXJ02200922-0542388&35.0383&-5.7109&0.091&                 &2.0&70&23.59&73&0.44&-21.35&0.87\\
  2MASSJ02215412-0440223&35.47534&-4.67313&0.081&&1.78&43&24.32&71&&&0.43\\
LEDA1044131&35.78158&-5.48433&0.075&                 &1.63&42&24.03&52&0.69&-20.45&0.28\\
  2MASXJ02240162-0234269&36.0065&-2.574&0.058&                 &1.24&70&23.31&69&0.62&-21.88&0.81\\
  2MASXJ02242997-0436135&36.125&-4.60353&0.069&                 &1.49&58&23.52&63&0.53&-21.84&0.80\\
  2MASXJ02243830-0041515&36.15946&-0.69766&0.097& E?              &2.14&61&23.97&63&0.55&-20.91&0.52\\
  UGC1876&36.2729&-0.6042&0.041& S0-a            &0.87&36&23.80&50&0.74&-20.53&0.53\\
    2MASXJ02254456-0546433&36.43575&-5.77867&0.053& Sab             &1.14&54&23.3&54&0.68&-21.47&0.36\\
     2MASXJ02272735-0438584&36.86392&-4.64953&0.071&                 &1.53&59&22.74&51&0.74&-21.39&0.90\\
  2MASXJ02274954-0526053&36.9565&-5.4348&0.046&                 &0.98&52&23.99&60&0.75&-20.35&0.71\\
  2MASXJ02285454-0223466&37.22731&-2.39615&0.054&                 &1.15&73&23.40&59&0.78&-21.35&0.96\\
  2MASXJ02294973-0228590&37.45721&-2.48306&0.053& Sab             &1.14&53&23.55&64&0.70&-21.68&0.55\\
  2MASXJ02305122+0047190&37.7135&0.7886&0.071& E               &1.53&53&23.62&58&0.64&-21.17&0.98\\
  LEDA1044960&38.0983&-5.4201&0.046&                 &0.98&50&24.97&73&0.75&-19.58&0.55\\
    UGC2061&38.6571&-0.9799&0.049& Sab             &1.05&53&23.51&51&0.76&-21.08&0.28\\
  LEDA1208506&39.02338&1.868&0.079&                 &1.71&51&23.32&60&0.55&-21.74&0.79\\
  2MASXJ02361204+0101289&39.0501&1.0242&0.069& E?              &1.5&51&24.70&62&0.81&-19.91&0.30\\
  2MASXJ02371005-0644137&39.29192&-6.73719&0.086&                 &1.88&67&23.67&69&0.60&-21.68&0.75\\
  LEDA1030608&39.35579&-6.63207&0.072&                 &1.56&41&23.80&51&0.68&-20.61&0.61\\
\hline
  2MASXJ01544433-0450111&28.68475&-4.83639&0.082&                 &1.79&67&19.51&36&0.59&-22.28&0.98\\
  LEDA174711&30.80108&1.93714&0.075&                 &1.64&52&20.94&35&0.61&-21.45&0.35\\
  2MASXJ02135007-0557541&33.4587&-5.9656&0.054&                 &1.15&53&22.35&44&0.61&-21.51&0.38\\
  MCG-01-07-012&36.7888&-2.8726&0.055& Sbc             &1.18&57&21.98&44&0.68&-21.54&0.23\\
  UGC2010&38.0297&-1.36213&0.038& SBbc            &0.79&53&21.91&42&0.62&-21.95&0.64\\
\hline 
 \end{tabular} 
 \end{center} 
 \end{table*}

\begin{table*}
\caption{ AGN (optical and X-ray selection) and the environment of giant LSB/HSB galaxies\label{agnenv_all}}
 \begin{center} 
 \begin{tabular}{lccl}
 \hline
Galaxy& BPT& $L_X$ (0.2--10~keV), $10^{40}$erg/s & Environment\\ 
 \hline 
  UGC1382& AGN (SDSS) & 0.773 $\pm$ 0.155 (XMM) &central galaxy in a group; close pair\\
  2MASXJ01550099+0004519& SF (SDSS) & <18.08 (XMM)&in a group of 8 galaxies,  companion in 216 kpc, $\Delta V=330$ \kms\\
  2MASXJ01584646-0533277& ? (6dF) & <671.7 (ROSAT)& isolated \\
  2MASXJ01590693+0115288& Composite (SDSS) & <428.4 (ROSAT) &companion at 100 kpc\\
  2MASXJ01593468-0249365& n/a & <121.8 (ROSAT)&two companions at 250 kpc\\
  WISEA J020019.56-012210.9& n/a & <807.1 (XMM)&companion at 150 kpc $\Delta V=500$ \kms\\
  Mrk584& n/a (Sy1.8) & 13720 $\pm$ 290 (SWIFT) &isolated\\
  2MASXJ02015377+0131087& n/a &<34.01 (SWIFT) &isolated \\
  2MASXJ02024375-0633453& AGN (6dF) & 23.32 $\pm$ 8.89 (XMM) & isolated \\
  2MASXJ02041915+0020395 & SF (SDSS) & <115.1 (SWIFT) &two companions in 400 kpc\\
  LEDA1144613& SF (SDSS) & <331.2 (ROSAT)&isolated within 500 kpc\\
  MCG-01-06-068& ? (FAST/6dF)  & <7.681 (XMM)&isolated within 500 kpc\\
   UGC1697&AGN (6dF) &6.763 $\pm$ 3.033 (SWIFT) &isolated within 500 kpc\\
  2MASXJ02132915-0539092& AGN (GAMA) &427.4 $\pm$ 64.4 (XMM) & isolated\\
  LEDA1084748 & n/a & <279.7 (ROSAT) &isolated\\
  LEDA1043747 & AGN (GAMA) & <9.815 (XMM) &isolated\\
  2MASXJ02171125-0421194 & ? (6dF) & 9.241 $\pm$ 4.621 (XMM) &isolated\\
  Mrk592& SF (SDSS) & <48.57 (ROSAT) &isolated within 500 kpc\\
  2MASXJ02200922-0542388& SF (GAMA) & <51.36 (XMM)&3 companions within 500 kpc\\
  2MASSJ02215412-0440223& AGN (GAMA) & <14.22 (XMM) &companion at 370 kpc\\
LEDA1044131& ? (GAMA) & <14.05 (XMM) &isolated\\
  2MASXJ02240162-0234269& ? (6dF) & <568.2 (XMM) &isolated within 500 kpc\\
  2MASXJ02242997-0436135& AGN Type-I (6dF/GAMA) &124.2 $\pm$ 83.6 (XMM) &isolated\\
  2MASXJ02243830-0041515& composite (SDSS) & <2419 (XMM) &isolated\\
  UGC1876& AGN Sy2 (SDSS/6dF) & <550.8 (XMM) &isolated\\
   2MASXJ02254456-0546433&SF (6dF) &8.764 $\pm$ 4.799(XMM) &in group of 2 galaxies,  companions in 270, 340 kpc\\
    2MASXJ02272735-0438584& Composite (6dF) & <9.042 (XMM) &companion at 180 kpc\\
  2MASXJ02274954-0526053& ? (6dF) & <4.469 (XMM) &isolated\\
  2MASXJ02285454-0223466& ? (6dF) & <223.5 (XMM) &in a group of 4 galaxies\\
  2MASXJ02294973-0228590& ? (6dF) & <289.0 (XMM) &isolated within 500 kpc\\
  2MASXJ02305122+0047190& Composite (SDSS) & <12.36 (XMM)&isolated within 500 kpc\\
  LEDA1044960&AGN (SDSS/GAMA) & <1.647 (XMM) &isolated \\
    UGC2061& Composite (SDSS/6dF) & <321.8 (ROSAT) &companion at 300 kpc\\
  LEDA1208506& Composite (LAMOST)& <438.0 (ROSAT)&isolated\\
  2MASXJ02361204+0101289& AGN (SDSS) & <272.8 (ROSAT) &isolated \\
  2MASXJ02371005-0644137& Composite (6dF) &<599.0 (ROSAT) &isolated (?) no redshift data for the adjacent galaxy\\
  LEDA1030608&Composite (SDSS) & <1247 (XMM) &isolated (?) no redshift data for the adjacent galaxy\\
\hline
  2MASXJ01544433-0450111& ? (6dF) & <215.8 (ROSAT)&isolated\\
  LEDA174711&Composite (LAMOST) & <1157 (XMM)&isolated\\
  2MASXJ02135007-0557541& ? (6dF) & 55.81 $\pm$ 13.51 (XMM) & isolated \\
  MCG-01-07-012& SF (SDSS) & <8.053 (XMM) & isolated \\
  UGC2010& AGN (6dF) & <595.9 (XMM) &isolated\\
\hline 
 \end{tabular} 
 \end{center} 
 \end{table*}
% Alternatively you could enter them by hand, like this:
% This method is tedious and prone to error if you have lots of references
%\begin{thebibliography}{99}
%\bibitem[\protect\citeauthoryear{Author}{2012}]{Author2012}
%Author A.~N., 2013, Journal of Improbable Astronomy, 1, 1
%\bibitem[\protect\citeauthoryear{Others}{2013}]{Others2013}
%Others S., 2012, Journal of Interesting Stuff, 17, 198
%\end{thebibliography}

%%%%%%%%%%%%%%%%%%%%%%%%%%%%%%%%%%%%%%%%%%%%%%%%%%

%%%%%%%%%%%%%%%%% APPENDICES %%%%%%%%%%%%%%%%%%%%%

%\section{Some extra material}

%If you want to present additional material which would interrupt the flow of the main paper,
%it can be placed in an Appendix which appears after the list of references.

%%%%%%%%%%%%%%%%%%%%%%%%%%%%%%%%%%%%%%%%%%%%%%%%%%

% Don't change these lines
%\bsp	% typesetting comment
\label{lastpage}
\end{document}